\RequirePackage{ifpdf}
\documentclass[letterpaper]{JHEP3}
\usepackage{epsfig}
\usepackage{comment}
\usepackage{amsmath}

\def\ba{\begin{eqnarray}}
\def\ea{\end{eqnarray}}
\def\be{\begin{equation}}
\def\ee{\end{equation}}
\def\nn{\nonumber}
\def\exd{{\rm d}}

\makeatletter
\def\x@arrow{\DOTSB\Relbar}
\def\xlongequalsignfill@{\arrowfill@\x@arrow\Relbar\x@arrow}
\newcommand{\xlongequal}[2]{%
    \ext@arrow 0099\xlongequalsignfill@{#1}{#2}}
\makeatother

\newcommand{\roughly}[1]{\mathrel{\raise.3ex\hbox{$#1$\kern-0.85em
\lower1ex\hbox{$\sim$}}}}

\newcommand{\gsim}{\roughly>}

\def\nott#1{\setbox0=\hbox{$#1$}                
   \dimen0=\wd0                                 
   \setbox1=\hbox{/} \dimen1=\wd1               
   \ifdim\dimen0>\dimen1                        
      \rlap{\hbox to \dimen0{\hfil/\hfil}}      
      #1                                        
   \else                                        
      \rlap{\hbox to \dimen1{\hfil$#1$\hfil}}   
      /                                         
   \fi}                                         %

\def\endignore{}
\def\ignore #1\endignore{} 

\def\be{\begin{equation}}
\def\beq\begin{equation}
\def\ee{\end{equation}}
\def\bea{\begin{eqnarray}}
\def\eea{\end{eqnarray}}

\def\eqa{\begin{eqnarray}}
\def\eeqa{\end{eqnarray}}
\def\eq{\begin{equation}}
\def\eeq{\end{equation}}

\def\nn{\nonumber}

\def\pref#1{(\ref{#1})}

\def\ol#1{\overline{#1}}

\def\exd{{\rm d}}
\def\nn{\nonumber}
\def\pref#1{(\ref{#1})}
\def\be{\begin{equation}}
\def\ee{\end{equation}}

\def\beq{\begin{equation}}
\def\eeq{\end{equation}}
\def\beqa{\begin{eqnarray}}
\def\eeqa{\end{eqnarray}}

\def\bfp{{\bf p}}

\def\mfa{\mathfrak{a}}
\def\mfg{\mathfrak{g}}
\def\mfh{\mathfrak{h}}
\def\mfs{\mathfrak{s}}
\def\mfK{\mathfrak{K}}
\def\mfF{\mathfrak{F}}

\def\cA{{\cal A}}

\def\cD{{\cal D}}

\def\cF{{\cal F}}
\def\cG{{\cal G}}
\def\cH{{\cal H}}

\def\cL{{\cal L}}

\def\cO{{\cal O}}

\def\cR{{\cal R}}

\def\cT{{\cal T}}

\def\ssA{{\scriptscriptstyle A}}
\def\ssB{{\scriptscriptstyle B}}

\def\ssD{{\scriptscriptstyle D}}
\def\ssE{{\scriptscriptstyle E}}
\def\ssF{{\scriptscriptstyle F}}

\def\ssH{{\scriptscriptstyle H}}
\def\ssI{{\scriptscriptstyle I}}

\def\ssL{{\scriptscriptstyle L}}
\def\ssM{{\scriptscriptstyle M}}
\def\ssN{{\scriptscriptstyle N}}
\def\ssP{{\scriptscriptstyle P}}
\def\ssQ{{\scriptscriptstyle Q}}
\def\ssR{{\scriptscriptstyle R}}
\def\ssS{{\scriptscriptstyle S}}
\def\ssT{{\scriptscriptstyle T}}

\def\ssW{{\scriptscriptstyle W}}
\def\ssX{{\scriptscriptstyle X}}

\def\DE{{\scriptscriptstyle DE}}

\def\UV{{\scriptscriptstyle UV}}

\def\SM{{\scriptscriptstyle SM}}

\def\QCD{{\scriptscriptstyle QCD}}

\def\EW{{\scriptscriptstyle EW}}

\newcommand{\bmat}{\left(\begin{array}}
\newcommand{\emat}{\end{array}\right)}

\def\-{\hphantom{-}}

\def\s2{\frac{1}{2}}

\def\IF{\relax{\rm I\kern-.18em F}}
\def\II{\relax{\rm I\kern-.18em I}}
\def\IP{\relax{\rm I\kern-.18em P}}
\def\IC{\relax{\rm I\kern-.48em C}}
\def\IR{\relax{\rm I\kern-.18em R}}
\def\IK{\relax{\rm I\kern-.20em K}}
\def\IM{\relax{\rm I\kern-.25em M}}

\def\GN{G_\ssN}

\def\GREFT{{\scriptscriptstyle GREFT}}
\def\EH{{\scriptscriptstyle EH}}

\def\y2{Y_{\ssM\ssN} Y^{\ssM\ssN}}
\def\Riem2{R_{\ssA\ssB\ssM\ssN} R^{\ssA\ssB\ssM\ssN}}
\def\Ricci2{R_{\ssM\ssN} R^{\ssM\ssN}}

\def\f2{F^{a}_{\ssM\ssN} F^{\ssM\ssN}_a}

\def\Asl{\hbox{/\kern-.7500em\it A}} 
\def\dsl{\hbox{/\kern-.5500em$\partial$}}
\def\pxpsl{\hbox{/\kern-.5600em$p$}}
\def\Dsl{\,\raise.15ex\hbox{/}\mkern-13.5mu D}
\def \one{\relax{\rm 1\kern-.26em I}}

\def\exd{{\rm d}}

\def\nn{\nonumber}

\def\({\left(}
\def\){\right)}

\title{Dark Energy and the Symbiosis\\ Between Micro-physics and Cosmology (Naturally)\footnote{Lectures given to the Les Houches Summer School ``Dark Universe,'' 7 July - 1 August 2025}}
 
\author{C.P. Burgess\\

Department of Physics \& Astronomy, McMaster University,\\
\qquad 1280 Main Street West, Hamilton, Ontario, Canada, L8S 4M1.\\
Perimeter Institute for Theoretical Physics,\\ \qquad 31 Caroline
Street North, Waterloo, Ontario, Canada, N2L 2Y5.\\
School of Theoretical Physics, Dublin Institute for Advanced Studies,\\ \qquad 10 Burlington Rd., Dublin, Co. Dublin, Ireland. }

\date{}

\abstract{These lectures aim to highlight the remarkable symbiosis that currently exists between the physics of the very small and the physics of the very large, using the unsolved puzzle of the nature of Dark Energy as a vehicle for so doing. The lectures first summarize what we know observationally about the properties of Dark Energy (and the Dark sector more broadly) and then discuss several approaches to explain them. Along the way this involves determining the types of interactions that would on general grounds be expected to be present in the low-energy limit of fundamental theories involving the many hierarchy of scales we see around us. This includes (but is not limited to) a discussion of technical naturalness (and `t Hooft naturalness) as well as the arguments for their use as a criterion for distinguishing amongst candidate theories. Some recent approaches I find promising are briefly summarized at the end.}

\begin{document}


\section{The Facts in the Sky}
\label{sec:Overview}

These notes use a discussion of Dark Energy as a vehicle for illustrating the peculiarly effective symbiosis that currently exists between our understandings of physics at the smallest and largest scales. This symbiosis is peculiar because it seems to fly in the face of an important fact of Nature: decoupling. 

Decoupling states that details of small-distance physics tend not to be important for understanding long-distance physics. This is indeed partially why science makes progress at all -- although Nature comes to us with many scales we are not required to understand them all at once. This is why it was possible to figure out how atoms work before also understanding the nature of the atomic nucleus. It turns out that atomic physics mostly depends only on a few nuclear properties -- its charge and mass and spin, for example -- but not on the rest of the nuclear nitty gritty. This is also why it is not that surprising that the Standard Model of particle physics gets right all of the details of (say) condensed matter physics or of quantum optics. {\it Any} theory of micro-physics that properly predicts Quantum Electrodynamics at low energies automatically gets all condensed matter and optical phenomena right for free. This is a good thing because it means our understanding of the properties of matter in bulk or of light in matter is robust to changes to our understanding of currently unknown microphysics. 

The situation is different in cosmology, where different micro-physical theories can differ radically in their cosmological implications and the observational success or failure of these implications are often used to constrain what might be possible at the shortest distances to which we have access. Perhaps even more interesting: many popular models in cosmology seem not to be obtainable from sensible micro-physics -- if true this might be a useful clue that allows us to choose amongst the very many models on the market.   

\subsection{Vanilla cosmology}

Let's start with a very brief recap of cosmology basics (a classic textbook for this is \cite{Peebles}). The vast majority of cosmological models start with the premise that the geometry of the Universe around us can be described by the classical solutions to Einstein equations of General Relativity (GR):\footnote{We denote spacetime coordinates by $x^\mu = \{ x^0, x^i \} = \{ x^0 = t, x^1 = x, x^2 = y, x^3 = z \}$ and choose the metric signature $(-+++)$ together with Weinberg's curvature conventions \cite{Wbg} (which differ from those of Misner, Thorne and Wheeler \cite{MTW} -- more commonly used in the relativity community -- only in the overall sign of the Riemann tensor.}
\be \label{GRDE}
  \cR_{\mu\nu} - \tfrac12 \, \cR \, g_{\mu\nu} + \kappa^2 T_{\mu\nu} = 0 \,.
\ee
Here $\cR_{\mu\nu}$ is the Ricci tensor built from the spacetime metric, $g_{\mu\nu}$, while $\cR = g^{\mu\nu} \cR_{\mu\nu}$ is its Ricci scalar and $T_{\mu\nu} = T_{\nu\mu}$ is the stress energy tensor of all of the forms of matter that are currently present (or were present in the past). The parameter $\kappa^2 = 8 \pi \GN$ denotes the gravitational coupling, where $\GN$ is Newton's constant of universal graviation. In fundamental units (for which $\hbar = c = 1$) its value defines the (reduced) Planck mass: $M_p = \kappa^{-1}$.  

The combination $\cG_{\mu\nu} =\cR_{\mu\nu} - \frac12 \, \cR \, g_{\mu\nu}$ satisfies a well-known {\it Bianchi identity}: $\nabla^\mu \cG_{\mu\nu} = 0$, where $\nabla_\mu$ is the covariant derivative built from the metric. Consistency requires that whatever the matter is that is present, its total stress energy must be covariantly conserved: $\nabla^\mu T_{\mu\nu} = 0$. In cosmology it often happens that the matter of interest is a homogeneous and isotropic fluid whose elements move through spacetime with 4-velocity $u^\mu(x)$. As is true for any 4-velocity, $u^\mu(x)$ must satisfy $g_{\mu\nu} u^\mu u^\nu = - 1$ and so the fluid rest frame is defined as the frame where the spatial components satisfy $u^i = 0$ (and so in this frame $u^0 = |g_{00}|^{-1/2}$). Denoting the fluid's rest-framepressure and energy density by $p$ and $\rho$ respectively, the fluid's stress-energy to be used in \pref{GRDE} is  
\be \label{FluidTmn}
   T_{\mu\nu} = p \, g_{\mu\nu} + (p + \rho) \, u_\mu u_\nu \qquad\qquad \hbox{(fluid)} \,.
\ee

In the special case where the universe is homogeneous and isotropic the spacetime metric can always be written in the Friedmann, LeMaitre, Robertson, Walker (FLRW) form 
\bea \label{FRWMetric}
    \exd s^2 &=& - \exd t^2 + a^2(t) \, \left[
    \frac{\exd r^2}{1 - \mfK
    r^2/R_0^2} + r^2 \, \exd\theta^2 + r^2 \sin^2\theta \,
    \exd \phi^2 \right] \\
    &=& - \exd t^2 + a^2(t) \, \left[
    \exd \ell^2 + r^2(\ell) \, \exd\theta^2 +
    r^2(\ell) \sin^2\theta \, \exd \phi^2 \right]
    \,, \nonumber
\eea
where $R_0$ is a constant and $\mfK$ can take one of the following three values: $\mfK = 1,0,-1$. The coordinate $\ell$ is related to $r$ by $\exd \ell = \exd r/(1 - \mfK r^2/R_0^2)^{1/2}$, and so
\be  \label{rvsell}
    r(\ell) = \left\{ \begin{matrix}
    R_0 \, \sin (\ell/R_0) & \hbox{if} \quad \mfK = +1 \\
    \ell & \hbox{if}\quad \mfK = 0 \\
    R_0 \, \sinh (\ell/R_0) & \hbox{if}\quad \mfK = -1 \,.
    \end{matrix}
    \right. 
\ee
The geometry at fixed $t$ is in this case a 3-sphere when $\mfK = 1$, flat space (when $\mfK = 0$) or a hyperbolic space ($\mfK = -1$). It is usually convenient to rescale $\ell \to R_0 \ell$ when $\mfK = \pm 1$. For $\mfK = 0$ it is convenient instead to rescale $\ell$ to ensure that the scale factor is unity at a particular time, $a(t_0) = 1$ (with the particular time chosen to be now). These rescalings amount to choosing convenient units of length  

With these choices the fundamental evolution equation \pref{GRDE} boils down to two independent differential equations relating $a(t)$ to $\rho(t)$ and $p(t)$. These may be chosen to be the {\sl Friedmann equation},
\be \label{FriedmannEqn}
    H^2 + \frac{\mfK}{a^2} =   \frac{8 \pi G}{3} \, \rho \,,
\ee
as well as the equation describing the {\sl Conservation of Stress-Energy} ($\nabla^\mu T_{\mu\nu} = 0$),
\be \label{EnergyConservationEqn}
    \dot{\rho} + 3 H
    (\rho + p) = 0 \,.
\ee
In these expressions over-dots denote differentiation with respect to $t$ and the Hubble function is defined by $H(t) = \dot a/a$. Equation \pref{EnergyConservationEqn} has an intuitive interpretation if it is rewritten $\exd ( \rho \, a^3 ) + p \, \exd (a^3) = 0$, which relates the rate of change of the total energy, $\rho\,  a^3$, to the work done by the pressure as the universe expands.  For a thermodynamic fluid this is consistent with the First Law of Thermodynamics when the evolution is at constant entropy.

Eqs.~\pref{FriedmannEqn} and \pref{EnergyConservationEqn} provide two differential equations for the three unknown functions $\rho(t)$, $p(t)$ and $a(t)$ and so can only be fully integrated after more information is provided. Typically this information comes from identifying the types of matter making up the fluid. Any specific type of fluid -- a gas of photons, for example, or nonrelativistic electrons -- has an equation of state: a relation relating $\rho$ to $p$. Once an equation of state is specified there is enough information to integrate eqs.~\pref{FriedmannEqn} and \pref{EnergyConservationEqn} to obtain the histories $a(t)$, $p(t)$ and $\rho(t)$.

For instance, if it happens that the equation of state has the commonly occuring form
\be \label{EqnofState}
    p = w \, \rho \,,
\ee
where $w$ is a $t$-independent constant then eq.~\pref{EnergyConservationEqn} integrates to give
\be \label{rhovsa}
    \rho = \rho_0 \left( \frac{a_0}{a} \right)^{\sigma}
    \qquad \hbox{with} \quad \sigma = 3(1+w) \,.
\ee
In the special case that $\mfK = 0$ this allows eq.~\pref{FriedmannEqn} to be integrated to give
\be \label{avst}
    a(t) = a_0 \left( \frac{t}{t_0} \right)^\alpha \qquad
    \hbox{with} \quad \alpha = \frac{2}{\sigma} = \frac{2}{3(1+w)}
    \,.
\ee

\subsubsection{$\Lambda$CDM}
\label{sssec:LCDM}

The core theory of Hot Big Bang cosmology postulates that all ordinary matter starts off in the remote past as a hot dense fluid, and then asks what evidence for this exists in the later universe. It turns out it does: as the universe expands it cools and bound states form as the temperature falls below the relevant binding energy. The formation of nuclei leads to the successful Big Bang nucleosynthesis prediction for the abundances of light elements; atom formation leads to the universe becoming transparent and the associated Cosmic Microwave Background (CMB) relic radiation, and so on. 

All told, a minimal successful description of cosmological observation requires four main types of components to the cosmic fluid, each of which (in the later universe at least) does not exchange energy with the others -- so their stress-energies are individually conserved and satisfy \pref{EnergyConservationEqn} separately.

\begin{itemize}
\item {\bf Radiation:} Photons (and neutrinos) are relativistic through (most of) the universe's history and so turn out to have a pressure-to-energy-density ratio of $w_{\rm rad} \simeq \frac13$ (we collectively call such species `radiation'). Eq.~\pref{rhovsa} then implies $\rho_{\rm rad}(a)/\rho_{{\rm rad}\,0} = (a_0/a)^4$.
\item {\bf Ordinary Matter (baryons):} Ordinary matter is nonrelativistic for much of the epoch to which we have observational access (electrons and neutrinos are the exception for earlier parts of the universal history). Since $p/\rho$ involves the ratio of some measure of particle kinetic energy (like temperature) over rest mass we have $w_b \simeq 0$ for nonrelativistic species and \pref{rhovsa} implies $\rho_b(a)/\rho_{b0} = (a_0/a)^3$. Because of electromagnetic interactions this fluid is only uncoupled from the radiation fluid in the relatively late universe (after neutral atoms are able to form).
\item {\bf Cold Dark Matter:} There is considerable evidence for the existence of another fluid that behaves gravitationally much like baryons do ({\it i.e.}~it clumps together in galaxies and clusters of galaxies due to gravitational attraction) but which does not otherwise interact with ordinary matter. If this fluid describes the bulk behaviour of a new type of matter then this matter must be moving slowly -- {\it i.e.}~be `cold' -- in order to clump sufficiently efficiently, and so is also well-described as a fluid with a nonrelativistic equation of state parameter $w_c \simeq 0$. As a result its density also falls in an expanding universe like $\rho_c(a)/\rho_{c0} = (a_0/a)^3$.
\item {\bf Vacuum Energy:} There is good evidence the vacuum is Lorentz invariant to high accuracy and so its stress energy tensor must be proportional to the metric:\footnote{The vacuum is usually meant as the lowest-energy state but there is no reason it has to have zero energy density, particularly for $\langle T_{\mu\nu} \rangle$ evaluated in a quantum vacuum state. More about quantum effects below.} 
\be \label{VacTmn}
T^{\mu\nu}_{\rm vac} = - \rho_{\rm vac} \, g^{\mu\nu} \,. 
\ee
Conservation of stress energy ($\nabla_\mu T_{\rm vac}^{\mu\nu} = 0$) then implies $\rho_{\rm vac}$ must be a constant. Because it is a constant it contributes to Einstein's equations \pref{GRDE} in the same way as would Einstein's cosmological constant term $\Lambda g_{\mu\nu}$. Comparing \pref{VacTmn} with the general fluid stress-energy \pref{FluidTmn} then shows that $p_{\rm vac} = - \rho_{\rm vac}$ and so the equation of state parameter is $w_{\rm vac} = -1$. In this case we have $\rho(a) = \rho_{{\rm vac}}$ is independent of $a$. Because either $\rho_{\rm vac}$ or $p_{\rm vac}$ must negative (observations say $\rho_{\rm vac} > 0$ and so it is $p_{\rm vac}$ that is negative) this is distinct from Dark Matter, and so is given its own name: Dark Energy.
\end{itemize}

The above fluids are part of the definition of the $\Lambda$CDM model of cosmology, and taken together they imply the relative abundance of the different fluid components changes as the universe expands (see Fig.~\ref{EnergyEvolutionFig}). In particular the total energy density and pressure have the form
\bea \label{MixedFluidrhoandp}
    \rho(a) &=& \rho_{\rm vac} + \rho_{{\rm m} 0} \left( \frac{a_0}{a}
    \right)^3 + \rho_{{\rm rad} \, 0} \left( \frac{a_0}{a} \right)^4 \nn\\
    p(a) &=& - \rho_{\rm vac} + \tfrac13 \, \rho_{{\rm rad}\, 0} \left(
    \frac{a_0}{a} \right)^4 \,,
\eea
if the energy exchange between fluids is neglible. Here $\rho_{{\rm m}0} := \rho_{b0} + \rho_{c0}$ sums the contributions of the two types of nonrelativistic fluids (which is only appropriate after the baryons have decoupled from the radiation). Using the above expression for $\rho(a)$ in the Friedmann equation \pref{FriedmannEqn} gives $H = \dot a/a$ as a function of $a$, which can be integrated to get $a(t)$. Comparing to \pref{avst} shows this implies in particular that $a \propto t^{1/2}$ when radiation dominates the energy density (`radiation domination') and $a \propto t^{2/3}$ when nonrelativistic matter dominates (`matter domination').

\FIGURE[tbh]{
\begin{tabular}{ll}
 \epsfig{file=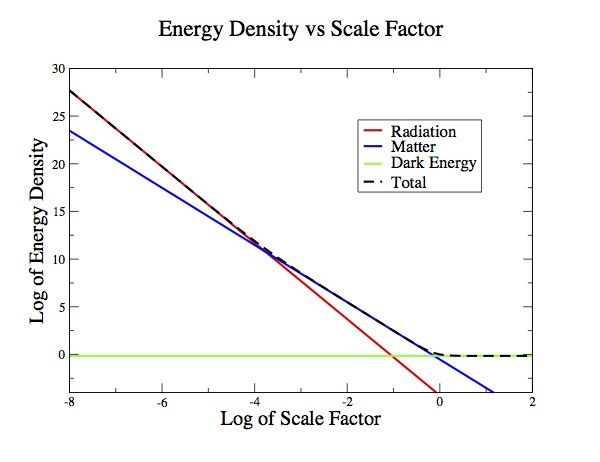,height=0.495\hsize}  
\end{tabular}
\caption{$\log\rho(a)$ as given in \pref{MixedFluidrhoandp} vs $a$ with realistic choices for the present-day energy densities (with present-day defined by $a(t_0) = 1$). \label{EnergyEvolutionFig}
}}

For a given Hubble parameter, $H$, it is conventional to define the critical density by $\rho_{\rm crit}(a) := 3 H^2/(8 \pi G_\ssN)$. Given the current measurement $H_0 \simeq 70$ km/sec/Mpc, the critical density's numerical value today becomes $\rho_{{\rm crit}0} \simeq 9 \times 10^{-30}$ g/cm${}^{3}$. $\rho_{\rm crit}$ is defined this way because the Friedmann equation becomes\footnote{It is sometimes convenient to put the $\mfK$ term on the right-hand side of the Friedmann equation and define $\Omega_\kappa := - \mfK/(aH)^2$ so that the Friemann equation becomes $\Omega_\Lambda + \Omega_{{\rm m}} + \Omega_{{\rm rad} } + \Omega_{\kappa} = 1$.}
\be \label{PresentDayHubble}
    H^2 + \frac{\mfK}{a^2}
    = \frac{8 \pi G_\ssN}{3} \, \rho \qquad \hbox{or} \qquad
    1 + \frac{\mfK}{(a H)^2} = \frac{\rho}{\rho_{\rm crit}}
    =: \Omega(a) \,,
\ee
and so if there should be a time $t_0$ when $\rho(t_0) = \rho_{\rm crit}(t_0)$ then $\mfK = 0$ and so $\rho = \rho_{\rm crit}$ for all times. Similarly if $\mfK = +1$ then we must have $\rho > \rho_{\rm crit}$ and if $\mfK = -1$ then $\rho < \rho_{\rm crit}$. The last equality of \pref{PresentDayHubble} defines $\Omega(a) := \rho(a)/\rho_{\rm crit}(a)$: the total energy density in units of this critical density. 

Normalizing densities in terms of $\rho_{\rm crit}$ proves to be useful because the best evidence currently is consistent with $\mfK = 0$ and so at present $\rho_0 \simeq \rho_{{\rm crit}0}$ and therefore $\Omega_0 \simeq 1$. When this is true the densities of the cosmic fluid components when normalized to $\rho_{\rm crit}$ -- {\it i.e.}~$\Omega_{{\rm rad}} := \rho_{\rm rad}/\rho_{\rm crit}$, $\Omega_{\Lambda} := \rho_{\rm vac}/\rho_{\rm crit}$ and $\Omega_{\rm m} := \Omega_b + \Omega_c$ (with $\Omega_b := \rho_b/\rho_{\rm crit}$ and $\Omega_c := \rho_c/\rho_{\rm crit}$) -- give their fraction of the total energy density and so should all sum to unity: $\Omega_{\rm vac} + \Omega_{\rm m} + \Omega_{\rm rad} \simeq 1$. 

Although we do not pursue this further here, there is also good theoretical reasons why $\Omega = 1$ should be true to a very good approximation: it is what would be expected if the much earlier universe were to have undergone a significant period of accelerated expansion, such as proposed by inflationary models \cite{firstINF} for which it is hypothesized that the scale factor evolves like $a(t) = a_0 \, e^{H_\ssI (t-t_0)}$ for roughly 50 $e$-foldings or more. Such an expansion would quickly drive $\mfK/(aH)$ to be extremely small even if $\mfK$ were nonzero. Inflationary models are attractive inasmuch as they provide a dynamical explanation for some of the initial conditions that are required for successful description of our later universe, including providing a mechanism for the origins and properties of the primordial density fluctuations that ultimately source the distribution of matter we now find around us \cite{Fluctuations}.  
 
\subsubsection{Observations}
\label{sssec:Data}

The above picture proves to be a spectacularly successful description of the universe we see around us. This agreement is even better than the above discussion might suggest because the assumption that the universe is exactly homogeneous and isotropic can be relaxed to follow how perturbations around homogeneity and isotropy evolve in time. Although a full description goes beyond the scope of this survey the result provides a beautifully accurate picture of how the much clumpier distribution of galaxies we see around us now arises from the gravitational attraction of initially very small deviations from homogeneity and isotropy (for textbook discussions see {\it e.g.}~\cite{CosmoTexts}). Agreement with observations determines the various parameters of the cosmology (such as $\Omega_{c0}$, $\Omega_{b0}$ and $\Omega_{\Lambda}$) to the percent level or better. In particular fits to observations confirm that $\Omega_{c0} \simeq 0.28$ and $\Omega_\Lambda \simeq 0.67$ are nonzero and $\Omega_\kappa$ is consistent with zero (see {\it e.g.}~Fig.~\ref{OmegaLambdaOmegakFig}). 

\FIGURE[tbh]{
\begin{tabular}{ll}
 \epsfig{file=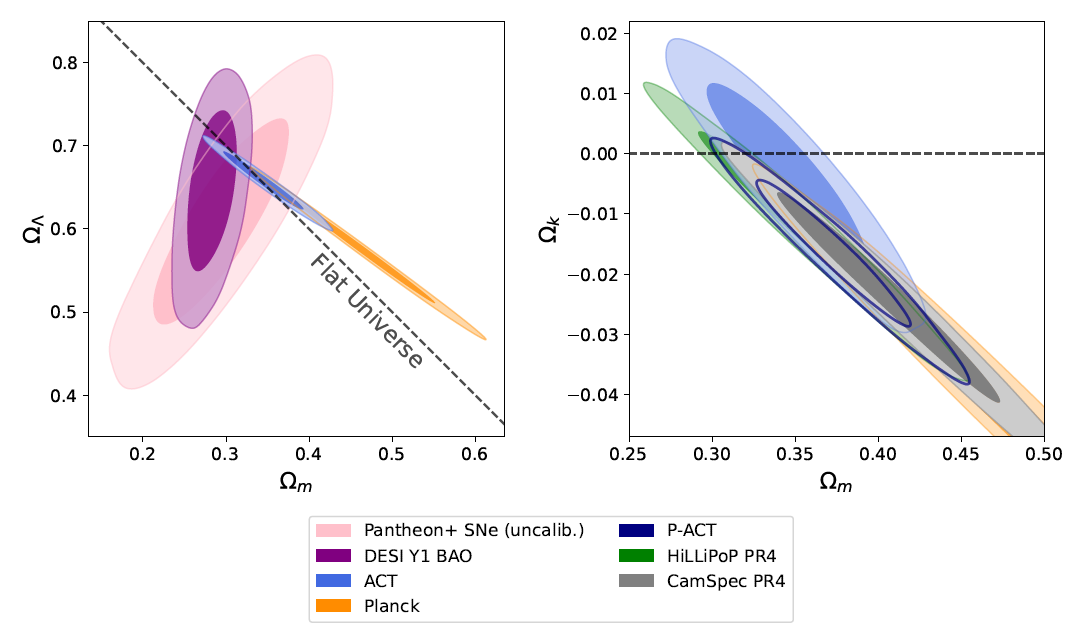,height=0.445\hsize}  
\end{tabular}
\caption{Left panel: Best-fit values for $\Omega_\Lambda := \Omega_{\rm vac}$ vs $\Omega_{\rm m}$ evaluated in the present day where the diagonal line corresponds to $\mfK = 0$ (a spatially flat universe). Right panel: Best fits for the present-day values of $\Omega_\kappa = - \mfK/(a H)^2$ vs $\Omega_{\rm m}$. Different colours correspond to fits to different data sets. Both figures taken from \cite{ACT:2025fju}\label{OmegaLambdaOmegakFig}
}}

Because $\Omega_\kappa$ is consistent with zero cosmologists define a 6-parameter model -- called $\Lambda$CDM -- for which $\kappa = 0$ is set by hand. The results for the six cosmological parameters is then obtained by fitting to observations and given these parameters many other observables can be computed. The results of such a process are listed in Fig.~\ref{ACTParams}, which shows the precision of agreement is currently better than the percent level.  

\FIGURE[tbh]{
\begin{tabular}{ll}
\epsfig{file=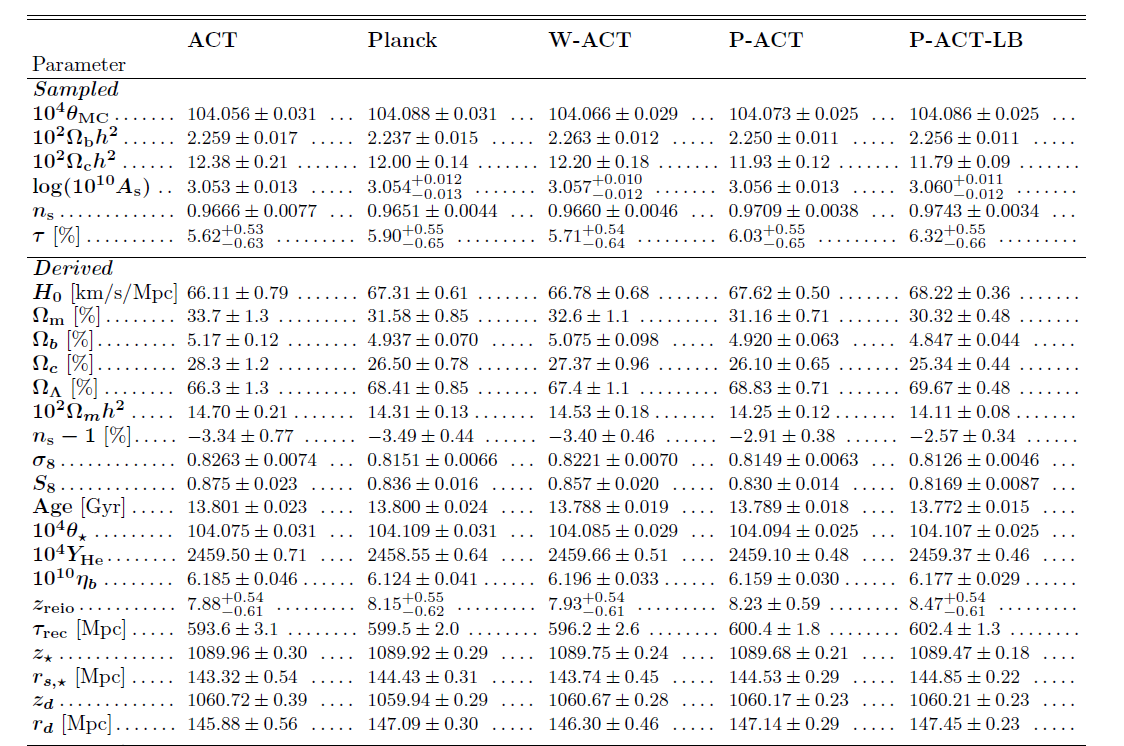,height=0.645\hsize}
\end{tabular}
\caption{The current (mid 2025) state of the art for cosmological parameters obtained by fitting observations to $\Lambda$CDM cosmology. The different columns fit to different datasets -- see \cite{ACT:2025fju} for details. \label{ACTParams}
}}

This successful description in particular tells us $\Omega_{b0} \simeq 0.05$ and $\Omega_{{\rm rad}\,0} \sim 10^{-4} \ll 1$, which allows at most only about 5\% of the current energy density to be matter we understand in detail. The remarkable fact that we can describe the universe so accurately while being almost completely ignorant about the fundamental nature of 95\% of what is in it is one of the central scientific puzzles of our times.\footnote{One attitude is that the vacuum having an energy density is not unexpected and so the evidence for Dark Energy is not really that mysterious. Even if so Dark Matter -- 28\% of what is out there -- is still a puzzle. The rest of these notes make the case that interpreting Dark Energy as a vacuum energy contains many puzzles.} We have direct observational evidence that we are missing something important but with not (yet) enough information to pin down decisively what is going on. Indeed an unusual opportunity. 

The rest of these notes explore this opportunity further, but before doing so it is worth reassessing the validity of the big picture. On one hand it is claimed that we have a detailed description of cosmology that is very accurate. On the other hand this detailed description requires the universe to have many unknown ingredients. Perhaps this is really tellng us our overall conceptual framework is flawed. Questions like these have stimulated much study of the foundations on which cosmology is based (some of which is summarized below).

Ultimately our confidence in the existence of things like Dark Matter relies on the redundancy of the evidence in its favour. Redundancy is convincing in two ways. First it provides protection from some of the observations simply being wrong (due to unknown mistakes). Redundant evidence survives even when some individual experiments are thrown away. Second, redundancy provides confidence in an overall picture. The situation is much like it was for the discussion of atoms at the turn of the 20th century: they could not be directly detected but the properties of bulk matter provided multiple independent lines of evidence for their existence. After all, if atoms did not exist there is no reason why independent inferences of their mass, size and abundance from the properties of bulk matter should all give the same answer. But they did and it is the agreement of multiple independent lines of evidence that is compelling. Although not explored in detail in these notes, the evidence for Dark Matter is similarly redundant, coming from several different sources within cosmology, but also from myriad observations of galaxies, galaxy clusters, and the large-scale distribution of matter. 

\subsection{Playing the field}

The purpose of these notes is to explore what we know about Dark Energy in two separate ways. We start here by summarizing the extent to which evidence is building that the Dark Energy density might not be constant in time which, if true, would mean the Dark Energy could not just be a vacuum energy. This evidence is one of a small set of `tensions' within $\Lambda$CDM cosmology. Tensions arise as degradation of the quality of the fit to observations if the agreement between observations and predictions starts to deteriorate as either or both become more accurate. We later explore the extent to which there are useful clues in demanding consistency of viable cosmological proposals with more microscopic physics. 

\subsubsection{Anomalies?}
\label{sssec:Anomalies}

\FIGURE[tbh]{
\begin{tabular}{ll}
\epsfig{file=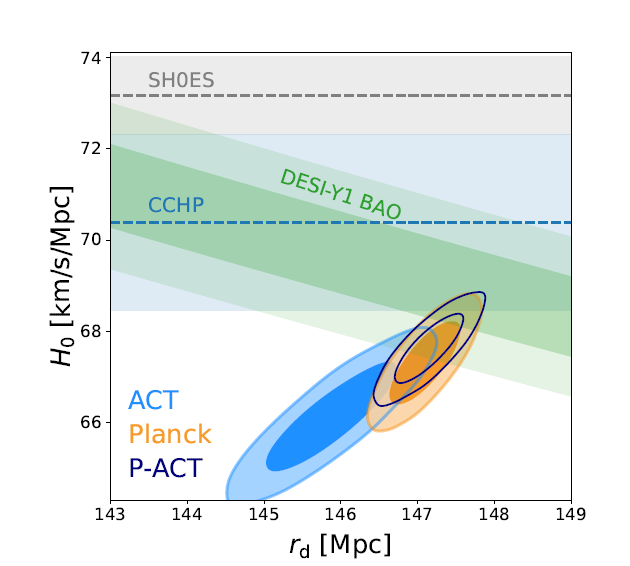,height=0.395\hsize} & 
\epsfig{file=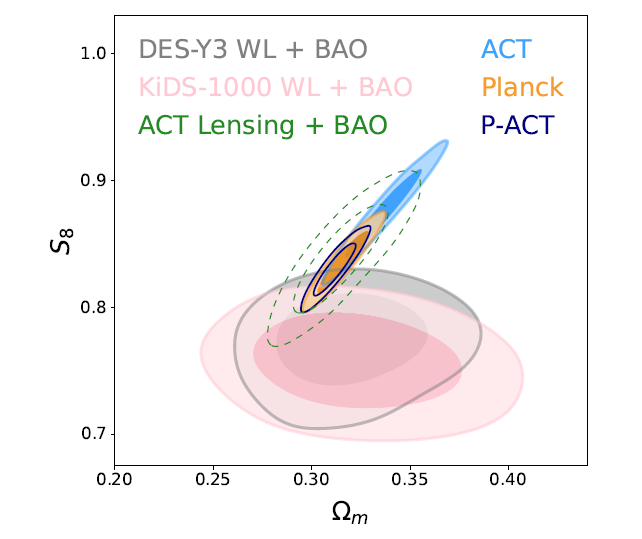,height=0.395\hsize} 
\end{tabular}
\caption{Left panel: Summary of the constraints on $H_0$ and the sound horizon $r_d$ coming from different kinds of observations, illustrating the Hubble tension. Right panel: Summary of the constraints on $S_8$ and the density of nonrelativistic matter $\Omega_{\rm m}$ coming from different kinds of observations, illustrating the $S_8$ tension. Both figures taken from \cite{ACT:2025fju}.\label{TensionsFig}
}}

Three types of tensions are normally discussed when asking about the ability of $\Lambda$CDM model to fit cosmological data. 
\begin{itemize}
\item {\bf Hubble tension:} The value of $H_0$ inferred from measurements of the Cosmic Microwave Background (CMB) seems to disagree with the value of $H_0$ obtained by measuring the luminosity of and distance to relatively nearby objects. Inferences based on the CMB (such as those in Fig.~\ref{ACTParams}) tend to prefer $H_0 \simeq 67$ km/sec/Mpc with an error of just under 1 km/sec/Mpc. Observations using relatively nearby supernovae instead give a value of $H_0 \simeq 73$ km/sec/Mpc with an error of around 1 km/sec/Mpc \cite{Murakami:2023xuy} (see Fig.~\ref{TensionsFig}). It is not yet clear whether this disagreement is telling us about non-$\Lambda$CDM physics or about the difficulty of performing the relevant measurements \cite{Freedman:2021ahq, CosmoVerse:2025txj}.
\item {\bf $S_8$ tension:} A similar tension has arisen for inferences of the size of clumping at particular scales --  parameterized by a quantity $S_8$ -- as measured in the distant and closer-by universe. CMB-based inferences (such as those in Fig.~\ref{ACTParams}) give $S_8 \simeq 0.85$ to within a few percent but those using more recent observables instead find $S_8 \simeq 0.77$ with similar errors \cite{DES:2021wwk}  (see Fig.~\ref{TensionsFig}). Again there is uncertainty as to how much of this discrepancy is associated with systematic errors \cite{CosmoVerse:2025txj}.
\item {\bf Time-dependent Dark Energy:} More recent than the previous two are tentative indications that the energy density of Dark Energy is time-dependent. This is illustrated in Fig.~\ref{DEEvoFig} which plots the equation of state parameter $w_{\rm vac}$ as a function of redshift (which is a proxy for scale factor). This result, if it survives scrutiny, would directly contradict the interpretation of Dark Energy as a constant vacuum energy.  
\end{itemize}

\FIGURE[tbh]{
\begin{tabular}{ll}
\epsfig{file=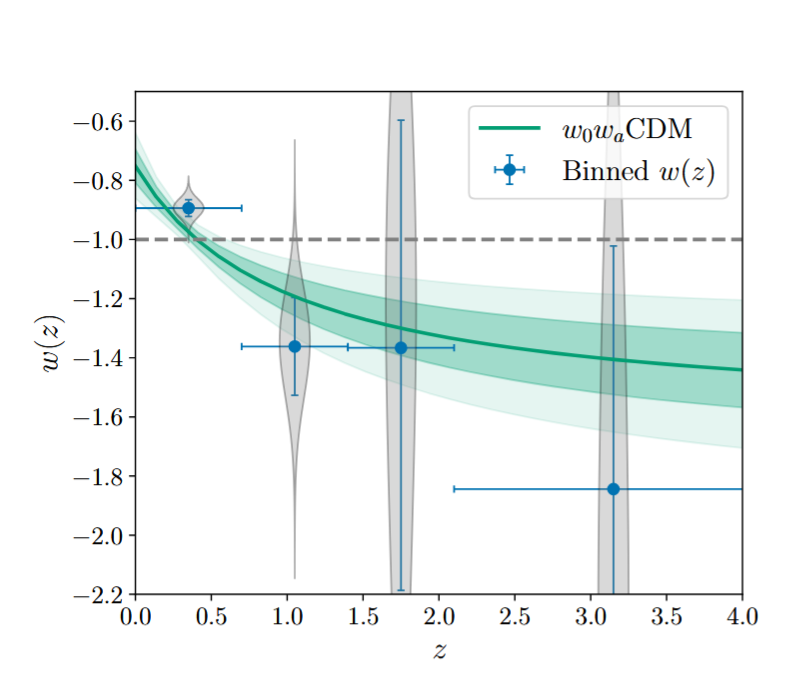,height=0.395\hsize} & 
\epsfig{file=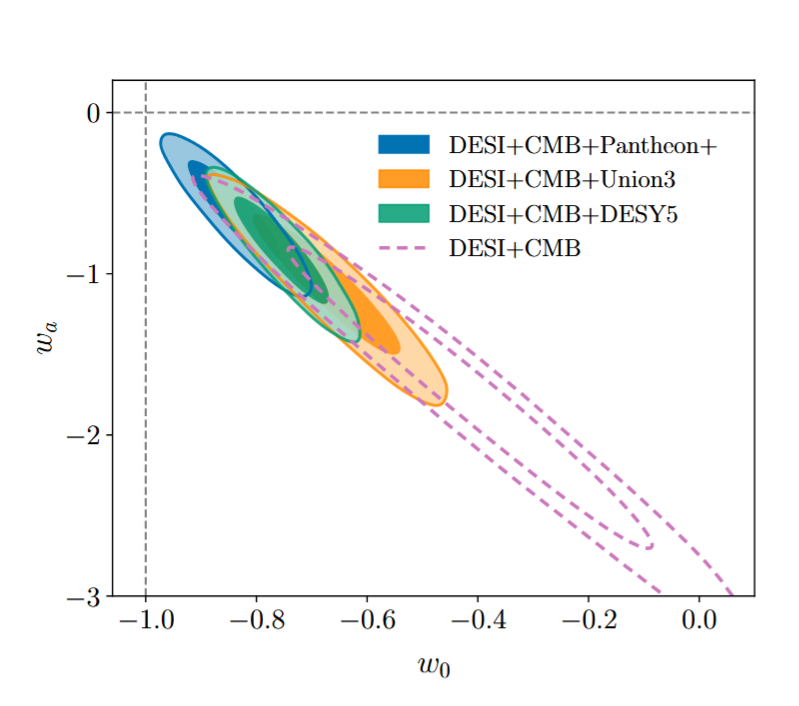,height=0.395\hsize} 
\end{tabular}
\caption{Left panel: constraints on the evolution of the Dark Energy equation of state parameter $w$ as a function of redshift (and so also of universal scale factor). The green swathe denotes the expected shape given a phenomenological parameterization $w(a) = w_0 + w_a \, a$. Right panel: best-fit values for the parameters $w_0$ and $w_a$. Both figures taken from \cite{DESI:2025zgx}.\label{DEEvoFig}
}}

Many cosmological models have been designed to explain these tensions in terms of new fundamental physics (for the Hubble tension see \cite{DiValentino:2021izs, Khalife:2023qbu} for recent reviews), or as solutions to other problems that happen also to have cosmological significance. The second direction these notes explore is the extent to which one can differentiate amongst the many cosmological models by asking whether they can plausibly emerge at low-energies given what we know about the physics of higher energies elsewhere in physics. This turns out to be fairly restrictive and it is argued that the criteria required to interface with higher-energy physics is an important clue when figuring out what is going on. We use the third anomaly -- evolution of Dark Energy density with time -- as a vehicle for having this discussion.

\subsubsection{Simple Scalar Model}
\label{ssec:SimpleModels}

To get things going it is useful to have a straw man: a concrete example of a model that could give a time-evolving Dark Energy density. This can be used to illustrate the kinds of difficulties that can arise. A vanilla starting point of this type postulates the existence of a scalar field $\phi$ that evolves homogeneously over cosmological time scales. This will look approximately like Dark Energy if its kinetic energy, $K$, is much less than its potential energy, $V$ -- becoming exactly like a vacuum energy in the limit $K \to 0$. 

Consider for instance supplementing the cosmological model with a scalar field whose action has the form
\be \label{VanillaScalarAction}
  S = - \int \exd^4x \, \sqrt{-g} \; \Bigl[ \tfrac12 \, g^{\mu\nu} \partial_\mu \phi \, \partial_\nu \phi + V(\phi) \Bigr]  \,,
\ee
where $V(\phi)$ is a scalar potential to be specified below. Such a field satisfies the classical field equations
\be \label{VanillaScalarDE}
    \Bigl[ - \Box + V'(\phi)  \Bigr] = 0 \,,
\ee
where $\Box = g^{\mu\nu} \nabla_\mu \nabla_\nu$, and contributes to Einstein's equations \pref{GRDE} by adding a new term to the stress energy. Applying the definition $\frac12 \sqrt{-g} \; T^{\mu\nu} = {\delta S}/{\delta g_{\mu\nu}}$ to \pref{VanillaScalarAction} one finds 
\be \label{VanillaScalarStressEnergy}
   T^{(\phi)}_{\mu\nu} = \partial_\mu \phi \, \partial_\nu \phi - g_{\mu\nu} \Bigl[ \tfrac12 \, g^{\lambda\rho} \partial_\lambda \phi \, \partial_\rho \phi + V(\phi) \Bigr] \,.
\ee

Working within an FLRW geometry \pref{FRWMetric} and assuming $\phi$ depends only on $t$ in the rest-frame of the cosmological fluid then reduces \pref{VanillaScalarDE} to an ordinary differential equation:
\be \label{VanillaScalarPhiEvo}
   \ddot \phi + 3H \, \dot \phi + V' (\phi) = 0 \,,
\ee
where $H = \dot a/a$ as usual and $V'$ is the derivative of $V(\phi)$ with respect to $\phi$. In the special case $V = \frac12 m^2 \phi^2$ \pref{VanillaScalarPhiEvo} is a linear equation that describes damped oscillations with a frequency set by $m$ and a damping rate set by $H$. When $m \gg H$ these oscillations are rapid on cosmological timescales and the damping ensures the energy density in these oscillations drops with universal expansion like $\rho_\phi \propto 1/a^3$. (Exercise: prove this.) As a consequence nontrivial scalar evolution within a potential is normally only important over long times in cosmology\footnote{An important exception to this is if the rapid oscillations themselves are the Dark Matter, whose energy also falls like $1/a^3$.} if the scalar mass is not large compared with $H$. For the present-day Hubble constant this is an extremely small scale, $H_0 \sim 10^{-32}$ eV, relative to microphysical scales. 

Homogeneity and isotropy also imply $T^{(\phi)}_{00} = \rho_\phi$, $T^{(\phi)}_{0i} = 0$ and $T^{(\phi)}_{ij} = p_\phi \, g_{ij}$ where
\be \label{VanillaScalarRhoP}
   p_\phi = \tfrac12 \dot \phi^2 - V(\phi) \qquad \hbox{and} \qquad
   \rho_\phi = \tfrac12 \dot \phi^2 + V(\phi)  \,,
\ee
so the Einstein equations governing homogeneous evolution remain \pref{FriedmannEqn} and \pref{EnergyConservationEqn}, but with the pressure and energy density of \pref{MixedFluidrhoandp} supplemented by adding \pref{VanillaScalarRhoP}, evaluated at the solution to \pref{VanillaScalarPhiEvo}. 

The ratio 
\be\label{VanillaScalarW}
  w_\phi := \frac{p_\phi}{\rho_\phi} = \frac{\tfrac12 \dot \phi^2 - V(\phi) }{\tfrac12 \dot \phi^2 + V(\phi) }
\ee
is in general time-dependent and so this kind of theory gives a cosmic fluid with a time-dependent equation-of-state parameter. It in general does not conform to the special case discussed in eqs.~\pref{EqnofState} through \pref{avst} apart from in a few special limits. For instance when the motion is rapid enough that $\frac12 \dot\phi^2 \gg V(\phi)$ -- the so-called {\it kination} regime -- then $p_\phi \simeq \rho_\phi$ and so $w_\phi \simeq +1$. In this case \pref{rhovsa} and \pref{avst} apply and give $\rho_\phi \propto a^{-6}$ and (if the scalar energy dominates) $a(t) \propto t^{1/3}$. 

Another limit for which eqs.~\pref{EqnofState} through \pref{avst} apply is the {\it slow-roll} regime, for which $\frac12 \dot\phi^2 \ll V(\phi)$. In this case \pref{VanillaScalarRhoP} implies $p_\phi \simeq - \rho_\phi$ and so $w_\phi \simeq -1$, mimicking a vacuum energy along which $\rho_\phi \simeq V(\phi)$ is approximately constant. This suggests that choosing $V(\phi)$ to be sufficiently shallow provides a candidate for Dark Energy where both $\rho_\DE$ and $w_\DE$ can vary with time. A possible difficulty with such a candidate is the observation that reproducing $\Omega_\Lambda \simeq 0.67$ requires $\rho_\phi \simeq V > 0$ during the slow roll. But for $V > 0$ eq.~\pref{VanillaScalarW} implies $-1 \leq w_\phi \leq +1$ and so in particular one should never enter the regime $w_\DE < -1$ seen in the left-hand panel of Fig.~\ref{DEEvoFig}.  We return to the question of how discouraging we should find this in \S\ref{ssec:NaturalRelaxation} below.
 
\section{Prior Knowledge}
\label{sec:UVInfo}

A great many models for Dark Energy can be (and have been) built in this way (see for example the reviews \cite{DEReviews}), and at first sight there seems to be little chance of being able to distinguish amongst so many models using only the limited data available to us from cosmology (wonderful though this data surely is). The next few sections step back and ask whether what we know about the rest of physics (outside of cosmology) can usefully constrain the search for phenomenologically successful models. 

It turns out that it can, and this might come as something of a surprise since experience in other areas of physics tells us that the details of short-distance physics usually are not important when computing long-distance properties --  a phenomenon called `decoupling' -- and cosmology deals with the longest distances of all. As we shall see, the theories that describe cosmology well rely heavily on the few things that do depend on what happens at higher energies. One of the things we learn along the way by asking this question is what controls the corrections to the basic semiclassical limit that lies behind the logic of solving equations like \pref{GRDE} and \pref{VanillaScalarDE} classically in the first place.

\subsection{Semiclassical methods in gravity (and why they work)}
\label{ssec:SemiclassicalGravity}

Since our goal is to ask what we can learn by thinking about cosmological models as the low-energy limit of some more fundamental theory, the first step is to systematize what kinds of things emerge in general for the low-energy limit of physical systems. The answer to this question is best answered using the tools of Effective Field Theories (EFTs) \cite{Weinberg:1978kz} and so we start with a brief digression to summarize these (leaning heavily on the reviews \cite{Burgess:2003jk, DonoghueEFTrev, Burgess:2020tbq}). Although quantum effects are often small in practical applications to gravity we nonetheless explore low-energy EFTs for quantum systems. For applications to gravity this will bring the later payoff of showing us what controls the semiclassical approximation in the first place.  

\subsubsection{EFT methods}
\label{sssec:EFTs}

Suppose we have a physical system with a characteristic scale $M$, such as a collection of `heavy' degrees of freedom with masses of order $M$ represented by fields collectively denoted $h(x)$. Suppose the theory also has very `light' degrees of freedom whose masses are much smaller than $M$, represented by fields collectively denoted $\ell(x)$.  Our interest is in observables, $\cA(E,M)$, involving the light fields that involve energies much smaller than $M$, such as scattering of $\ell$ particles with centre-of-mass energy $E \ll M$. These observables inevitably simplify once Taylor expanded in powers of $E/M$. The hard way to find the simple $E \ll M$ limit is to compute $\cA(E,M)$ in all of its glory and then Taylor expand. EFT methods seek instead to do the Taylor expansion as early as possible in a calculation in order to exploit the simplicity as effectively as possible.

A conceptually simple way to do so is to write out the path-integral expression for $\cA$ and then `integrate out' the heavy degrees of freedom once and for all early in the calculation. For instance, suppose $\cA$ has the path-integral representation 
\be \label{LEObs}
  \cA = \int \cD \ell\, \cD h \; \cO(\ell) \, \exp \left[ i \int \exd^4x \, \cL(\ell, h) \right]  \,,
\ee
where $\cO$ is some operator built only from the light degrees of freedom. (This would be true in particular for correlation functions from which a great many observables can be derived.) There is a great deal of freedom in choosing precisely how to separate the path-integral into integrals $\cD \ell$ and $\cD h$ over light and heavy degrees of freedom -- for instance on flat space one might imagine dividing up all field modes in momentum space as heavy or light depending on whether or not\footnote{It is usually most convenient to do this in Euclidean signature.} $p^2 + m^2 > \Lambda^2$ or $p^2 + m^2 < \Lambda^2$, where $p$ is the corresponding particle momentum and $m$ is its mass. $\Lambda$ here is an arbitrary cutoff chosen to be much smaller than the heavy scale but much larger than the energies of interest in $\cA$: that is $E \ll \Lambda \ll M$. 

Although the details of the heavy-light split can affect intermediate steps in the calculation, these choices must drop out of the final physical predictions because they are just an artefact of how we organize the calculation. They do not appear at all in the original integral \pref{LEObs} before trying to separate the fields into high and low energy parts. So one is free to use calculational convenience as a guide when making this split. 

Now comes the main point: the observables (by assumption) do not depend on the heavy degrees of freedom and so the integration over $h$ does not depend on the choice for $\cO$ and can be evaluated once and for all right at the very beginning, with the efficiency of performing an expansion in powers of $1/M$ reaped very early on. When this is done the influence of heavy fields on any dynamics at low energies is completely encoded in the following {\it effective
action}:
\begin{equation}   \label{leffdef} 
  e^{i S_\mathrm{eff}[\ell,\Lambda]} :=
  \int_\Lambda {\cal D}h \; \exp\left[ i \int d^4x \, {\cal L}(\ell,h) \right].
\end{equation}
The dependence of $S_{\rm eff}$ on $\Lambda$ is a shorthand for a dependence on all of the details of precisely how one makes the low-energy/high-energy split.

Physical observables at low energies are now computed by performing the remaining path integral over the light degrees of
freedom only:
\be \label{LEObs2}
  \cA = \int^\Lambda \cD \ell  \; \cO(\ell) \, \exp \Bigl[ iS_{\rm eff}(\ell, \Lambda) \Bigr]  \,,
\ee
showing that the integration over light fields is weighted by $S_\mathrm{eff}(\ell)$ in precisely the same way as the classical action $S = \int \exd^4x \, \cL(\ell,h)$ does for the original integral over both heavy and light degrees of freedom. This derivation also makes clear that any dependence on $\Lambda$ ({\it i.e.}~on the details of the high/low-energy split) must cancel between the explicit $\Lambda$-dependence of $S_{\rm eff}(\Lambda)$ and the implicit dependence in the definition of the low-energy integration $\int^\Lambda \cD \ell$.

Although $S_{\rm eff}$ obtained in this way is in general a hot mess, it simplifies dramatically once it is expanded to a finite order in $1/M$, in which case it become local in spacetime,\footnote{In nonrelativistic systems the low-energy expansion makes $S_{\rm eff}$ local in time but locality in space depends on whether or not short distance degrees of freedom can also be low-energy degrees of freedom.} so
\be
   S_{\rm eff} = \int \exd^4x \; \cL_{\rm eff}(\ell, \Lambda) \,,
\ee
where $\cL_{\rm eff}$ is (to any finite order in $1/M$) a simple function -- usually a polynomial -- of $\ell$ and its derivatives all evaluated at the same spacetime point. This happens in detail because the low-energy expansion of massive particle propagators is local
\be \label{LocalProp}
    \Bigl(- \Box + M^2 \Bigr)^{-1} = \frac{1}{M^2} \sum_{n=0}^\infty \left( \frac{\Box}{M^2} \right)^n \,,
\ee
when truncated to any finite order. Physically this has its roots in the uncertainty principle: high-energy states can only get into low-energy predictions by violating energy conservation, which the uncertainty principle allows\footnote{More precisely, energy conservation can be violated in old-fashion Rayleigh-Schr\"odinger perturbation theory, but is conserved in Schwinger-Feynman perturbation theory (in which case the same conclusions follow from off-shell contributions as described in \pref{LocalProp}).} provided they are only done over times $\Delta t \ll 1/M$, making them effectively local for low-energy observers who cannot resolve such small intervals. 

The upshot of all of this is that {\it all} low-energy contributions of the heavy degrees of freedom are encoded in an effective lagrangian (or Wilson action) that is a product of powers of the light field $\ell$ and derivatives. Because the integral in \pref{leffdef} defining $S_{\rm eff}$ involves only high-energy states any dimensionful parameters in this lagrangian will involve the heavy scale $M$ (such as is true for each additional power of $\Box$ in \pref{LocalProp}). On dimensional grounds any additional powers of derivatives and fields generically cost additional powers of $1/M$ and so become negligible once one restricts to a finite order. 

Only a very small number of terms can involve absolutely no suppression by powers of $1/M$ and the lagrangian obtained by keeping all such unsuppressed terms is called {\it renormalizable}. We expect renormalizable theories to describe the dominant physics at low energies, and this is indeed what we find in successful theories like Quantum Electrodynamics, Quantum Chromodynamics and the Weinberg-Salam model of electroweak unification. This is the modern understanding of why the renormalizable theories like the Standard Model of particle physics (which includes the other three mentioned) work so well: what we are looking at in practice in Nature is consistent with being the low-energy tip of an iceberg: some more fundamental theory describing physics at much higher energies. 

\subsubsection{GREFT}
\label{sssec:GREFT}

This is all very nice, but although cosmology involves the longest distances to which we have access (and so the lowest energies of all) the theory most relevant to it is General Relativity, which is {\it not} renormalizable. Why is the above EFT discussion relevant? 

In the EFT picture renormalizable interactions usually dominate nonrenormalizable ($1/M$-suppressed) interactions {\it when renormalizable interactions exist}. But sometimes no renormalizable interactions are possible and in such cases nonrenormalizable interactions can dominate. An example of this is when a renormalizable theory contains an `accidental' symmetry (like baryon number $B$ or lepton-number $L$ conservation in the Standard Model for instance).  Accidental symmetries are symmetries not built in as assumptions; they instead emerge as accidental consequences of renormalizability for a given field content. If the Standard Model emerges as the low-energy limit of a more fundamental theory in which baryon number is not conserved (such as a Grand Unified Theory, or GUT) then the leading rates for $B$- or $L$-violating processes at low energies would be described by nonrenormalizable interactions because the renormalizable ones of the Standard Model preserve baryon number. 

An even more informative example -- for which both the low-energy and high-energy theories are well understood -- is the low-energy effective Fermi theory of the weak interactions that are responsible for many radioactive decays. In this case the fundamental high-energy theory is the Standard Model itself and the low-energy theory is obtained once the $W$ boson (with mass $M_\ssW \simeq 80$ GeV) is integrated out (together with other, heavier, particles). In this effective theory the renormalizable interactions preserve particle flavours (like charm, strangeness, up-ness or down-ness {\it etc}) and so radioactive decays mediated by the weak interactions that violate the conservation of these quantities (like $\pi^+ \to e^+ \nu$ or nuclear $\beta$-decays) are well described by the nonrenormalizable Fermi theory. In this theory the effective Fermi coupling constant, $G_\ssF$, has dimension $M^{-2}$ for a scale $M$ much larger than the energies in the decays -- the characteristic suppression by inverse powers of a heavy scale typical of nonrenormalizable interactions. But because we also understand the more fundamental high-energy theory we can in this case explicitly relate the size of $G_\ssF$ to the scale of the heavy physics that was integrated out (in this case the $W$-boson mass, $M_\ssW$), with $G_\ssF \simeq g^2/M_\ssW^2$ (where $g \sim 10^{-1}$ is a measure of the coupling strength of the $W$ boson).

A similar story applies to gravity: it turns out there are no renormalizable couplings possible for the graviton and so its interactions must be nonrenormalizable. There is even a candidate for a more fundamental theory -- string theory -- that produces General Relativity in its low-energy limit with Newton's constant $G_\ssN \simeq g_s^2/M_s^2$ calculable in terms of the couplings $g_s$ and masses $M_s$ of very high-energy states.\footnote{Although we do not yet know whether string theory correctly describes nature, much of the attention it receives hinges on it being a rare example of a consistent fundamental UV completion for General Relativity, including quantum effects.}  As we shall see, even without such a UV completion the EFT framework also seems to be required if the theoretical error associated with quantum effects in gravity are to be reliably estimated.

The main practical consequence of regarding GR as part of a low-energy EFT within a more fundamental theory (much as we do for the Standard Model) is that the action can no longer be limited to the vanilla Einstein-Hilbert action. The Einstein-Hilbert action should instead just be regarded as the leading term in an expansion in powers of derivatives of the metric divided by some heavy scale $1/M$. 

Recall for these purposes that the field relevant for GR is the metric, $g_{\mu\nu}$, of spacetime itself, and that its action is required to be invariant under general covariance and local Lorentz invariance. Invariance under these symmetries dictate the metric can appear in the action only through curvature invariants built from the Riemann tensor,
\be
 {R^\mu}_{\nu\rho\lambda}  =  \partial_\lambda \Gamma^\mu_{\nu\rho} + \Gamma^\mu_{\lambda\alpha} \Gamma^\alpha_{\nu\rho} - (\lambda \leftrightarrow \rho) \quad\hbox{with}\quad
 \Gamma^\mu_{\nu\lambda}  =  \tfrac12 \, g^{\mu\beta} \Bigl( \partial_\nu g_{\beta\lambda} + \partial_\lambda g_{\beta\nu} - \partial_\beta g_{\nu\lambda} \Bigr) \,,
\ee
and its contractions -- such as the Ricci curvature $R_{\mu\nu} = {R^\alpha}_{\mu\alpha\nu}$ and Ricci scalar $R = g^{\mu\nu} R_{\mu\nu}$ --  and their covariant derivatives. What is important in what follows about these definitions is that the curvature tensors involve precisely two derivatives of the metric.
 
The low-energy EFT for the metric (called GREFT) is defined as the local action involving all possible powers of derivatives of the metric, which general covariance then requires must be built from powers of the curvature tensors and their derivatives,
\bea
\label{gravaction}
 - \, {{\cal L}_{\GREFT} \over \sqrt{- g}} &=& \lambda
+\tfrac12 M_p^2  \, R  + c_{41} \, R_{\mu\nu} \, R^{\mu\nu} + c_{42} \, R^2
+  c_{43} \, R_{\mu\nu\lambda\rho} R^{\mu\nu\lambda\rho} + c_{44} \, \Box R \\
 && \qquad \qquad\qquad + \frac{c_{61} }{ M^2}\; R^3 + \frac{c_{62}}{M^2} \partial_\mu R \, \partial^\mu R +  \cdots \,.\nn
\eea
The first two terms here are the only ones possible involving just the metric and two or fewer derivatives and these agree with the Einstein-Hilbert action of General Relativity with cosmological constant $\lambda$. The rest of the first line includes all possible terms involving precisely four derivatives, and (for brevity) the third line includes only two representative examples of the many possible terms involving six or more derivatives. 

The constants $c_{dn}$ appearing in \pref{gravaction} are labelled using the convention that $d$ counts the number of derivatives of the corresponding effective operator and $n = 1, \cdots, N_d$ runs over the number of such couplings. These couplings are dimensionless because the appropriate power of a high-energy mass scale $M$ has been extracted to ensure this is so (assuming 4 spacetime dimensions). Although it is tempting to use $M \simeq M_p$ everywhere for this high-energy mass scale (given that this is what appears in front of the Einstein-Hilbert term) this would in general be a mistake. 

To see why, imagine generating a contribution to these effective couplings by integrating out a heavy particle of mass $M$. 
All of the terms listed are generically generated when doing so, with $M$ appearing in each coefficient as required on dimensional grounds (leading to terms like $M^2 R$ and $R^2$ and $R^3/M^2$ and so on). The complete coefficient of any one term in $\cL_\GREFT$ would then be obtained by summing over all of the possibly many particles appearing in the fundamental theory, making the coefficient of $R$ in this lagrangian a sum of the schematic form $\sum_n k_n M^2_n$ while the coefficient of $R^3$ would instead be something like $\sum_n \tilde k_n M^{-2}_n$. 

Here comes the point: although it is the largest mass that dominates in any sum over positive powers of $M_n$, it is the {\em smallest} mass that dominates a sum over negative powers of $M_n$. Consequently we are not surprised at all to find a large coefficient like $M_p \sim 10^{18}$ GeV appearing in front of the Einstein-Hilbert term, but this does {\it not} provide evidence for the scale $M$ appearing in the curvature-cubed and higher terms in \pref{gravaction} also being this large. Instead one should expect $M$ in any given application to be of order the lightest of the heavy particles whose integrating out generates $\cL_\GREFT$. For instance, for applications to the solar system $M$ might be the electron mass; for applications to post-nucleosynthesis Big-Bang cosmology $M$ might be of order the QCD scale, and so on.) Of course, contributions like $M^2 R$ or $R^3/M_p^2$ could also exist, but when $M \ll M_p$ these are completely negligible compared to the terms displayed in eq.~(\ref{gravaction}). 

The first, cosmological constant, term in eq.~(\ref{gravaction}) is the only one with no derivatives and the alert reader will notice that we did not write its coefficient as $\lambda = c_{01} M_p^4$. This was not done because (as discussed in \S\ref{sssec:LCDM}) it contributes to observables in the same way as does the vacuum energy and so plays the role of $\Lambda$ in the $\Lambda$CDM model. As a consequence its value has already been measured, with observations implying $\lambda \simeq \rho_{\rm vac} \simeq 0.67 \rho_c \sim (3 \times 10^{-3} \; \hbox{eV})^4$ and so is roughly 122 orders of magnitude smaller than $M_p^4$. Since $c_{01} \sim 10^{-122}$ it is an extremely good approximation for most applications to neglect it completely when asking for the implications of GREFT in noncosmological settings.  Much of the rest of this review will be devoted to how puzzling we should find it that $\lambda$ should be so small, which is called the cosmological constant problem \cite{Weinberg:1988cp, Burgess:2013ara, CCrevs}. (We return to this issue below when discussing implications for cosmology.) 

\subsubsection*{Redundant interactions}

The attentive reader might notice that not all of the interactions listed in \pref{gravaction} are equally important, even when only comparing interactions having the same number of derivatives. For instance, the freedom to drop total derivatives\footnote{These cannot be dropped if one cares about boundary information or topology, so we when making these arguments we have in mind the vast majority of other local effects for which surface terms are irrelevant.} from the lagrangian allows us to ignore the coupling $c_{44}$,  because $\sqrt{-g} \, \Box R = \partial_\mu (\sqrt{-g} \, \partial^\mu R)$ is a total derivative. A similar argument applies as well (in 4 dimensions) to $c_{43}$ since the quantity
\begin{equation}
 \sqrt{-g} \; X = \sqrt{-g} \Bigl( R_{\mu\nu\lambda\rho} R^{\mu\nu\lambda\rho} -4 R_{\mu\nu}
 R^{\mu\nu} + R^2\Bigr)  \,,
\end{equation}
is locally also a total derivative (it integrates to give a topological invariant in 4 dimensions). Dropping total derivatives allows us to replace, for example, $R_{\mu\nu\lambda\rho} R^{\mu\nu\lambda\rho}$ with the linear combination $4 \, R_{\mu\nu}R^{\mu\nu} - R^2$, with no consequences for any observables (provided these observables are insensitive to the overall topology of spacetime, as are the classical equations or perturbative particle interactions). 

It is also possible to ignore any effective interactions in \pref{gravaction} that involve the Ricci tensor $R_{\mu\nu}$ (and so also its trace $R = g^{\mu\nu} R_{\mu\nu}$), provided we work only perturbatively in powers of $1/M$. This is because the variation of the leading Einstein-Hilbert action under a field redefinition $\delta g_{\mu\nu}(x)$ is (dropping total derivatives)
\be
    \delta S_\EH = \int \exd^4 x\; \left( \frac{\delta S_\EH}{\delta g_{\mu\nu}}  \right) \delta g_{\mu\nu} =  \tfrac12 M_p^2 \int \exd^4 x\; \sqrt{-g} \, \Bigl( R^{\mu\nu} - \tfrac12 R g^{\mu\nu} \Bigr) \delta g_{\mu\nu} 
\ee
This means that any term in the GREFT action that vanishes for a Ricci-flat geometry, like
\be
    S_\GREFT \ni - \int \exd^4 x \sqrt{-g} \, \Delta^{\mu\nu} R_{\mu\nu} \,,
\ee
can be removed at leading order by choosing 
\be
   \delta g_{\mu\nu}  =  2 \kappa^2 \Bigl( \Delta_{\mu\nu} - \tfrac12 g^{\lambda\rho} \Delta_{\lambda\rho} \, g_{\mu\nu} \Bigr)  \,.
\ee
This argument is a special case of a more general statement that also applies when matter is present: any effective interaction that vanishes when the lowest-order equations of motion are used can be similarly removed by performing an appropriate field redefinition.  

Any interaction that is a surface term or can be removed using a field redefinition in this way is called a {\it redundant} interaction because most observables (except perhaps those sensitive to boundary terms) cannot depend on their coefficients. It is useful to remove all such interactions from the effective theory because carrying them around is not wrong but is needlessly time-consuming since they have no effects. 

In practice this means that for pure gravity (no other fields, like matter) {\it all} of the effective interactions beyond the Einstein-Hilbert term that are written explicitly in \pref{gravaction} are redundant (in 4 spacetime dimensions) because they are either total derivatives or they vanish when $R_{\mu\nu} = 0$ (or both). The first nontrivial non-redundant effective interaction involves cubic or higher powers of the Riemann tensor. 

\subsubsection{Power counting (gravity only)}

We see there can be a large number of interactions in an EFT -- potentially arbitrarily large if one works to arbitrary fixed order in $1/M$. How can a theory with so many effective couplings ever be predictive? This is a central question whose general answer is given by power-counting \cite{Weinberg:1978kz} (as we describe for gravity in this section).

In any EFT we imagine expanding all observables in powers of $q/M$ where $q$ is a typical energy scale of interest in the low-energy sector (perhaps a centre-of-mass scattering energy or the Hubble expansion rate) and so a very important question asks which interactions are relevant when computing observables at a specific order in powers of $q/M$ (and $q/M_p$ in the case of the lagrangian \pref{gravaction}). We here briefly recap the result without repeating the details (see however \cite{Burgess:2003jk}).  

To see how various interactions contribute to physical processes consider using the lagrangian \pref{gravaction} to calculate a correlation function or a scattering amplitude involving a path integral like in \pref{LEObs2}. For simplicity we ignore here the cosmological constant term $\lambda$, but return to it when we consider cosmology in the next sections. The integral is evaluated semiclassically by expanding around some classical background spacetime $\ol g_{\mu\nu}$ that we assume to be a stationary point of the action built from \pref{gravaction}. We then write the full metric as\footnote{Strictly speaking a factor of $2$ would pre-multiply $h_{\mu\nu}$ if fluctuations were to be canonically normalized, but our focus here is on how the scales $M$ and $M_p$ appear.} $g_{\mu\nu} = \ol g_{\mu\nu} + h_{\mu\nu}/M_p$ and do a double expansion of the action $S_\GREFT$ in powers of both $h_{\mu\nu}$ and of derivatives, keeping in mind that the curvature involves all possible powers of $h_{\mu\nu}$, but precisely two derivatives. 

One finds in this way the expansion
\be
 S_\GREFT[\ol g + h] = S_\GREFT[\ol g] + S_{\EH(2)}[\ol g, h] + S_{\rm int}[\ol g, h]   \,,
\ee
where 
\be
   S_\EH[g] = -   \tfrac12 M_p^2 \int \exd^4x \sqrt{-g} \;  R  \,,
\ee
is the Einstein-Hilbert lagrangian -- or equivalently the terms in \pref{gravaction} involving precisely two derivatives -- and $S_{\EH(2)}$ contains those terms in the expansion of the Einstein-Hilbert action arising at quadratic order\footnote{Any linear term in the expansion of $S_\EH$ simply contributes to cancellation of the `tadpole' graphs (those with one external leg) that determine how the background metric changes from the solution to Einstein's equations once higher-derivative terms are included.} in $h_{\mu\nu}$.  The `interaction' term contains everything else:
\be
  S_{\rm int}[\ol g, h]   =  S_{\EH\,{\rm int}}[\ol g, h] + S_{{\rm eff} \, {\rm int}}[\ol g + h] \,,
\ee
where $S_{\EH\,{\rm int}}$ contains two-derivative terms coming from the expansion of $S_\EH[\ol g + h]$ that are cubic or higher in $h_{\mu\nu}$ and $S_{{\rm eff}\,{\rm int}}$ contains terms involving any number of powers of $h_{\mu\nu}$ but with no fewer than 4 derivatives -- {\it i.e.}~the higher-derivative terms in \pref{gravaction}.  

The integrand of the path integral is then written perturbatively in $S_{\rm int}$
\be
   e^{iS_\GREFT[\ol g + h]} 
   = e^{iS_\EH[\ol g] +i S_{\EH(2)}[\ol g, h]} \sum_{r=0}^\infty \frac{1}{r!} \Bigl[i S_{\rm int}[\ol g, h] \Bigr]^r \,,
\ee
so that the path integration becomes gaussian and can be evaluated using the standard Feynman procedure (including covariant gauge fixing and ghosts in the usual way, the details of which do not change the arguments to be made below). Evaluating the gaussian integrals can still be hard in practice because we so far make no assumptions about the nature of the background metric $\ol g_{\mu\nu}$, but it can be done explicitly for simple spacetimes like Minkowski space or anti-de Sitter space, say. 

Our goal here is less ambitious than full evaluation, however. We wish only to perform a power-counting exercise to identify what must be small in order for this expansion to be a good approximation. This involves identifying how an arbitrary Feynman graph depends on the scales $M_p$ and $M$ appearing in the lagrangian \pref{gravaction}, which can be done in great generality in some circumstances. In particular, it can be done in situations when there is only one scale of interest in the low-energy theory\footnote{A nontrivial example of this might be if we follow fluctuations about de Sitter space and focus only on fluctuations whose physical momenta $k/a$ are roughly the same size as the background curvature scale $H$. In this case we could choose $q \sim H$.}  -- call it $q$ say -- since in this case it boils down to a dimensional argument.\footnote{Dimensional arguments become more complicated if UV divergences are regularized using a cutoff but go through as expected naively if one instead uses dimesional regularization, for instance.} 

Consider an arbitrary graph that contributes at $L$ loops to the amputated\footnote{Amputation means that the graphs have no external lines, such as might be encountered when computing the size of coefficients in a low-energy effective action itself.} $E$-point $h_{\mu\nu}$ correlation function, ${\cal A}_E(q)$, performed with all external background curvatures and mode numbers characterized by a single low-energy scale $q \ll M \ll M_p$. Suppose the graph contains $V_{id}$ vertices involving $d$ derivatives and $i$ factors of the fluctuation field $h_{\mu\nu}$. The dependence of ${\cal A}_E(q)$ on the scales $M$ and $M_p$ can be read off from the Feynman rules that determine the propagators and vertices of the graph in question and then all of the remaining dimensions are taken to be captured by the appropriate power of the low-energy scale $q$. This leads \cite{Burgess:2003jk} to the following prediction for the $q$, $M$ and $M_p$ dependence of $\cA_E(q)$:  
\begin{equation}
\label{GRcount1a}
 {\cal A}_\ssE(q) \sim q^2 M_p^2 \left( {1
 \over M_p} \right)^{E}
 \left( {q \over 4 \pi M_p} \right)^{2
 L} {\prod_{i} \prod_{d>2}} \left[{q^2 \over M_p^2}
 \left( {q \over M} \right)^{(d-4)}  \right]^{V_{id}} \,.
\end{equation}
Notice that since $d$ is even for all of the interactions, the condition $d > 2$ in the product implies there are no negative powers of $q$ in this expression. The argument leading to \pref{GRcount1a} is sketched out in a bit more detail in the next section.

Eq.~(\ref{GRcount1a}) is this section's main result, and it contains lots of information. 
\begin{itemize}
\item First, the appearance of only positive powers of $q$ verifies that it is indeed self-consistent to organize calculations using a derivative expansion when computing using \pref{gravaction}. The weakness of gravitational self-couplings comes purely from the low-energy approximations $q \ll M_p$ and $q \ll M$. 
\item For a fixed process ({\it i.e.}~for a fixed number, $E$, of external lines) each additional loop costs a factor of $q^2/(4\pi M_p)^2$. But it is the number of loops that also counts the factors of $\hbar$ that premultiply the action in non-fundamental units $e^{iS/\hbar}$, making the loop expansion also the semiclassical expansion. Why is the classical approximation good in GR? We see it is ultimately the hierarchy $q \ll 4\pi M_p$ that justifies the use of semiclassical methods: the semiclassical approximation {\em is} the low-energy approximation.
\item Notice that there is no low-energy penalty for using as many 2-derivative interactions as we like. This shows that there is nothing in the low-energy limit that allows us to neglect the full nonlinearity of GR.
\item Even though the ratio $q/M$ could be much larger than $q/M_p$, it only arises in ${\cal A}_{E}$ together with a factor of $q^2/M_p^2$, making it hard in practice to exploit the hierarchy $M \ll M_p$ to obtain surprisingly large effects.
\end{itemize}

Eq.~\pref{GRcount1a} can be used to identify the dominant contributions to any low-energy process (graviton scattering amplitude or correlation function) that is characterized by a single scale $q \ll M \ll M_p$. Eq.~\pref{GRcount1a} shows that the least suppressed contributions come from graphs with $L = 0$ and $V_{id} = 0$ for all $d > 2$. That is to say, using only tree graphs ($L = 0$) constructed purely from the Einstein-Hilbert ($d=2$) action. As might have been expected, it is classical General Relativity that dominantly governs the low-energy dynamics of gravitational fluctuations. 

For instance, the above estimate applies in particular to graviton-graviton scattering, in which case we take $E = 4$. Specializing eq.~\pref{GRcount1a} to this case (with $L = 0$ and $V_{id} = 0$ for all $d > 2$) then says that at low energies we have $\cA_4 \simeq (q/M_p)^2$, which agrees well with the result obtained by explicit calculation \cite{DeWitt:1967uc}, which for 2-body graviton scattering on flat space gives (at tree level)
\be
   \cA_4 \simeq 8\pi i \GN \left( \frac{s^3}{tu} \right) \,,
\ee
where $s$, $t$ and $u$ are the Mandelstam invariants for 2-body scattering, defined in terms of the initial 4-momenta $p_i^\mu$ and final 4-momenta ${p'_i}^\mu$ by $s = - (p_1 + p_2)^2$, $t = - (p_1 - p_1')^2$ and $u = - (p_1 - p_2')^2$. What is important for comparing with the estimate $(q/M_p)^2$ is that $8\pi \GN = M_p^{-2}$ and $s$, $t$ and $u$ when evaluated in the centre-of-mass frame are $s = 4 E_{cm}^2$, $t = -2 E_{cm}^2(1-\cos \vartheta)$ and $u = - 2 E_{cm}^2(1+ \cos\vartheta)$, where $E_{cm}$ is the center-of-mass energy and $\vartheta$ is the angle between the incoming momentum $\bfp_1$ and the outgoing momentum $\bfp'_1$ in this frame. These imply $\cA_4 \propto (E_{cm}/M_p)^2$ in agreement with \pref{GRcount1a} with $E_{cm}$ playing the role of the low-energy scale $q$.

But \pref{GRcount1a} also identifies which graphs give the next-to-leading contributions. These come in one of the following two ways:
\begin{itemize}
\item $L = 1$ and $V_{id} = 0$ for any $d\ne 2$ but $V_{i2}$ is arbitrary, or
\item $L = 0$, $\sum_i V_{i4} = 1$, $V_{i2}$ is arbitrary, and all other $V_{id}$ vanish.
\end{itemize}
That is to say: the next to leading contribution comes from one-loop graphs constructed using only the interactions of General Relativity, or by working to tree level and including precisely one insertion of a curvature-squared interaction in addition to any number of interactions from GR. Both of these are suppressed compared to the leading term by a factor of $(q/M_p)^2$. The next-to-leading tree graphs provide precisely the counter-terms required to absorb the UV divergences in the one-loop graphs. And so on to any desired order in the expansion. Despite being nonrenormalizable the theory is  predictive {\it provided} one works only to a fixed order in $q/M$ and $q/M_p$.

\subsection{Power-counting in cosmology}
\label{ssec:EFTs}

We are now ready for the main event: asking more systematically whether the observation that any successful theory of cosmology must emerge as the low-energy EFT for some more fundamental theory carries any practical consequences.

To this end -- and partly with scalar models of Dark Energy and/or Dark Matter in mind -- we repeat the power-counting estimates made above for GREFT but this time do so for gravity coupled to a collection of $N$ dimensionless scalar fields, $\theta^i$. A generic EFT containing these low-energy fields can be expanded in a derivative expansion, leading to a lagrangian that extends \pref{gravaction} to include new scalar interactions:
\bea \label{Leffdef}
 - \frac{ \cL_{\rm eff}}{\sqrt{-g}} &=& v^4 U(\theta) + \tfrac12 
 \, g^{\mu\nu} \Bigl[ M_p^2 \,  W(\theta) \, R_{\mu\nu}
 + f^2 \, G_{ij}(\theta) \, \partial_\mu \theta^i
 \partial_\nu \theta^j \Bigr] \\
 && \quad + A(\theta) (\partial \theta)^4 + B(\theta)
 \, R^2 + C(\theta) \, R \, (\partial \theta)^2
 + \frac{E(\theta)}{M^2} \, (\partial \theta)^6
 + \frac{F(\theta)}{M^2} \, R^3 + \cdots \,,\nn
\eea
where all terms involving up to two derivatives are written explicitly in the first line, with the rest written schematically on the second line.\footnote{These higher derivative terms are schematic inasmuch as $R^3$ collectively represents all possible independent curvature invariants involving six derivatives, and so on.} 

The explicit mass scales $M_p$ and $M$ are explicitly written, as before, so that the functions $W(\theta)$, $G_{ij}(\theta)$, $A(\theta)$, $B(\theta)$ {\it etc}, are dimensionless. The functions $W(\theta)$ and $G_{ij}(\theta)$ are positive definite and there can be positivity conditions on some of the other functions as well \cite{Positivity}. Here $M$ is the {\it lowest} scale integrated out to obtain $\cL_{\rm eff}$ (since this is what dominates in the denominator). We allow the scalar kinetic term to be normalized differently, with some new scale $f$ appearing there instead of $M_p$, but our interest here is scalars that interact with gravitational strength for which $f = M_p$. We assume $M \ll M_p$ and we take $f < M_p$ in situations where $f \neq M_p$. 

A new scale, $v$, is also added so that the scalar potential $V(\theta)$ is order $v^4$ when the dimensionless function $U(\theta)$ is order unity.  With cosmological applications in mind we imagine $v$ to be a low-energy scale and take $v \ll M \ll M_p$. Notice that if we were to use a more canonical normalization where $\phi^i \sim f\,  \theta^i$ then if $U(\theta) = \lambda_0 + \lambda_i \theta^i + \lambda_{ij} \theta^i \theta^j + \cdots$ is order unity for $\theta^i$ order unity then this choice for the scalar potential implies
\be \label{assumedVform2}
 V(\phi) = v^4 \left[ \lambda_0 + \lambda_i \frac{\phi^i}{f} + \lambda_{ij} 
 \frac{\phi^i \phi^j}{f^2}  + \lambda_{ijk} 
 \frac{\phi^i \phi^j \phi^k}{f^3}   + \cdots \right] \,,
\ee
which shows that $V$ changes by order $v^4$ as $\phi^i$ ranges through values of order $f$. This form captures qualitative features of many explicit cosmological models when $f \sim M_p$ and $\lambda_{ijk\cdots} \sim \cO(1)$, but for future purposes it is worth keeping in mind that these choices make $V$ remarkably shallow compared to most scalar potentials considered in particle physics (more about which in later sections). In this language a typical particle physics potential would change by order $v^4$ when $\phi$ goes through scales much smaller than $M_p$ (which amounts to taking $f \ll M_p$ in \pref{assumedVform2}).\footnote{For instance the Higgs potential in the Standard Model would correspond to choosing $v \sim f \sim 100$ GeV while the QCD axion potential has $v \sim \Lambda_\QCD \sim 0.2$ GeV and $f \sim 10^{10}$ GeV.}

As usual, there is considerable freedom to simplify the action \pref{Leffdef} by performing field redefinitions, and we use this freedom to Weyl rescale the metric to set $W(\theta) = 1$ ({\it i.e.}~go to Einstein frame). The function $G_{ij}(\theta)$ can often similarly be simplified through redefinitions of the form $\theta^i \to f^i(\theta)$, but it cannot be made $\theta$-independent if the Riemann tensor ${\cR^i}_{jkl}$ built in the usual way from the `target-space metric' $G_{ij}$ is flat.\footnote{$G_{ij}(\theta)$ transforms like a covariant tensor under redefinitions of the $\theta^i$ fields, and since it also is positive and symmetric it can be regarded as a metric on the `target space' ({\it i.e.}~the range of the function $\theta^i(x)$).}  

\subsubsection{Power counting}
\label{sssec:PowerCounting}

From here the argument proceeds much as in the earlier section discussing pure gravity (for details see \cite{Burgess:2009ea, Adshead:2017srh, Burgess:2017ytm}). To this end, as above, we expand $S_{\rm eff} = \int \exd^4 x \, \cL_{\rm eff}$ about a classical solution using fields that have canonical dimension, $\theta^i(x) = \bar \theta^i(x) + {\phi^i(x)}/{f}$ and $g_{\mu\nu} (x) = \bar g_{\mu\nu} (x) + {h_{\mu\nu}(x)}/{M_p}$. As above we keep track of the scales appearing in the action \pref{Leffdef} by reading them off from the Feynman rules for each vertex and propagator, and we assign the dependence on any low energy scale purely on dimensional grounds.

The dependence of a Feynman graph on the scales $M$, $M_p$, $f$ and $v$ are found by expanding the lagrangian \pref{Leffdef} in powers of the fluctuation fields $\phi^i$ and $h_{\mu\nu}$, leading to a sum of interactions of the form
\be \label{Leffphih}
 \cL_{\rm eff}(\theta, g_{\mu\nu}) = \cL_{\rm eff}(\bar\theta,\bar g_{\mu\nu}) + M^2 M_p^2 \sum_{n}
 \frac{c_{n}}{M^{d_{n}}} \; \cO_{n} \left(
 \frac{\phi^i}{f} , \frac{ h_{\mu\nu}}{M_p} \right)
\ee
where the functions, $\cO_{n}$, are monomials involving $N_n = N^{(\phi)}_n + N^{(h)}_n \ge 2$ powers\footnote{Terms linear in the fluctuations only arise at subdominant order, where they cancel the `tadpole' graphs (those with exactly one external line), since this is the condition that defines the background fields. At leading order this makes the background fields solve the classical equations of motion, with corrections order by order in perturbation theory.} of the fields $\phi^i$ and $h_{\mu\nu}$ and their derivatives. The parameter $d_{n}$ counts the number of derivatives appearing in $\cO_n$, and the coefficients $c_n$ are dimensionless and calculable in terms of the functions $U(\theta)$, $G_{ij}(\theta)$ and so on (and their derivatives) evaluated at the background fields. 

The prefactor, $M^2 M_p^2$, ensures the kinetic terms ({\it i.e.}~the terms with $N_n^{(h)} = d_n = 2$) for $h_{\mu\nu}$ are independent of $M$ and $M_p$, and so the same is also true for its propagator. The same is true for the kinetic term for $\phi^i$ if $f = M_p$. For more general $f$ the expansion of the `sigma-model' term $f^2 \, G_{ij} \,\partial \phi^i \partial \phi^j$ term of \pref{Leffdef} gives operators of the form \pref{Leffphih} that are proproportional to $f^2/M_p^2$, so
\be \label{cnkindef}
  c_n^{\sigma\,{\rm mod}} = \hat c_n \left( \frac{f^2}{M_p^2} \right) \,,
\ee  
where $\hat c_n$ is order unity. This ensures that the $\phi^i$ propagators are also scale independent.

Similarly, the lagrangians of \pref{Leffphih} and \pref{Leffdef} only make equivalent predictions for the $M$ and $M_p$ dependence in Feynman graphs if the coefficients $c_n$ for all of the higher-derivative terms in \pref{Leffdef} are proportional to $M^2/M_p^2$, so
\be \label{cndgt2h}
 c_n  = \left( \frac{ M^2}{ M_p^2} \right) g_n
 \qquad \hbox{(if $d_n > 2$)} \,,
\ee
where $g_n$ is at most order unity. For terms with no derivatives --- {\it i.e.}~those coming from the scalar potential, $V(\theta)$ --- one instead finds that agreement requires
\be \label{cndeq0}
 c_n = \left( \frac{v^4}{M^2 M_p^2} \right) \lambda_n
 \qquad \hbox{(if $d_n = 0$)} \,,
\ee
where the dimensionless couplings $\lambda_n$ are also independent of $M_p$ and $M$ (up to logarithms).

Consider first the case where $f = M_p$ and ask how the various scales enter into an amputated $E$-point correlator of $h_{\mu\nu}$ and $\phi^i$ fields at $L$ loops. As before we keep track of the coupling for each vertex to see how the scales $M$, $M_p$ and $v$ appear in the graph. Using dimensional analysis -- with dimensional regularization to remove the need for a confounding cutoff scale -- then gives the dependence on the (assumed) single low-energy scale (call it $H$ this time since it is often the Hubble scale in cosmology): 
\be  \label{PCresult2}
 \cA_\ssE (H)  \simeq  M_p^2 H^2 \left( \frac{1}{M_p} \right)^E
 \left( \frac{H}{4 \pi \, M_p}
 \right)^{2L}  \mfF_{d = 0} \; \mfF_{d = 2} \;\mfF_{d > 2}   \,,  
\ee
where vertices coming from the $d_n = 0$ terms of the lagrangian (the scalar potential) contribute the factor 
\be  \label{PCresult2x}
\mfF_{d = 0} = \prod_{n} \left[ \lambda_n \left( \frac{v^4}{H^2 M_p^2}
 \right) \right]^{V_n} \qquad \hbox{(if $f = M_p$)}   \,,  
\ee 
while the 2-derivative terms and higher-derivative terms contribute
\be  \label{PCresult2y}
 \mfF_{d=2} = \prod_{n}  c_n^{V_n}  \qquad \hbox{(if $f = M_p$)}  \,, \ 
\ee 
and
\be  \label{PCresult2z}
  \mfF_{d>2} = \prod_{n} \left[ g_n \left( \frac{H}{M_p}
 \right)^2 \left( \frac{H}{M} \right)^{d_n-4} \right]^{V_n}   \qquad \hbox{(if $f = M_p$)}   \,.
\ee 
In all three of these expressions the product only runs over those values of $n$ that correspond to vertices that have the number of derivatives as indicated (so $d_n = 0$ for \pref{PCresult2x}, $d_n = 2$ for \pref{PCresult2y} and $d_n > 2$ for \pref{PCresult2z}).

Eqs.~\pref{PCresult2} through \pref{PCresult2z} are key results because they quantify very explicitly the size of corrections to semiclassical methods.   
\begin{itemize}
\item The appearance of only positive powers of $H$ (except for within $\mfF_{d=0}$ -- more about which below) again verifies that the derivative expansion controls perturbative calculations made using \pref{Leffdef}, with perturbation theory relying on there being a hierarchy $H \ll M_p$ and $H \ll M$. 
\item For a fixed number of external lines $E$ each loop costs a factor of $H^2/(4\pi M_p)^2$, so this is again what controls the corrections to the semiclassical expansion. 
\item Once again there are no low-energy penalties for using as many 2-derivative interactions as we like, either from the Einstein-Hilbert lagrangian and from the sigma-model term $M_p^2 \, G_{ij} \, \partial \phi^i \partial \phi^j$. At low energies the full nonlinearity of GR remains important and the two-derivative interactions in the sigma-model term like to compete with GR.
\item  The scalar-potential terms appearing in $\mfF_{d=0}$ are generically dangerous because in them $H$ appears in the {\it denominator} rather than the numerator. This acts to undermine the validity of the semiclassical approximation because including zero-derivative interactions in a graph can amplify its size and make it no longer subdominant to other graphs that were naively bigger. This can make semiclassical methods suspect in surprising ways. This issue does {\it not} pose a problem for cosmological models if the low-energy scale $H$ is the Hubble scale and if the potential is what generates the Hubble curvature, since in this case the Friedmann equation implies $H \simeq v^2/M_p$, and so connects the size of $H$ to the scale $v$ in the potential. When this is true the potentially dangerous factor $\mfF_{d = 0}$ becomes
\be \label{CosmoOKd0}
 \mfF_{d=0} = \prod_{n} \left[ \lambda_n \left( \frac{v^4}{H^2 M_p^2}
 \right) \right]^{V_n} \simeq \prod_{d_n=0} \lambda_n^{V_n} \,,
\ee
\end{itemize}

As before, for any low-energy process that is characterized by a single scale $H \ll M \ll M_p$ the dominant contributions to observables come from graphs with $L = 0$ and $V_{id} = 0$ for all $d > 2$; {\it i.e.}~tree graphs constructed using just from the $d\leq 2$ terms in the action: the Einstein-Hilbert term with the sigma model and scalar-potential terms.  The leading corrections are again generated by loops involving these interactions plus tree graphs that include precisely one 4-derivative interaction, and so on.
 
Notice that there is no particular low-energy penalty working with large fields $\phi \gsim M_p$ {\it provided} the functions $U(\theta)$, $G_{ij}(\theta)$ and the like remain order unity when $\theta$ is order unity. That is, if large fields do not also imply large energy then they need not cause difficulties with the low-energy limit.  

\subsubsection{More general $f$}
\label{sssec:GeneralF}

Before continuing we pause here briefly to record how power-counting formulae like \pref{PCresult2} change if we relax the condition $f \simeq M_p$, typically with scales $f \ll M_p$ in mind. Rather than setting things up from scratch again it is easier to just flag the changes relative to the estimate made above when $f = M_p$. 

Inspection of \pref{Leffphih} and \pref{cnkindef} shows that there are two sources of change. One of these is the explicit factor of $f^2/M_p^2$ that now appears in the Feynman rule for any vertex coming from expanding the sigma-model interaction $G_{ij} \, \partial \phi^i \, \partial \phi^j$ about the background. This has the effect of multiplying the factor $\mfF_{d=2}$ given in \pref{PCresult2y} by a factor $\prod_n ( {f^2}/{M_p^2} )^{V_{\sigma n}}$ where $V_{\sigma n}$ counts the number of vertices of type `$n$' that come specifically from the sigma-model interaction. For 2-derivative interactions coming from the Einstein-Hilbert term there is no change and $\hat c_n = c_n$ (rather than their being related by \pref{cnkindef}).

The other change comes because the scalar fields appear as $\phi/f$ in \pref{Leffphih} rather than as $\phi/M_p$ as was used earlier when studying the limit $f = M_p$. This is corrected by taking every appearance of $\phi$ in any interaction and rescaling $\phi \to  (M_p/f)\, \phi$. This introduces a new factor into the amplitude $\cA_\ssE(H)$ of the form $\prod_n \left( {f}/{M_p}\right)^{- \mfs_n V_n}$, where the product is over all vertices (not just those coming from the sigma-model interaction) and $\mfs_n$ counts the number of scalar lines that converge on vertex $n$ (and $\mfh_n$ similarly counts the number of $h_{\mu\nu}$ lines that converge on this vertex). 

Combining both sources of $f$-dependence again leads to expression \pref{PCresult2} -- repeated again here:
\be  \label{PCresult22}
 \cA_\ssE (H)  \simeq  M_p^2 H^2 \left( \frac{1}{M_p} \right)^E
 \left( \frac{H}{4 \pi \, M_p}
 \right)^{2L}  \mfF_{d = 0} \; \mfF_{d = 2} \;\mfF_{d > 2}   \,,  
\ee
but with \pref{PCresult2x} through \pref{PCresult2z} replaced by
\be  \label{PCresult2xf}
\mfF_{d = 0} = \prod_{n} \left[ \lambda_n \left( \frac{v^4}{H^2 M_p^2}
 \right) \right]^{V_n} \left( \frac{f}{M_p}\right)^{- \mfs_n V_n}   \qquad \hbox{(general $f$)}   \,,  
\ee 
while the 2-derivative terms and higher-derivative terms contribute
\be  \label{PCresult2yf}
 \mfF_{d=2} = \prod_{n}  \hat c_n^{V_n}  \left( \frac{f}{M_p}\right)^{2V_{\sigma n} -  \mfs_n V_n}  \qquad \hbox{(general $f$)}   \,, \ 
\ee 
and
\be  \label{PCresult2zf}
  \mfF_{d>2} = \prod_{n} \left[ g_n \left( \frac{H}{M_p}
 \right)^2 \left( \frac{H}{M} \right)^{d_n-4} \right]^{V_n}   \left( \frac{f}{M_p}\right)^{- \mfs_n V_n}  \qquad \hbox{(general $f$)}     \,.
\ee 
As before the products run over the vertices that have the number of derivatives given by $d$.

The power of $f/M_p$ can be rewritten using three very useful identities, which hold for any graph consisting of $I_s$ scalar internal lines and $I_h$ tensor internal lines. The first of these is simply the definition of the number of loops in a graph:\footnote{This definition gives the intuitive number of loops for any graph that is topologically a disc -- {\it i.e.}~can be drawn on a page (draw some graphs and check) -- but is the definition of $L$ regardless of the graph's topology.}
\be \label{Loopdef}
   1 = L - I + \sum_n V_n \qquad \hbox{(definition of $L$)} 
\ee
where $I = I_s + I_h$ is the total number of internal lines. The other two identities express `conservation of ends' (which says the number of ends of external and internal lines must equal the number of ends appearing in all vertices, separately for both scalar and tensor lines):
\be \label{ConservationOfEnds}
    E_s + 2I_s = \sum_n \mfs_n V_n \qquad \hbox{and} \qquad
    E_h + 2I_h = \sum_n \mfh_n V_n \,.
\ee
Recall here that $\mfs_n$ and $\mfh_n$ respectively count the number of scalar and tensor lines that converge at vertex `$n$'. These last two identities can be used to eliminate $I_s$ and $I_h$ from any expression, and after this is done \pref{Loopdef} implies the number of vertices and external lines are related to the number of loops by the following expression:
\be
    E + 2(L-1) = \sum_n \Bigl[ (\mfs_n + \mfh_n) - 2 \Bigr] V_n \,,
\ee
where $E = E_s + E_h$. 

This is useful because it allows the power of the product of the factors $(f/M_p)^{-\mfs_n V_n}$ appearing in \pref{PCresult2xf} through \pref{PCresult2zf} to be written as 
\be
   - \sum_n  \mfs_n V_n  = 2(1-L) - E + \sum_n \Bigl( \mfh_n - 2 \Bigr) V_n \,.
\ee
This in turn allows \pref{PCresult22} to be cast in a way where the factors of $f/M_p$ mostly have positive powers:
\be   \label{PCresult222}
 \cA_\ssE (H)   \propto   f^2 H^2 \left( \frac{1}{f} \right)^E  \left( \frac{H}{4 \pi  f}
 \right)^{2L}  \widetilde \mfF_{d = 0} \; \widetilde \mfF_{d = 2} \; \widetilde \mfF_{d > 2}   \,,  
\ee
where  
\be  \label{PCresult2xfg}
 \widetilde\mfF_{d = 0} = \prod_{n} \left[ \lambda_n \left( \frac{v^4}{H^2 f^2}
 \right) \right]^{V_n} \left( \frac{f}{M_p}\right)^{\mfh_n V_n}   \qquad \hbox{(general $f$)}   \,,  
\ee 
while the 2-derivative terms and higher-derivative terms contribute
\be  \label{PCresult2yfg}
 \widetilde\mfF_{d=2} = \prod_{n}  \hat c_n^{V_n}  \left( \frac{f}{M_p}\right)^{(\mfh_n-2)V_{\EH n} + \mfh_n V_{\sigma n}}  \qquad \hbox{(general $f$)}   \,, \ 
\ee 
and
\be  \label{PCresult2zfg}
 \widetilde \mfF_{d>2} = \prod_{n} \left[ g_n \left( \frac{H}{f}
 \right)^2 \left( \frac{H}{M} \right)^{d_n-4} \right]^{V_n}   \left( \frac{f}{M_p}\right)^{ \mfh_n V_n}  \qquad \hbox{(general $f$)}     \,.
\ee 
Eq.~\pref{PCresult2yfg} uses that 2-derivative interactions must either come from the sigma-model term or the Einstein-Hilbert term -- {\it c.f.}~the lagrangian \pref{Leffdef} -- so for them $V_n = V_{\sigma n} + V_{\EH n}$.  

As a check, consider evaluating a graph with $L$ loops and $E_s$ external scalar lines and $E_h$ external tensor lines, for which {\it all} of the vertices come from the sigma-model interaction $G_{ij} \, \partial \phi^i \, \partial \phi^j$). In this case the only vertices present have $d_n = 2$ derivatives so $\mfF_{d=0} = \mfF_{d>2} = 1$. Since every vertex comes from the sigma-model interaction we also have $V_n = V_{\sigma n}$ for all nonzero $V_n$'s. Under these assumptions the dependence on $M$, $f$, $v$, and $M_p$ comes from using \pref{PCresult2yf} in \pref{PCresult22}. Combining everything implies either \pref{PCresult22} or \pref{PCresult222} can be written
\be \label{PCresult22sigma}
 \cA_\ssE (H)  \propto  f^2 H^2   \left( \frac{1}{f} \right)^{E}   \left( \frac{H}{4 \pi  f}
 \right)^{2L}  \left( \frac{f}{M_p}\right)^{\sum_n \mfh_n V_n}  \,.
\ee 
In the special case where $h_{\mu\nu}$ does not appear in the graph we have $\mfh_n = 0$ and this reproduces the standard sigma-model power-counting expression \cite{Weinberg:1978kz} -- which is basically \pref{PCresult2} with $\mfF_{d=0} = \mfF_{d>2} = 1$ and $M_p \to f$. More generally $\sum_n \mfh_n V_n \geq E_h$ since any external metric line must end somewhere, so an amplitude with $E_s$ external scalars and $E_h$ external metric fluctuations always has a proportionality constant of at least $(1/M_p)^{E_h} (1/f)^{E_s}$.

A second useful special case is the result for graphs with no metric external or internal lines: $E_h = I_h = 0$ and so from \pref{ConservationOfEnds} we also have $\mfh_n = 0$. Because we work in Einstein frame the only interactions coming from the Einstein-Hilbert term involve the metric and so for a scalar-only graph we can take $V_{\EH n} = 0$. In this case eqs.~\pref{PCresult222} through \pref{PCresult2zfg} imply
\be    \label{PCresult2fScalarOnly}
 \cA_\ssE (H)    \propto  f^2 H^2 \left( \frac{1}{f} \right)^{E_s}  \left( \frac{H}{4 \pi  f}
 \right)^{2L}  \prod_{d_n = 0} \left[ \lambda_n \left( \frac{v^4}{H^2 f^2}
 \right) \right]^{V_n}     \prod_{d_n > 2} \left[ g_n \left( \frac{H}{f}
 \right)^2 \left( \frac{H}{M} \right)^{d_n-4} \right]^{V_n} \,.
\ee 
Recall that the scalar potential depends on the coefficients $\lambda_n$ through \pref{assumedVform2}.

\subsection{Lessons for cosmology}
\label{sec:Naturalness}

Nothing {\it requires} UV physics to provide new light fields for us to discover, but the fact that cosmology only makes sense if well-understood matter is supplemented by Dark Energy and Dark Matter seem to suggest that new light fields might well be present in Nature. It is also true that the few theories we have for which gravitational interactions make sense at the quantum level {\it at high energies} (such as string theory) usually predict the existence of a host of new particles that couple very weakly -- often only with gravitational strength -- to ordinary matter, of which the graviton is only one example. If any of these were to be light they could also be around to be discovered in cosmology or tests of GR (and might indeed play a role in Dark Matter or Dark Energy). For such new fields the previous section sets the general stage for what should be expected for them at low energies.  

It already tells us something interesting. First, it tells us is that what cosmologists like the most about scalar field phenomenology -- the scalar potential -- is also the thing that is the most dangerous at low-energies. The potential is important for several reasons. A key property for any field relevant to cosmology or tests of GR is that it is very light compared with other scales in particle physics, and for scalar fields this is encoded in the scalar potential. Similarly, we saw in \S\ref{ssec:SimpleModels} that for simple scalar models of Dark Energy the equation of state parameter tells us that the energy density is currently dominated by the scalar potential. 

The scalar potential actually contains several types of dangers at low energies. One of these is the appearance of inverse powers of the low-energy scale $H$ in expressions like \pref{PCresult2xf}, which work to undermine the low-energy expansion that underpins the entire strategy of analyzing the model semiclassically. This becomes more of a problem the stronger the zero-derivative scalar interactions are. But we've also seen how this need not be a problem for potentials with the structure $V = v^4 U(\phi)$ for generic order-unity functions $U(\phi)$, when the low-energy scale $H$ is the Hubble scale dominated by $V$, since in this case the dangerous factor in \pref{PCresult2x} cancels out because $H \sim v^2/M_p$, as in eq.~\pref{CosmoOKd0}. 

In this section we start, in \S\ref{ssec:Multiscalar}, by assuming the scalar potential is suppressed to the point that it does not overwhelm the two-derivative terms at low energies, returning in \S\ref{ssec:Naturalness} to how difficult this regime is to achieve and to discuss more broadly the problems raised at low energies by zero-derivative interactions.  

\subsubsection{Two-derivative interactions: more is different}
\label{ssec:Multiscalar}

If the potential is somehow suppressed at low energies then it is the sigma-model term $G_{ij} \, \partial \phi^i \, \partial\phi^j$ that provides the next most important interactions for scalars at low energies. These scale at low-energies in precisely the same was as do the interactions in the Einstein-Hilbert action of GR, which also involve precisely two derivatives. Two-derivative interactions can compete with one another at all energies without undermining the underlying derivative expansion, and with it the validity of semiclassical methods themselves. 

But (as mentioned above) these cannot have physical implications if they can be removed by a field redefinition, and this can always be done if the target-space metric $G_{ij}(\phi)$ is flat. An important example where the target space is flat is the case of a single scalar field. For a single field it is always possible to redefine $\phi = \phi(\psi)$ such that $G(\phi) \, (\partial \phi)^2 = (\partial \psi)^2$, with $\phi(\psi)$ found by integrating $G(\phi) \, \exd \phi = \exd \psi$. Because of this, in the special case of a single scalar field the leading nonminimal couplings are the higher-derivative interactions appearing in the second line of \pref{Leffdef}. 

A large effort has been made to characterize the kinds of higher derivative interactions that can arise for a single scalar field, and how these can contribute to classical observables (see for example \cite{Horndeski}).\footnote{It is sometimes argued that it is necessary to restrict to a subset of higher-derivative interactions in order to avoid the Ostrogradski instability \cite{Woodard:2015zca} that is generic to theories with higher derivative interactions. This turns out not to be necessary when working with low-energy EFTs because of the perturbative nature of the $1/M$ expansion \cite{Burgess:2014lwa}, although in practice it turns out that the distinction between these points of view only arises at relatively high order in $1/M$ \cite{Solomon:2017nlh}.} There is an important conceptual problem with these analyses, however: to the extent that an interaction with 4 or more derivatives starts to compete with GR the power-counting argument that culminated in eqs.~\pref{PCresult2} through \pref{PCresult2z} shows that it is necessarily true that the derivative expansion itself must be breaking down. But \pref{PCresult2} also shows that when this is true there is no justification for analyzing these systems classically while neglecting loop effects.

But this unfortunate conclusion is specific to the restriction to a single scalar field (or to multiple fields with a flat target space metric since for these $G_{ij} = \delta_{ij}$ can be arranged using a field redefinition). There is a premium for exploring models with at least two low-energy scalar fields because these are the only ones that allow the nontrivial sigma-model interactions that scale at low energies in the same way as do the interactions of GR itself. 

If we work within the class of effective theories with scalar potentials of the form $V = v^4 U(\theta)$, where $U(\theta)$ is a generic order-unity function, then there is also an argument why it is not especially unlikely to have more than one field be very light. After all, the scalar mass matrix implied by this form for the potential is
\be
  M^2_{ij} := \frac{\partial^2 V}{\partial \phi^i \partial \phi^j}  = \left( \frac{v^4}{f^2} \right) \frac{\partial^2 U}{\partial \theta^i \partial \theta^j} \,,
\ee
so if all of the derivatives of $U$ are order unity the mass eigenvalues are all of order $m \simeq v^2/f$. In particular, for gravitationally coupled scalars $f \sim M_p$ and so these masses are all the same order as $H \sim v^2/M_p$. As we shall see, the puzzle with potentials with this kind of structure is why $v$ should be so small, but once it is any gravitationally coupled scalar appearing in it is necessarily very light. 

Minimal multi-scalar models of this type will be encountered again in \S\ref{sec:WaysForward} below.

\subsubsection{Lots of potential}
\label{ssec:Potential}

Returning to the scalar potential, we next ask how generic it is to have a potential of the form $V = v^4 U(\theta)$ with small $v$. The problem is that this form seems not to be easy to obtain at low energies from theories like the ones we believe describe nature on much shorter length scales. The good news is that it is not completely impossible either, and trying to find how they can arise seems to be an important clue when searching for descriptions of cosmology.

The basic problem can already be seen in the lagrangian \pref{Leffdef}, in which it was argued that the various terms are generically generated by integrating out multiple heavy particles. This leads to the expectation that an interaction $g \cO \in \cL_{\rm eff}$ with operator, $\cO$, with mass-dimension $n$ arises with a coefficient $g$ that has mass-dimension $4 - n$ (in 4 spacetime dimensions) so that $g\cO$ has dimension (mass)${}^4$, just like $\cL_{\rm eff}$. For instance, an operator like $\cO = R^3$ has mass-dimension $n = 6$ and so its coupling has dimension (mass)${}^{-2}$, as indicated in \pref{Leffdef}.

Intuitively, integrating out a particle of mass $M$ should generate a contribution to each effective operator proportional to the appropriate power of $M$ dictated by dimensional analysis, so $\delta g \propto M^{4-n}$. If $g$ is obtained by summing many such contributions then the dominant mass $M$ controlling the size for each $g$ depends on the sign of $4-n$: it is the smallest mass that dominates in $g$ when $n > 4$ and it is the largest mass that dominates when $n < 4$. 

This intuition can be made more precise using the power-counting arguments of the previous sections by applying them to the graphs used to compute contributions to $\cL_{\rm eff}$ when a light particle of mass $m$ is integrated out. To this end 
	suppose we have a scalar potential $V = m^4 U(\phi/m)$ and kinetic term $(\partial \phi)^2$, as might reasonably be chosen to describe a scalar with mass $m$. (The Higgs potential in the Standard Model, for example, has a potential and kinetic term of this type with $m \sim 100$ GeV.) Integrating out this type of particle gives a contribution to $\cL_{\rm eff}$ and its size turns out to be the found by computing the generating functional for amputated 1-particle irreducible (1PI) correlation functions,\footnote{1PI correlation functions are computed by evaluating amputated Feynman graphs that cannot be broken into two disconnected graphs by cutting a single internal line.} and so can be estimated by specializing the power-counting argument given above to the choices $f \sim v \sim m$. 

A term in \pref{Leffdef} involving $E$ factors of fields $\phi^i$ depends on scales like $M$, $M_p$, $f=v$ in same way as does $\cA_\ssE$ where the low-energy scale $H$ can either be $m$ or an external momentum (or derivative) $q$. The power counting estimate for the size of the $L$-loop coefficient of $\phi^2$ in the scalar potential is therefore $\delta m^2 \sim \cA_2(m)$ (since there are no derivatives). An estimate of the size of the correction to the kinetic term $\cG (\partial \phi)^2$ is order $(q/m)^2 \cA_2(m)$ because it must be precisely quadratic in $q$, and so $\delta\cG \sim m^{-2} \cA_2(m)$. The contribution to the coefficient of $\cH \phi^2 (\partial \phi)^2$ is similarly order $\delta \cH \sim m^{-2} \cA_4(m)$, and so on. 

Using $f \sim v \sim H \sim m$ in \pref{PCresult2fScalarOnly} therefore gives (in order of magnitude)
\be \label{massshift}
  \delta m^2 \sim   m^2   \left( \frac{1}{4 \pi }
 \right)^{2L} \left(  \prod_{d_n = 0}  \lambda_n^{V_n}    \right)    \prod_{d_n > 2} \left[ g_n \left( \frac{m}{M} \right)^{d_n-4} \right]^{V_n} 
\ee
and
\be
  \delta \cG \sim  \left( \frac{1}{4 \pi }
 \right)^{2L} \left(  \prod_{d_n = 0}  \lambda_n^{V_n}    \right)    \prod_{d_n > 2} \left[ g_n \left( \frac{m}{M} \right)^{d_n-4} \right]^{V_n} 
\ee
and
\be
  \delta \cH \sim \frac{1}{m^2}  \left( \frac{1}{4 \pi}
 \right)^{2L} \left(  \prod_{d_n = 0}  \lambda_n^{V_n}    \right)    \prod_{d_n > 2} \left[ g_n \left( \frac{m}{M} \right)^{d_n-4} \right]^{V_n} 
\ee
which show the expected dimensional dependence on the mass $m$ of the particle that is integrated out (plus corrections in powers of $m/M$). What we considered above as a light field of mass $m$ would also be regarded as a heavy field from  the point of view of any other, even lighter, field. From the point of view of this lighter field the particle of mass $m$ is just another heavy field and so $m$ would be lumped among the $M$'s, which after all were masses of particles that had previously been integrated out. Because $m \ll M$ the contribution (say) $\delta \cH \sim m^{-2}$ is much bigger than $\delta \cH \sim M^{-2}$, showing in more detail why in practice we can take $M$ to be the mass of the lightest UV field that was integrated out.  

What is important is that the mass appearing on the right-hand side of an expression like \pref{massshift} need not be the same as the mass being corrected on the left-hand side. For instance if we had two scalars, a scalar $\sigma$ that is massless and a scalar $\phi$ with mass $m$ that are coupled to one another by a term in $V$ like $g \sigma^2 \phi^2$ then the potential would start off looking like 
\be
  V = \tfrac12 m^2 \phi^2 + g \, \sigma^2 \phi^2 = m^4 \left( \frac{\phi^2}{2m^2} +  \frac{g \,\phi^2 \sigma^2}{m^4} \right) \,,
\ee
and the leading correction to the $\sigma$ mass due to integrating out $\phi$ would still be given by \pref{massshift}, with $L = 1$ and $\lambda_n =  g$, implying $\delta m_\sigma^2 \sim g \, m^2/(4\pi)^2$. 

From this point of view, when a sequence of heavy particles is integrated out it seems reasonable to find an enormous coefficient like\footnote{One might ask why the above argument ever stops -- {\it i.e.}~why should the Planck scale emerge as the `heaviest' scale rather than something even bigger? Heaviest here means heaviest until one of the assumptions going into this EFT power-counting fails. This can happen if one reaches an energy scale above which a tower of states emerges, such as Kaluza-Klein (KK) modes if the high-energy theory is higher-dimensional. EFT methods never capture the power-counting appropriate to infinite towers of states because there is no longer a hierarchy of scales to exploit \cite{Burgess:2020tbq, Burgess:2023pnk}. If the tower encountered is a KK tower the effective description transitions to a higher-dimensional field theory, in which similar estimates would apply up to a scale where this description also breaks down (such as the onset of a tower of string states). This is why calculations in string theory usually have the form given above but with the string scale playing the role of the largest mass scale (which becomes powers of $M_p$ once factors of the extra-dimensional size and the string coupling are included).} $M_p^2$ in front of the Einstein-Hilbert term given that for it the mass-dimension is $n = 2$ and so loops of the heaviest fields should dominate. {\it But all other things being equal it is not generic to find small contributions to interactions with mass-dimension $n < 4$.} This is true in particular for the vacuum energy (a field-independent piece in $V$ for which $n= 0$) and for scalar masses ($V=\tfrac12 m^2 \sigma^2$ for which $n=2$). Cosmological applications require both of these to be small, whereas they naively should appear with prefactors $M_p^4$ and $M_p^2$ respectively. 

How can this be reasonable? As the rest of these notes argue: the key part of the above italicized phrase is `all other things being equal'. All other things need {\it not} be equal and understanding how cosmology can emerge from the low-energy limit of something more fundamental provides a crucial clue for unravelling what is really going on.

\section{Naturalness: Tomorrow's Hope or Yesterday's News?}
\label{ssec:Naturalness}

One attitude to take about the corrections to $\cL_{\rm eff}$ estimated above is that they do not matter. This section discusses this attitude and argues why -- despite being a self-consistent point of view -- it has not put the discussion to rest.

\subsection{The issue}
\label{ssec:Issue}

Why should we care if the coefficient of the Einstein-Hilbert term arises as a sum over the masses of heavy particles when it is in any case only their sum $M_p^2$ that we directly measure? 

This question becomes even more pointed when it is recalled that the Feynman graphs performed when generating this sum actually diverge in the UV.\footnote{Recall that we ignored these divergences when making dimensional estimates because we chose to regularize them using dimensional regularization (rather than some sort of a UV cutoff). We imagine renormalizing using a dimensionless scheme like minimal subtraction, and all masses in this section are renormalized in this way (so our expressions are not UV divergent).} Such divergences represent the contributions of the shortest wavelengths of the theory and we normally do not angst too much about their size since they are in the end of the day absorbed into a renormalization of otherwise unknown parameters in $\cL_{\rm eff}$ (such as $M_p^2$). It is only the final renormalized combination that is measurable. If divergences can be ignored like this, why should finite-but-large contributions proportional to heavy masses be any more worrisome?

Another way to phrase this attitude is to observe that renormalized parameters in any lagrangian run as a function of scale, and are typically only measured at a particular scale. All that matters is the value of a parameter (like the mass of a light field) at the scale where it is measured to be small. Why do we care if it is not small at some other scale?

To answer this it helps to have a concrete example in mind. Suppose we have a fundamental theory with two types of massive particles involving a hierarchy of mass scales $m \ll M$. We imagine there being a fundamental theory describing physics at energies $E > M$ for which the scalar potential contains terms like
\be \label{VTheory1}
   V_\UV = \widetilde V_0 + \tfrac12 \Bigl( \widetilde m^2 \phi^2 + \widetilde M^2 \psi^2 \Bigr) + g \, \phi^2 \psi^2 + \lambda_\phi \phi^4 + \lambda_\psi \psi^4 + \cdots \,.
\ee 
The physical masses of these particles (as measured by experiments say) are related to the parameters in the lagrangian (including the leading loop correction) by formulae like
\be \label{mphysfull}
  m^2_{\rm phys}  =  \widetilde m^2 + \hbox{($\psi$-loop)} + \hbox{($\phi$-loop)}  + \cdots
  =  \widetilde m^2 +\frac{C_1 \, g }{(4\pi)^2}\, \widetilde M^2 +  \frac{C_2 \lambda_\phi }{(4\pi)^2} \, \widetilde m^2 + \cdots\,,
\ee 
and
\be  
  M^2_{\rm phys} = \widetilde M^2 + \hbox{($\psi$-loop)} + \hbox{($\phi$-loop)} + \cdots 
  = \widetilde M^2+\frac{C_3 \lambda_\psi }{(4\pi)^2}\, \widetilde M^2  +   \frac{C_4 \, g}{(4\pi)^2} \, \widetilde m^2+ \cdots \,,
\ee
in which $C_i$ are order-unity dimensionless constants. 

For applications to low energies we can integrate out $\psi$ and work with the EFT involving only $\phi$, whose scalar potential includes
\be \label{VTheory2}
   V_{\rm eff} =  V_0 + \tfrac12 m^2  \phi^2   + \lambda_\phi \phi^4 + \cdots \,.
\ee 
Explicitly performing the integral over $\psi$ allows the new parameters to be computed in terms of the old ones, leading to
\be \label{matching}
    V_0 \simeq \widetilde V_0 + \frac{C_5}{(4\pi)^2}\, \widetilde M^4 \,, \qquad 
   m^2 \simeq \widetilde m^2 + \frac{C_1 \,g}{(4\pi)^2} \,   \widetilde M^2  
\ee
for another dimensionless constant $C_5$. The calculation of the loop-corrected physical mass in the low-energy theory then is
\be 
   m^2_{\rm phys} =  m^2   +  \frac{C_2 \lambda_\phi }{(4\pi)^2} \,   m^2 + \cdots \,,
\ee
which agrees with \pref{mphysfull} (to the order we work) because of \pref{matching}.

Now comes the main point. Weak coupling and the absence of very heavy particles in the low-energy theory imply $m_{\rm phys}$ and $m$ are approximately equal. But $m_{\rm phys}$ is an observable and so must have the same value for both the full theory or the low-energy EFT, so if $\widetilde M$ is sufficiently large then $\tilde m$ cannot be similar in size to $m_{\rm phys}$. Instead it must be large and approximately opposite in sign to the loop correction $C_2 \, g \widetilde M^2/(4\pi)^2$ so that these terms mostly cancel to leave a small value for $m^2$. The required cancellation can be extremely accurate. For instance if $g/(4\pi)^2 \sim 10^{-3}$, $\widetilde M \sim 10^{15}$ GeV and $m_{\rm phys} \sim 10^2$ GeV -- reasonable choices within a Grand Unified Theory (GUT) \cite{GUT}, say -- the cancellation must occur to more than 23 decimal places.

\subsection{Technical naturalness}
\label{sssec:TechNat}

The kind of cancellation described above is remarkable for several reasons. First, while it is true that parameters in the lagrangian run and it is true that parameters that are small at some scales need not be small at other scales what is odd about the above is not that the parameter $\widetilde m^2$ is much larger than $m^2$ in the high-energy theory. The odd thing is the extreme precision with which the value of $\widetilde m^2$ must be chosen. The basin of attraction for the flow of $\tilde m^2$ that leads to acceptably small values for $m^2$ is extraordinarily narrow. Couplings for which the high-energy couplings must take extraordinarily accurate values to reach acceptably small sizes at low energies are called `fine-tuned'. 

A second, more telling, remarkable feature of the fine-tuning described above is the fact that it is unusual. That is, there are many hierarchies of scale known in nature and none of the ones we understand actually work this way. Normally we never are required to deal with all of the degrees of freedom in the universe all at once (thank God) so our description is cast in terms of some sort of EFT. But there is not a unique choice of effective lagrangian since different EFTs apply at different scales. Normally whenever a system has a hierarchy of scales -- like $m \ll M$ in the above example -- the hierarchy can be understood in {\it any} of these EFTs, and not just in the EFT describing the scales where the smaller scale is measured.  

A famous example of a hierarchy is the large size of atoms relative to nuclei: $a_\ssB \sim 10^5 r_\ssN$. We can describe this in the EFT below 100 MeV in which the basic particles are protons and electrons and in this EFT $r_\ssN \sim m_\pi^{-1}$ is a given parameter (hadron substructure has been integrated out) that is of order the inverse of the pion mass (since pion exchange mediates nuclear interactions at long distances), while atomic radii are of order $a_\ssB \sim (\alpha m_e)^{-1}$ where $\alpha \sim 0.01$ is the electromagnetic fine-structure constant and $m_e$ is the electron mass. Atoms are larger than nuclei because $\alpha$ is small (electromagnetic interactions are weak) and the electron is much lighter than the pion. 

But the same question can be asked in the effective theory (for instance the Standard Model) applicable at energies much larger than 100 MeV. In this theory protons are described as bound states of quarks and gluons and the size and mass of the nucleus are set by the size of the QCD scale $\Lambda_\QCD$ while atomic radii are again given by $(\alpha m_e)^{-1}$. In this theory one can in principle compute $r_\ssN$ in terms of $\Lambda_\QCD$ and can also compute how the values of parameters like $\alpha$ and $m_e$ in the lagrangian change once the physics above 100 MeV is integrated out. In practice $r_\ssN \sim \Lambda_\QCD^{-1}$ and the change in $\alpha$ is order $\alpha^2$ and the change in the electron mass is proportional to the electron mass, $\delta m_e \propto m_e$ and so these changes are not large. No fine-tuning is needed to ensure that nuclei are small compared to atoms; there is a clear reason in each EFT we choose to ask the question. 

The atom/nucleus example is the rule not the exception: with the exception of the ones arising in cosmology (such as why the vacuum energy is small -- more about which below) all the well-understood examples of scale hierarchy satsify two properties:
\begin{itemize}
\item There is an understanding within the fundamental theory at high scales (such as within the Standard Model) why the hierarchy holds in the first place (such as because some ratio of parameters like $m/M$ is small).
\item There is an understanding of why the parameter choices necessary for the hierarchy {\it stay} small as successive layers of physics are integrated out to reach the lower energies where the parameters are measured. The hierarchy has an understanding in {\it all} the EFTs describing scales in between.
\end{itemize}
A hierarchy that satisfies both of these criteria is called {\it technically natural}.\footnote{The qualifier `technically' is needed here to distinguish from many other things that are often called `natural', some of which are only aesthetic.} Our understanding of the relative size of nuclei and atoms is technically natural in this sense, while the understanding of why $m \ll M$ in the two-scalar model discussed above is not technically natural. The question of how to understand the small size of the vacuum energy density in a technically natural way is widely known as the {\it cosmological constant problem}.\footnote{This problem pre-dated the discovery of evidence for Dark Energy because it was equally puzzling why the vacuum energy could be consistent with zero, given the much larger scales arising in particle physics. The question of why the Dark Energy density takes precisely the value it is observed to have rather than being for some reason exactly zero is sometimes called the `new' cosmological constant problem.}

It is conservative to ask that our understandings of other more poorly understood hierarchies should also be technically natural, since this just extrapolates what we know to be true in all other instances we do understand. But it is not compulsory. It might be that Nature does not care and theories like the two-scalar model are self-consistent even if fine-tuned. Technical naturalness is a very useful clue however because the ingredients needed to make a theory technically natural cannot just involve particles at inaccessibly high energies, like the Planck scale. If this were true we could integrate them out at low energies and the problem with technical naturalness problem is back. The mechanism that keeps the small parameter small in a natural way usually has other consequences at low energies and this tends to make these theories easier to test than the alternatives.   

The next few sections describe approaches that have been tried to make the scalar field theories of interest to cosmology technically natural. This is a restrictive criterion because it is not generic that the things that are appealing for cosmological models (like very small scalar masses, $m^2$, and small vacuum energies, $V_0$, in a scalar potential) are technically natural. One hopes in this way to identify a well-motivated subset of the very many cosmological models on the market. Although observations alone cannot yet distinguish amongst the many models the hope is that observations together with technical naturalness can be a much more efficient filter. As we shall see, it is much easier to understand why scalar masses can be small in a technically natural way than it is to do the same for the vacuum energy density.

\subsection{The Usual Suspects (symmetries)}
\label{sec:UsualApproach}

Experience with other (noncosmological) hierarchies teaches that there is an important general mechanism for making small parameters technically natural: symmetries. Symmetries are usually preserved by quantum corrections (anomalies are the exceptions) so symmetry breaking term in a lagrangian are not generated by loops if the initial theory doesn't have them  (and so respects the symmetry). 

This also means that if a symmetry is only approximate -- {\it i.e.}~is broken by interactions with small coupling parameters, $\epsilon_i \ll 1$ -- then loop corrections to these parameters satisfy:\footnote{Strictly speaking $\delta \epsilon$ need not be linear in $\epsilon$. All that is required is that it vanish as $\epsilon \to 0$.}
\be
   \delta \epsilon_i \simeq {C_i}^j \epsilon_j \,, 
\ee
for some matrix of coefficients ${C_i}^j$ since the corrections must vanish if the parameters themselves vanish (because then the symmetry is then unbroken). Having $\delta \epsilon$ be proportional to $\epsilon$ forbids getting large contributions to otherwise small parameters (like $\delta m^2 \propto M^2$ where $M \gg m$) and so help understand why small parameters can be technically natural.

For example suppose we have two scalar fields, $\phi$ and $\psi$, and the action for them is invariant under a symmetry of the form 
\be \label{rotation}
     \left( \begin{matrix} \phi \cr \psi \end{matrix} \right) \to  \left( \begin{matrix} \cos \omega & \sin \omega \cr - \sin \omega & \cos \omega \end{matrix} \right)  \left( \begin{matrix} \phi \cr \psi \end{matrix} \right) 
\ee 
for arbitrary constant $\omega$. The mass term allowed by this symmetry is $\frac12 m^2 (\phi^2 + \psi^2)$ and so the symmetry requires both particles to have equal masses. Imagine now the theory is supplemented by a symmetry-breaking term of the form $\delta V = \frac12 \mu^2 (\phi^2 - \psi^2)$ that splits these masses, with $\mu^2 \ll m^2$ and all other interactions are invariant under \pref{rotation}. Then $\delta \mu^2$ must be proportional to $\mu^2$ -- as opposed to, say $m^2$ -- and the parameter $\mu^2$ is technically natural. Importantly this is true even though corrections to scalar masses are otherwise {\it generically} expected to be dominated by the largest mass scales. The largest mass still wins, but the symmetry argument only allows contributions from the largest mass that breaks the symmetry.

This is what actually happens for the electron mass in the Standard Model (the electron mass $m_e \simeq 0.5$ MeV is much smaller than are generic SM masses, which are more like 100 GeV). If the electron mass is set to zero then the Standard Model acquires a new `accidental' symmetry, under which its left- and right-handed parts rotate differently -- {\it i.e.}~invariant under a chiral rotation $\delta \chi = i \gamma_5 \chi$ of the electron field. This is why integrating out heavy degrees of freedom with mass $M \gg m_e$ only corrects the electron mass by $\delta m_e \propto m_e$ (as opposed to $\delta m_e \propto M$).

If a lagrangian has the property that it has more symmetry when a parameter is set to zero then corrections to that parameter tend to preserve its small size automatically. A parameter of this type is called {\it `t Hooft natural} \cite{tHooft}. If a small parameter is `t Hooft natural in this way it is also technically natural because the symmetry protects its corrections.\footnote{Although `t Hooft naturalness is sufficient for technical naturalness there are examples in supersymmetric theories that show that it is not strictly necessary. Supersymmetric theories (more about which below) have nonrenormalization theorems that can forbid quantum corrections in some circumstances even if the theory does not have a symmetry that makes it `t Hooft natural.}

Since symmetries can help understand why small scalar masses can be technically natural, we next list the symmetries that can do so. The first observation is that finding such a symmetry is harder for scalars than it is for fermions (like the electron example above). It is possible to have any nonzero value for a fermion be `t Hooft natural because fermion kinetic terms have a larger symmetry group than do fermion mass terms. That is, whereas the electron mass term $m_e \ol\chi \chi$ is invariant under a $U(1)$ transformation $\delta \chi = i \omega \chi$ for arbitrary $\omega$ the electron kinetic term $\ol \chi \dsl \chi$ is invariant under a $U(1) \times U(1)$ symmetry $\delta \chi = i \omega \chi + i \tilde \omega \gamma_5 \chi$ where both $\omega$ and $\tilde \omega$ are arbitrary. It is this extra $\tilde \omega$ symmetry that protects a small electron mass. For $N$ real scalars, $\phi^i$, however, the kinetic term $\frac12 \partial_\mu \phi^\ssT \partial^\mu \phi$ is invariant under arbitrary orthogonal $O(N)$ rotations amongst the scalars. But this is also the symmetry of a mass term $\frac12 m^2 \phi^\ssT \phi$ for any nonzero $m^2$. The kinetic term's $O(N)$ symmetry can force different scalars to have the same mass, but does not force their masses to be zero.   

\subsubsection{Shift symmetries}
\label{sssec:Shifts}

An example of a symmetry that {\it can} require a scalar mass to vanish is a `shift' symmetry: 
\be \label{shiftabelian}
   \phi \to \phi + \omega 
\ee
where $\omega$ is an arbitrary constant. Although this transformation is a symmetry of the kinetic term $\frac12 (\partial \phi)^2$ the only scalar potential that is invariant under \pref{shiftabelian} is a constant $V = V_0$ (independent of $\phi$). Scalars with a symmetry of this type are Goldstone bosons and their presence flags the existence of a spontaneously broken symmetry\footnote{Invariance under shifts is the smoking gun for spontaneous symmetry breaking because it is impossible for any particular classical background, $\phi = \phi_0$, to be invariant under \pref{shiftabelian}.}  ({\it i.e.}~a symmetry of the action but not the ground state) \cite{GBs}.

Evidently any scalar with this symmetry must be massless and cannot have any zero-derivative interactions at all (a special case of Goldstone's theorem). This does not mean such scalars do not interact at all, however. For instance they can couple to other fields $\psi$ through derivative interactions like $\partial_\mu \phi \, J^\mu(\psi)$. 

If there are multiple dimensionless scalars, $\theta^i$, then they can also interact amongst themselves (even at the two derivative level) if \pref{shiftabelian} is generalized to 
\be \label{shiftnonabelian}
  \delta \theta^i = \omega^\alpha \xi_\alpha^i(\theta) \,,
\ee
and there is no value $\theta^i_0$ for which all of the $\delta \theta^i$'s vanish, then although it is still true that the only invariant scalar potential must be a constant, the two-derivative $\sigma$-model interactions of the form $G_{ij}(\theta) \, \partial \theta^i \, \partial \theta^j$ can be invariant if for each $\alpha$  the following equation is satisfied
\be \label{KVF}
  D_i \xi_{\alpha j} + D_j \xi_{\alpha i} = 0 \qquad \hbox{where} \qquad \xi_{\alpha i} := G_{ij} \, \xi_\alpha^j
\ee
and $D_i\xi_j = \partial_i \xi_j - \Gamma^k_{ij} \xi_k$ is the covariant derivative built using the Christoffel symbol $\Gamma^k_{ij}$ for the target-space metric $G_{ij}$. Any solution $\xi_\alpha^i$ to \pref{KVF} is called a Killing vector field for the metric $G_{ij}$ and corresponds to a symmetry direction of the metric $G_{ij}$. Although not all metrics have such symmetries there is a broad class of nonflat metrics that do: the metrics on coset spaces of the quotient of two Lie groups: $G/H$ (such as spheres). These describe the interactions of Goldstone bosons for systems where the action has a symmetry group $G$ but the ground state is only invariant under a subgroup $H$. Because these metrics are not flat they contain nontrivial two-derivative interactions, consistent with the symmetry \pref{shiftnonabelian} \cite{CCWZ}.

These shift symmetries are overkill if our goal is only to have a technically natural scalar mass that is small but not exactly zero. The way to protect small nonzero masses is to have the symmetry \pref{shiftabelian} or \pref{shiftnonabelian} only be an {\it approximate} symmetry of the action. Scalars transforming as \pref{shiftnonabelian} under an approximate symmetry are called pseudo-Goldstone bosons. They can be systematically light when the symmetry breaking in the action is small because they must become honest-to-God massless Goldstone bosons in the limit that the symmetry breaking terms go away \cite{pGBs}. 

The low-energy lagrangian for pseudo-Goldstone bosons has the form
\be
   \cL_{pGB} = - \sqrt{-g} \Bigl\{ V_0 + \epsilon V_1(\theta) + f^2 \Bigl[ G_{ij}(\theta) + \epsilon H_{ij}(\theta) \Bigr] \partial^\mu \theta^i \, \partial_\mu \theta^j + \cdots \Bigr\} \,,
\ee
where $\epsilon$ represents a small symmetry-breaking parameter, $V_0$ is a constant (the only potential invariant under the symmetry) and $G_{ij}$ is an invariant target-space metric, but $V_1$ and $H_{ij}$ are {\it not} restricted to be invariant (and so are why the symmetry is only approximate).  The symmetry-breaking terms involving $V_1$ and $H_{ij}$ are `t Hooft natural because any corrections to them must be proportional to the small symmetry breaking parameter $\epsilon$. 

Notice in particular that if $V_1(\theta) = m^4 U(\theta)$ for some dimensionless function $U(\theta)$ then all of the field-dependence in the scalar potential has the form $V(\theta) = v^4 U(\theta)$ assumed earlier when power-counting, with $v^4 = \epsilon m^4 \ll m^4$. Small symmetry breaking $\epsilon$ for a group of pseudo-Goldstone bosons can provide a technically natural explanation for why $v^4$ can be systematically small compared with the other larger mass scales in the problem.

Notice also that the field-independent term $V_0$ is always allowed by the symmetry and so is not similarly suppressed by $\epsilon$. So although shift symmetries provide a good way to make small scalar masses technically natural, it does not do the same for the vacuum energy $V_0$. 

\subsubsection{Supersymmetry}
\label{sssec:SUSY}

Supersymmetry is a symmetry that relates bosons to fermions, which (when not spontaneously broken) to be present requires a theory to have equal numbers of bosonic and fermionic degrees of freedom  (for a textbook treatment of supersymmetry see \cite{SUSYreview}). For example, in 4 dimensions a left-handed Weyl fermion $\chi_\ssL$ (which describes the two fermionic spin states of a spin-half particle) can transform into a complex scalar $\Phi$ (which describes the two states of two spinless bosons) with a transformation of the schematic form 
\be
   \delta \Phi = \ol \varepsilon  \chi_\ssL \qquad \hbox{and} \qquad \delta \chi_\ssL = \gamma_\ssL \gamma^\mu \varepsilon \; \partial_\mu \Phi 
\ee
where $\varepsilon$ -- the symmetry parameter -- is a fermionic spinor (rather than a bosonic scalar) that has dimension (length)${}^{1/2}$.  

When not spontaneously broken the bosons and fermions related in this way have precisely equal masses and couplings. For example, a lagrangian describing the supersymmetric interactions of $\Phi$ and $\chi$ on flat space has the form
\be \label{Lsusy}
    \cL_{susy} =- \tfrac12 \ol\chi \dsl \chi  - (\partial_\mu \Phi)^* (\partial^\mu \Phi) - \tfrac12 \left[ \frac{\partial^2 W}{\partial \Phi^2} \Bigl(\ol \chi \gamma_\ssL \chi \Bigr)+ \hbox{c.c.} \right] - \left| \frac{\partial W}{\partial \Phi} \right|^2 \,,
\ee
where $W(\Phi)$ is an arbitrary holomorphic function of $\Phi$ (and not $\Phi^*$) and $\gamma_\ssL = \frac12(1+\gamma_5)$ projects onto left-handed states. The choice $W = \frac12 m \Phi^2 + \frac16 g \, \Phi^3$ gives renormalizable interactions and gives a theory where both bosons and fermions have mass $m$ and both the scalar-spinor Yukawa interactions and the cubic and quartic interaction terms in the scalar potential are controlled by the single coupling parameter $g$ \cite{Wess:1974tw}.

The reason this matters for a discussion of naturalness is this: fermions and bosons contribute to corrections to the lagrangian with opposite signs. For instance, if the vacuum energy obtained by integrating out $\chi$ is $\delta \rho_{{\rm vac}\, \chi} = C m^4$ for some function $C(g)$ then the vacuum energy obtained by integrating out the complex scalar $\Phi$ is $\delta \rho_{{\rm vac}\,\Phi} = - C m^4$ if $\Phi$ and $\chi$ have the equal masses and couplings dictated by \pref{Lsusy}. This makes their contributions to $\rho_{\rm vac}$ completely cancel. Integrating out a heavy pair (or supermultiplet) of supersymmetric particles similarly cancels out in the contributions to the mass terms of scalar fields in other light supermultiplets.

\subsubsection*{The supersymmetric dark}

This is all very nice, but we know that if the world is supersymmetric it must be spontaneously broken because none of the known elementary particles (like the electron) has a bosonic partner with precisely equal mass and coupling. Spontaneous breaking of supersymmetry can occur and when it does the bosons and fermions within a supermultiplet acquire different masses. But once the masses in a supermultiplet differ the cancellation in $\rho_{\rm vac}$ or in low-energy scalar masses no longer cancels. 

This needn't stop supersymmetry from being part of the explanation for the electroweak hierarchy, which asks why the enormous hierarchy between the Higgs mass, $m_\ssH \sim 100$ GeV, (and so also the masses of all Standard Model particles) and the Planck scale, $M_p \sim 10^{18}$ GeV, can be technically natural. Supersymmetry can help provided the mass differences within supermultiplets are not too much larger than $m_\ssH$, but it is less useful if the mass splittings are much larger than this. This makes it an attractive proposal because the assertion that supersymmetry helps understand the electroweak hierarchy comes with testable predictions: the super-partners required to enforce the cancellations cannot be too far out of reach of current accelerators. (Sadly these predictions have not described well what was actually seen in experiments to date, where there is no evidence for super-partners for ordinary particles.)

At face value it seems less useful as a proposal for understanding the small size, $v_{\rm eff} \sim 10^{-2}$ eV of the cosmological constant, since this is much smaller than the mass splittings that can exist between ordinary particles and their hypothetical super-partners. As we shall see, although this is true at face value it need not imply that supersymmetry has no role to play in the final story. In particular, although we do know that the particles we produce at colliders are not supersymmetric little is known about whether or not the gravitationally coupled dark sector is supersymmetric. 

Indeed there are good reasons to believe that any low-energy gravitationally coupled dark sector arising in a fundamental theory with supersymmetry at very high energies (such as string theory) could well be much more supersymmetric than are the particles of everyday experience (see for instance \cite{Burgess:2021juk}). This is because in supersymmetric theories the splitting of masses within any particular supermultiplet is given by an expression of the form
\be
    \Delta m^2 \sim g \, \cF \,,
\ee
where $\cF$ is the expectation of the field that breaks supersymmetry spontaneously and $g$ is the coupling of that field to the supermultiplet whose mass splitting is of interest. A supermultiplet whose couplings are all gravitational in strength is usually among the most weakly coupled supermultiplets in the theory, so it is not uncommon for their masses to be split by much less than other more strongly interacting sectors (like those containing the ordinary particles we see around us).

Suppose, for example, the supersymmetry breaking mass scale $M_s$ is set by $\cF = M_s^2$ and that Standard Model particles couple to this with a strength $g_\SM \sim \alpha \sim 0.01$ not unusual for ordinary particles but the Dark sector couples only with gravitational strength: $g_\ssD \sim M_s/M_p$. Then observations require that ordinary particles must be split from their superpartners by at least 10 TeV or so. The relation $\Delta m_\SM^2 \sim g_\SM \cF \sim \alpha M_s^2$ then implies $M_s \gsim 100$ TeV. But for $M_s \sim 100$ TeV masses within a gravitationally coupled dark sector would be split by $\Delta m_\ssD^2 \sim M_s^2/M_p \sim 0.1$ eV, not so different than the scale of Dark Energy density. In such a world the Dark sector could just include the graviton and gravitino, but it might equally well include a variety of other supermultiplets coupled to one another in an approximately supersymmetric way, and this need not contradict experience with colliders. Supersymmetric Large Extra Dimensional (SLED) scenarios \cite{SLED, SLEDrevs} provide concrete extra-dimensional realizations of this wherein ordinary particles are localized on a non-supersymmetric brane embedded in an otherwise supersymmetric bulk. (See \cite{Burgess:2013ara} for discussions of this scenario in a previous iteration of this school.)

\subsubsection{Classical scaling}
\label{sssec:Scaling}

There is a closely related type of transformation that can also (in some circumstances) suppress $V_0$ as well as scalar masses (for early attempts to exploit this see \cite{EarlyScaling}). A simple example of a transformations that can do so is obtained when the transformation \pref{shiftabelian} is also accompanied by a rescaling of the metric:
\be \label{scaling0}
    \sigma \to \sigma + \omega \qquad \hbox{and} \qquad g_{\mu\nu} \to e^{\omega} g_{\mu\nu} \,,
\ee
for constant parameter $\omega$. There is no loss of generality in choosing $e^{\omega}$ rather than $e^{2\omega}$ (or another power) because we can always rescale $\phi$ to make \pref{scaling0} true. Under this type of transformation we have $\sqrt{-g} \to e^{2\omega} \sqrt{-g}$ and ${R^\mu}_{\nu\lambda\rho} \to {R^\mu}_{\nu\lambda\rho}$ and so $R = g^{\mu\nu} R_{\mu\nu} \to e^{-\omega} R$ and so in particular the Einstein-Hilbert action scales as $S_\EH \to e^\omega S_\EH$. 

Since the Einstein-Hilbert action is not invariant this type of transformation is not a {\it bona fide} symmetry in the usual sense. But if $S[\sigma + \omega, e^\omega g_{\mu\nu}] \to e^{c\, \omega} S[\sigma, g_{\mu\nu}]$ for some constant $c$ then this transformation takes a stationary point of $S$ to another stationary point of $S$ and so is a symmetry of the equations of motion. This can be good enough inasmuch as the transformation becomes an approximate symmetry, at least within the semiclassical expansion. This makes the field $\sigma$ a pseudo-Goldstone boson for an approximate scaling (or dilatation) symmetry, which is why it is called the {\it dilaton}.
 
The kinetic energy $\sqrt{-g} \, g^{\mu\nu} \partial_\mu \sigma \, \partial_\nu \sigma$ scales in precisely the same way as does the Einstein-Hilbert action, and the same is true for the sigma-model interaction $\sqrt{-g} \, g^{\mu\nu} G_{ij}(\theta) \, \partial_\mu \theta^i \, \partial_\nu\theta^j$ provided the scalars appearing in $G_{ij}$ do not also transform,  {\it i.e.}~$\theta^i \to \theta^i$, as $\sigma$ and $g_{\mu\nu}$ are scaled.\footnote{If any of the $\theta$ fields also shift under the symmetry we can always redefine $\tilde \theta := \theta - p\sigma$ with $p$ chosen to ensure $\tilde \theta$ does not shift. So there is no loss of generality in assuming $\sigma$ is the only scalar field that transforms.}  The potential energy also scales the same way provided that the potential depends on $\sigma$ in a specific way: 
\be \label{VscaleU}
    V(\sigma,\theta) = e^{-\sigma} U(\theta) \,.
\ee  

We are led to an effective action of the following form, expanded to the 2-derivative level:
\be \label{LeffScale}
   \cL_{\rm eff} = - \sqrt{-g} \Bigl[ v^4 e^{-\sigma} U(\theta) + \tfrac12 M_p^2 R + \tfrac12 Z(\theta) (\partial \sigma)^2 + \tfrac12 f^2 G_{ij}(\theta) \, \partial \theta^i \partial \theta^j + \cdots \Bigr] \,,
\ee
by the requirement $\cL_{\rm eff}[\sigma + \omega,\theta^i,e^\omega \, g_{\mu\nu}] = e^{\omega} S[\sigma,\theta^i,g_{\mu\nu}]$. Notice in particular that the scalar potential is always minimized at $V = 0$ provided only that $U$ is non-negative. Minimization can happen in one of two ways. If $U(\theta)$ is minimized at some values $\bar\theta^i$ for which $\ol U = U(\bar \theta)$ is nonzero then the minimum occurs for $\sigma \to \infty$. If $\ol U = U(\bar \theta)$ is instead zero then the minimum is really a flat direction along which $\sigma$ can take any value and for which the potential vanishes. 

For nonzero $\ol U$ the potential can be made arbitrarily small just by making $\sigma$ large enough, with the effective scale for the potential being 
\be
   v_{\rm eff} = v \, e^{-\sigma/4} \,.
\ee
In particular there is always a value of $\sigma$ that is large enough that $V$ is the right size to be the Dark Energy density. But because $V$ has no minimum for finite $\sigma$ this field in general rolls down the potential. For kinetic term $Z \sim M_p^2$ the evolution of $\sigma$ occurs over cosmological time scales since $v_{\rm eff}^2/M_p = v^2 e^{-\sigma/2}/M_p$ is of order the Hubble scale. Whether this is acceptable depends on whether this evolution can be consistent with what we know about cosmology (more about which below). The required value for $\sigma$ to do so would be very large for any particle-physics value choices for $v$. 

Large $\sigma$ can also be a good thing from another point of view. Quantum corrections will not preserve the form of the lagrangian \pref{LeffScale} and so the size of these corrections is important to estimate. Their dependence on $\sigma$ can be determined quite generally because it is always possible to rescale the metric from Einstein frame to a Jordan frame $\hat g_{\mu\nu} = e^{-\sigma} \, g_{\mu\nu}$, defined so $\hat g_{\mu\nu}$ does not transform under the transformation \pref{scaling0}. Once this is done the only place where $\sigma$ appears undifferentiated in \pref{LeffScale} is as an overall factor $\cL_{\rm eff}[\sigma, g_{\mu\nu}] = e^\sigma \cL_{\rm eff}[\partial \sigma, \hat g_{\mu\nu}]$, as is indeed required to ensure $\cL_{\rm eff} \to e^\omega \cL_{\rm eff}$ in these variables. But this means that $e^{-\sigma}$ appears in the path integral integrand $e^{iS/\hbar}$ in the same way as does $\hbar$, and so repeating the power-counting arguments of previous sections shows each loop comes with a factor of $e^{-\sigma}$. 

The upshot is that loop corrections to the action come as a series of the form
\be \label{dilatonloopexpansion}
   \cL_{\rm eff} = e^\sigma \cL_{\rm tree}(\hat g_{\mu\nu})  + \cL_{\rm 1-loop}(\hat g_{\mu\nu})   +e^{-\sigma}  \cL_{\rm 2-loop}(\hat g_{\mu\nu})   + \cdots \,,
\ee
where $\cL_{\rm tree}$ is given by all terms in \pref{LeffScale} plus any others with higher derivatives that scale like $\cL_{\rm tree} \to e^\omega \cL_{\rm tree}$.. Each of the $\cL$'s here is a function only of $\partial_\mu \sigma$, $\theta^i$, $\hat g_{\mu\nu}$ and other scale-invariant combinations of fields, so the $L$-loop contribution transforms as $\cL_{L{\rm -loop}} \to e^{(1-L)\omega} \cL_{L{\rm -loop}}$ under \pref{scaling0}. Although the one-loop term seems to re-introduce a $\sigma$-independent potential -- seemingly again allowing constant contributions to the potential, like $V_0$ -- this is an illusion because \pref{dilatonloopexpansion} is written in terms of the Jordan-frame metric $\hat g_{\mu\nu}$ (which does not satisfy the usual Einstein equations). In terms of the Einstein-frame metric (which does) any potential appearing in the one-loop term becomes
\be
    - \sqrt{-\hat g} \; U_{\rm 1-loop}(\theta) = - \sqrt{-g} \; e^{-2\sigma} U_{\rm 1-loop}(\theta) \,,
\ee
as required for the one-loop term to be invariant under \pref{scaling0}.    

The widely read reader might notice that having a field play the role of a loop-counting parameter, as in \pref{dilatonloopexpansion}, is reminiscent of the role played by the 10-dimensional dilaton field in string theory. This is not an accident. A strong motivation for exploring classical scaling symmetries like \pref{scaling0} is precisely that they are generic to all known string vacuua \cite{Burgess:2020qsc}. They arise ubiquitously there because in string theory there are no parameters, only fields. So any expansion -- be it weak coupling or low energy -- is always an expansion in powers of fields like in \pref{dilatonloopexpansion}. Each term in such an expansion (and in particular the first term) by construction scales in a specific way under appropriate rescalings of the fields. 

We return below to whether these quantum corrections can be acceptably small, but it is encouraging that they are smallest in the regime of most interest: large $\sigma$ (which makes the scalar potential small)

\subsection{Desperate measures}
\label{ssec:Desperate}

The program for finding a technically natural understanding for why the cosmological constant is so small has so far not borne convincing fruit. It seems very hard to reconcile the known particle content in the Standard Model with having a technically natural vacuum energy as small as the observed Dark Energy density. This has caused some to doubt the utility of the criterion of technical naturalness altogether. This section describes the three options on offer by those who do so, leaning heavily (for two of them) on the review \cite{Burgess:2013ara}.

\subsubsection*{Head in the Sand}
\label{sssec:Ostrich}

The most common attitude about the cosmological constant problem is pragmatic despair. Since there are no good theories on offer, one works on other areas of physics forlornly hoping that whatever solves the cosmological constant problem is not going to be important to this other physics.

This is an easier point of view to adopt the further one's field is from cosmology, since in cosmology a commitment must be made as to whether the dark energy clusters or evolves with time.

\subsubsection*{Anthropic arguments}
\label{sssec:Anthropics}

A more sophisticated point of view interprets the absence of a compelling solution to the cosmological constant problem as evidence that quantum corrections to the vacuum energy need not be small after all \cite{Weinberg:1988cp, Anthropic}. That is, one denies that both of the questions given in \S\ref{sssec:TechNat} must be answered for the cosmological constant, and simply accepts that there is a very precise cancellation that occurs between the renormalized cosmological constant and the quantum contributions to $\rho_{\rm vac}$. As emphasized earlier, this is a logically consistent point of view, though it is radical in the sense that would be the first example at easily accessible energies where this occurs for a parameter in the Wilson action.\footnote{There are examples of coincidences of scale that do not require a fundamental explanation, such as the apparent sizes of the Sun and the Moon as seen from the Earth. However I do not know of any examples of this type that involve the smallness of parameters in a Wilson action.}

There is a better face that can be put on this cancellation if the microscopic theory has three features. First, the microscopic theory could have an enormous number of candidate vacua, with the effective cosmological constant differing from vacuum to vacuum. (This is actually likely to be true of a UV complete theory of quantum gravity if string theory is any guide.) Second, the microscopic theory might have a reason to have sampled many of these vacua somewhere in space at some time over the history of the universe. (This is also not far-fetched in theories that allow long periods of cosmic inflation within a complicated potential energy landscape, such as seems likely for string theory.) Third, it might be true that observers can only exist in those parts of the universe for which the vacuum energy has a very small range, not much different from the observed dark energy density.

With these conditions in place one might expect the universe to be populated with an enormous number of independent regions, in each of which a particular vacuum (and cosmological constant) is selected. The vast majority of these vacua do not have observers within them whose story needs telling, but those that do can only have a small cosmological constant since this is required for the observers to exist in the first place. Since we live in such a world we should not be surprised to find evidence for dark energy in the range observed.

Although this may well be how things work, most (though not all) of its proponents would prefer to have a technically natural solution to the problem (satisfying the two criteria of \S\ref{sssec:TechNat}) if only this were to exist. There are two dangers to adopting this kind of anthropic approach. One is that it becomes a dogma that stops people searching for a more traditional solution to the problem. Another is that it is difficult to know how to falsify it, and what the precise rules are that one should use when making predictions. (Of course this is partly the point: it is not clear how one makes predictions more generally in theories having an enormous landscape of possible vacua, and it is important that this gets thought through to see if a sensible formulation can be found.)

My own view on this is to accept that there is an important issue to be resolved to do with making predictions in theories (like string theory) that have a complicated landscape. But (to my understanding) so far no unambiguous framework for making predictions and deciding which parameters must be understood anthropically has been found, so it is hard to assess how useful the new anthropic framework really is. 

In practice the problem right now is not that we know of too many acceptable vacua of UV complete theories. The real issue is it is hard to find any good vacua at all given the large number that must be sorted through. Once we have two examples that include the Standard Model and everything else we find around us (and nothing else) we can start worrying about their statistics. One thing that might help in this search is to have `modules' that build in features we know to be true of the world around us. These modules include the Standard Model particle content and symmetries, some candidate for dark matter, and hopefully could include a technically natural description of dark energy if this could be found.

\subsubsection*{Swampy {\it vs}\, Solid ground}
\label{sssec:SwampyGround}

The Swampland program \cite{Vafa:2005ui} provides a much more recent form of naturalness denial. In essence this program asserts that there exist otherwise reasonable effective field theories for which no UV completion including gravity exists.\footnote{Since nobody knows what the real UV completion is for gravity in practice this assertion is taken to mean that the EFT cannot be obtained as the low-energy limit of some sort of string vacuum.} Effective theories for which UV completions cannot be found within the landscape of possible vacua are said instead to lie in the swampland. If this picture were correct there would be a great premium on knowing which EFTs are not in the swampland because only those would be embeddable into a sensible theory of all scales. 

There is even some evidence that a swampland like this might exist, if we assume (as people do in practice) that the UV completion is a string theory. For instance there are good arguments that perturbative string theories cannot contain any global symmetries \cite{Banks:1988yz, Burgess:2008ri} and if so then any EFT with an exact global symmetry must be in the swampland. 

But this example also exposes a real difficulty in actually using this observation: when using EFTs one only ever works to some finite order in $1/M$ and it is easy to arrange a global approximate symmetry that only appears to be exact at some fixed order in $1/M$. The Standard Model is the poster child for this: if the low-energy world consists only of Standard Model fields then the most general possible interactions allowed at zeroeth order in $1/M$ is the Standard Model itself. But the Standard Model famously has several accidental symmetries -- like baryon number and lepton number -- that are automatic conseqences of renormalizability and so are broken once nonrenormalizable interactions at nonzero order in $1/M$ are included. 

At low energies it is in practice incredibly difficult to tell the difference between an exact global symmetry and a `fake' accidental approximate global symmetry \cite{Burgess:2008ri}. A similar observation seems also to apply to the other lines of reasoning that support the existence of the swampland: the more sure we are that a low energy property is really required by a UV completion the easier it seems to be to fake at a fixed order in $1/M$ and so the less useful it is in constraining our options when describing the low-energy world. The difficulty in finding a criterion for the swampland that is both reliable and useful has been called the {\it Principle of Swamplementarity}. 

Another difficulty is that nobody really knows everything that is possible within string theory. This drives people instead to propose conjectures about what is possible and what is not, and then to see if these conjectures are informative. One such a conjecture is that de Sitter solutions should only be possible in EFTs that lie in the swampland \cite{Obied:2018sgi}. This in turn has led to a preference for quintessence like models and (more recently) to a resurgence of interest in large extra dimensions \cite{DarkDims} when trying to describe the Dark Energy density. 

In my opinion a key challenge for these models is their awkward relationship to decoupling and the general utility of EFT methods at low energies. To the extent that EFTs not in the swampland obey the usual rules, it should be possible to understand the naturalness issues that come with {\it any} EFT description of UV physics. But a key part of the swampland program is that there is low-energy information that is {\it not} captured by standard EFT methods, and a full assessment of its value will require understanding when EFT rules can be dropped. With the present state of the art there seems to be no new insights on the cosmological constant problem, apart from a belief that consistency with some of the conjectured behaviour of UV physics must eventually save the day in a way that cannot yet be explicitly articulated. 

\section{Ways forward (naturally)}
\label{sec:WaysForward}


This last section closes on a more optimistic note, focussing on the main directions that I think are the most promising ways to achieve a technically natural Dark Energy. Although this is a difficult thing to achieve I do not think that enough avenues have yet been sufficiently thoroughly explored to justify despair. This section aims to explain why, and to describe the predictions these directions make (some of which might be starting to bear fruit). 

There are two main directions that I believe deserve further exploration, each of which is briefly described here. They differ on whether the focus is on electroweak and higher energy scales or the much lower energies relevant to cosmology. In truth the two directions are likely two sides of the same coin. 

\subsection{Above eV scales: Supersymmetric Extra Dimensions}
\label{ssec:ExtraDs}

Let us start with the higher energies: the electroweak scales of everyday particle physics. In this energy range extra-dimensions provide a uniquely promising approach to dynamically evading the cosmological constant problem. This section reviews why this is so and what the challenges are (leaning heavily on lectures given at earlier versions of this school \cite{Burgess:2013ara}).  

To motivate the relevance of extra dimensions for the cosmological constant problem, recall what the essence of the problem is: we believe quantum fluctuations generate a large vacuum energy density, and the vacuum's Lorentz invariance automatically gives this the $w=-1$ equation of state of a cosmological constant: $T_{\mu\nu} = - \rho_{\rm vac} g_{\mu\nu}$. But when cosmologists measure the acceleration of the universe's expansion they are essentially detecting a very small curvature for 4D spacetime. The conundrum is that these are directly equated in Einstein's equations -- eqs.~\pref{GRDE} -- with the measured curvature much smaller than what would be expected for typical vacuum energies.

We wish to break this direct link between the energy of quantum fluctuations and the curvatures measured in cosmology. Moreover, we must do so only for very slow processes (involving the timescales of cosmology) and not also for fast ones (involving the timescales of {\it e.g.}~atoms) \cite{Bandpass}. Fast quantum fluctuations should gravitate in an unsuppressed way because we know that such fluctuations actually do contribute to energy levels in atoms. We know that these contributions gravitate because precision tests of the equivalence principle show that gravity couples to the entire energy of a source, regardless of its origin. The equivalence principle is tested to a part in $10^{15}$ or so \cite{LLR, Will} and we know quantum fluctuations contribute to atomic energies by more than this (and so their absence would be missed if they did not contribute). 

\subsection*{The extra-dimensional loophole}

The good news is there is a loophole within which it is possible to break the link between vacuum energy and curvature, without doing violence to everything else we know at accessible energies. This loophole is based on the observation that Lorentz invariance plays an important role in formulating the problem, because it so severely restricts the form of the vacuum stress energy. 

The situation would be different in an extra-dimensional world because then we would only know that the vacuum must be Lorentz invariant in the four dimensions that we can see. We also would only really know that the curvatures must be small in these same dimensions since these are the ones we access in cosmology. Although the vacuum stress energy must curve something, in extra-dimensional models it need not curve the dimensions we see.

The gravitational field of a cosmic string in four dimensions illustrates this loophole more concretely. Consider a string whose world-sheet sweeps out the $z-t$ plane, transverse to the $x$ and $y$ directions. The stress energy of a relativistic string is Lorentz invariant in the $z-t$ directions, $T_{ab} = - \cT g_{ab} \, \delta^2(x)$, where $\cT$ is the string's mass per unit length, and $a,b$ denotes the $z-t$ directions parallel to the string world-sheet. The gravitational field sourced by this stress energy is known \cite{CosmicString} and the spacetime away from the string's position is flat. More precisely, the two dimensions transverse to the string have the geometry of a cone whose apex is located at the string's position. The tension on the string gives rise to a curvature singularity, with the transverse 2D geometry having a curvature scalar $R \propto \kappa^2 \cT \, \delta^2(x)$ that is singular at the string's position. What is important for the present purposes is that the geometry along the two Lorentz-invariant on-string directions remains perfectly flat,\footnote{More precisely, it is flat if the string sits within an asymptotically flat geometry. It would not be flat if the string were sitting in a curved space like de Sitter space.} regardless of the precise value of $\cT$. The consistency of a large $\cT$ with a small spacetime curvature in the $z-t$ directions might appear to be a `cosmological constant problem' to a 2D cosmologist unable to see off the surface of the string.

This suggests trying similar examples having two more dimensions (six dimensions in total) with the 2-dimensional string world-sheet being replaced by the world-volume of a 4-dimensional Lorentz-invariant brane, situated at specific points within two compact extra dimensions. In the simplest examples the two extra dimensions have the geometry of a sphere and there is a brane located at both the sphere's north and south poles. The transverse curvature at these poles also has conical singularities, like for a cosmic string, and this gives the overall geometry more of a rugby ball (or American football) shape. All of the elementary particles we know are imagined to be confined to one of these branes, whose tension ({\em i.e.} vacuum energy per unit volume) is not particularly small relative to known particle-physics scales --- of order $(10 \; \hbox{TeV})^4$. The hope is that the geometry seen by an observer on the brane (us) can remain flat regardless of the size of the brane vacuum energy density.

The simplest models try to do so by simply assuming away the extra-dimensional cosmological constant \cite{NonSUSYRugby}, though this simply moves the underlying cosmological constant problem into the higher-dimensional theory. There is a better chance if the extra-dimensional physics is supersymmetric \cite{SLED}, however, because in six (and higher) dimensions supersymmetry forbids a cosmological constant (much as would more than one supersymmetry in four dimensions). Interestingly, they do so because extra-dimensional supergravities typically have scaling symmetries like \pref{scaling0} described above \cite{6Dflatnophi, FlatBcsScaling}. The generic appearance of scaling symmetries in extra dimensions also turns out to have a plausible explanation in their generic presence in string theory \cite{Burgess:2020qsc}.   

Notice that we do {\em not} also require the physics on the brane to be supersymmetric, and one might simply choose only the Standard Model to live on the brane. Such a brane can nonetheless be coupled consistently to supergravity using the `St\"uckelberg trick'; that is, promoting the non-supersymmetric brane to something supersymmetric, but with supersymmetry nonlinearly realized by coupling a Goldstone fermion --- the Goldstino --- in the appropriate way \cite{SUSYNLR}. It remains consistent to regard extra-dimensional fields to be supersymmetric despite them coupling to nonsupersymmetric matter on the brane because brane-bulk couplings are weak; they are gravitational in strength. From an extra-dimensional point of view the brane provides a non-supersymmetric boundary condition for bulk modes that splits bosons from fermions by the KK scale, $\Delta m \sim 1/L$ (where $L$ is the linear size of the extra dimensions). Because the 4D and 6D Planck scales are related by\footnote{The 6D gravitational coupling is $\kappa_6^2 = 8 \pi G_6 = 1/M_6^4$ where $G_6$ is the 6D Newton constant.} $M_p \sim M_6^2 L$ this shows that bulk mass splittings are Planck-suppressed: $\Delta m \sim M_6^2/M_p$.

It turns out that extra dimensions can be large enough to allow $1/L \sim \hbox{eV}$ without running into conflict with observations \cite{ADD, ADDpheno} and so can allow bulk supersymmetry to play a role suppressing the vacuum energy right down the the Dark Energy scale \cite{SLED, SLEDrevs, TNCC}. Remarkably extra dimensions can only be this large if there are at most two of them, providing another reason for liking six dimensions.\footnote{6D models with radii this large can also be ruled out if KK modes dominantly decay into observable particles like photons, but whether such decays dominate is a model-dependent issue \cite{MSLED}.} In any such a framework the 6D Planck scale is not too far above the electroweak scale, so from the 6D point of view the tension on any brane involving standard model particles would not be fantastically small relative to the Planck scale.

This picture leads to a novel kind of supersymmetric phenomenology \cite{MSLED, SLEDpheno}: a very supersymmetric gravity (or extra-dimensional, bulk) sector whose supersymmetry breaking scale is of order $1/L \sim 1$ eV or less, coupled to a particle (brane) sector that is not supersymmetric at all. In particular, the nonlinear realization of supersymmetry on the brane implies that a supersymmetry transformation of a brane particle like the electron gives the electron plus a Goldstino (or, equivalently, a gravitino) rather than a selectron. One does not expect to find a spectrum of superpartners for the Standard Model, despite the very supersymmetric gravity sector.\footnote{This particular prediction was made \cite{MSLED} before the LHC results showed it to be a huge success.}

Within this kind of picture the cosmological constant problem is a special case of the general problem of back-reaction: how does the spacetime geometry react to microscopic changes, such as to the vacuum energy. In a higher-dimensional context this requires also understanding what stabilizes the size of the extra dimensions, since this is also part of the general issue of gravitational back-reaction. To pin these issues down precisely it is useful to work within a concrete example, solving explicitly the equations of a specific higher-dimensional supergravity  \cite{SLED}. 

There are a variety of 6D supergravities from which to choose when formulating such an example, but a particularly convenient choice uses the Nishino-Sezgin chiral gauged supergravity \cite{NS}, for which a simple stabilization mechanism for the extra dimensions has long been known \cite{SS}. This involves the following 6D bosonic fields: the metric, $g_{\ssM\ssN}$, a scalar dilaton, $\phi$, and a specific $U(1)_\ssR$ gauge potential, $A_\ssM$ and a Kalb-Ramond 2-form gauge field $B_{\ssM\ssN}$ (with 3-form field strength $H_{\ssM\ssN\ssP}$). 

To lowest orders in the derivative expansions, the action is the sum of bulk and brane contributions, $S = S_\ssB + \sum_b S_b$, with the supersymmetric bulk contribution being
\bea \label{SSUSY}
 S_\ssB &=& - \int \exd^6x \sqrt{-g} \; \left[ \frac{1}{2\kappa_6^2} \; g^{\ssM\ssN} \left( \cR_{\ssM\ssN} + \partial_\ssM \phi \, \partial_\ssN \phi \right)  + \frac{2 \mfg^2}{\kappa_6^4} \; e^\phi  \right. \nn\\
 && \qquad\qquad\qquad\qquad\qquad \left.  + \frac{1}{12} \, e^{-2\phi} H_{\ssM\ssN\ssP} H^{\ssM \ssN \ssP} + \frac14 \, e^{-\phi} F_{\ssM\ssN} F^{\ssM \ssN} \right] \,,
\eea
while the contribution of each brane is
\bea \label{SSUSYb}
 S_b = - \int_{W_b} \exd^4 x \sqrt{-\gamma} \; \left( \cT_b +\tfrac{1}{4!} \cA_b \, \varepsilon^{\mu\nu\lambda\rho} \cF_{\mu\nu\lambda\rho} + \cdots \right) \,.
\eea
Here $W_b$ denotes the brane's world-volume and $\varepsilon_{\mu\nu\lambda\rho}$ is the volume form built from its induced metric $\gamma_{\mu\nu}$. $\cF_{\ssM\ssN\ssP\ssQ} = \frac12 \epsilon_{\ssM\ssN\ssP\ssQ\ssR\ssS} \, e^{-\phi} F^{\ssR\ssS}$ is the 6D dual of the Maxwell field, where $\epsilon_{\ssM\ssN\ssP\ssQ\ssR\ssS}$ is the volume form built from the 6D metric. $\mfg$ is the gauge coupling for the field $F_{\ssM\ssN}$ and the parameter $\cT_b$ denotes the brane tension. The quantity $\cA_b$ measures the amount of Maxwell flux that is localized on the brane \cite{LargeDimsCurvature} in a way that is made more precise below.

The simplest situation is when the two branes are identical, in which case there is a rugby ball solution to these field equations \cite{SS, SLED} for which $H_{\ssM\ssN\ssP} = 0$, $\phi = \phi_0$ and
\be \label{2DmetricSS}
 \exd s^2 = \hat g_{\mu\nu} \, \exd x^\mu \, \exd x^\nu + \ell^2 \Bigl( \exd \theta^2 + \beta^2 \sin^2 \theta \, \exd \xi^2 \Bigr) e^{-\phi_0} \quad \hbox{and} \quad
 F_{\theta\xi} = Q \beta \ell^2 \sin \theta \,,
\ee
where $\hat g_{\mu\nu}$ is a maximally symmetric 4D geometry with curvature scalar $\widehat R$ and $\phi_0$, $Q$, $\beta$ and $\ell$ are constants. With this ansatz the field equations boil down to
\be \label{SUSYsoln}
 \frac{1}{\ell^2} = \kappa_6^2 Q^2 = \left( \frac{2 \mfg}{\kappa_6} \right)^2 \,, \quad   1-\beta = \frac{\kappa_6^2 \cT}{2\pi} \quad \hbox{and} \quad \widehat R = 0
 \,.
\ee

Inspection of \pref{2DmetricSS} shows that the physical radius of the extra dimensions is $L = \ell \, e^{-\phi_0/2}$, and so eqs.~\pref{SUSYsoln} imply
\be \label{Lvsphi}
 L^2 e^{\phi_0} = \ell^2 = \left( \frac{\kappa_6}{2 \mfg} \right)^2 \,,
\ee
is fixed in terms of parameters in the lagrangian. Several features of this solution are noteworthy:

\subsubsection*{Flat direction and scaling}

The value of $\phi_0$ is not determined by any of the field equations. This `flat direction' is a consequence of a classical scale invariance of extra-dimensional supergravity, along the lines described in \S\ref{sssec:Scaling}. The scaling symmetry in this case applies to the 6D lagrangian, with 
\be \label{6DScalingSym}
   g_{\ssM \ssN} \to \zeta \, g_{\ssM\ssN} \quad \hbox{and} \quad e^{-\phi} \to \zeta \, e^{-\phi} \quad \hbox{with} \quad H_{\ssM\ssN\ssP} \; \hbox{and} \;  F_{\ssM\ssN} \; \hbox{held fixed.}
\ee
(This is why $\phi$ is called the 6D dilaton.) Under this $S_\ssB \to \zeta^2 S_\ssB$ and $S_b$ scales the same way, but only if both $\cT_b$ and $\cA_b$ are assumed to be $\phi$-independent. 

\subsubsection*{Localized flux and flux quantization}

As mentioned above, the $\cA_b$ term changes the Maxwell equation to become
\be
   \partial_m \Bigl[ e^{-\phi} \Bigl( \sqrt{g_2} \, F^{mn} - \sum_b \cA_b \, \epsilon^{mn} \, \delta^2(x-x_b) \Bigr) \Bigr] = 0 \,,
\ee
where $\epsilon_{mn}$ is the volume form for the extra-dimensional 2D geometry. This introduces localized flux into the solution at the position of each brane, and changes the flux quantization condition into
\be \label{FluxQ2}
     \int_{M_2} F_{mn} +  \sum_b \cA_b \, \frac{\epsilon_{mn}}{\sqrt{g_2}}   = \frac{n}{\mfg} \,,
\ee
where $n$ is an integer. Notice that this condition does not break scale invariance (because $F_{\ssM\ssN}$ doesn't scale and the transformations of $\epsilon_{mn}$ and $\sqrt{g_2}$ cancel one another) {\it provided} that $\cA_b$ is independent of $\phi$. 
    
For the solution given in \pref{2DmetricSS} and \pref{SUSYsoln} the integer must be $n = \pm 1$ and the flux localized on the branes is
\be \label{FluxQ2a}
    \Phi_{\rm tot} := \sum_b  \cA_b   = \pm \frac{1 - \beta}{\mfg} \,.
\ee
Notice also that \pref{FluxQ2} would not have any solutions at all if the $\cA_b$'s were all assumed to be zero without also assuming $\cT_b = 0$, as was often done in early studies of this system \cite{GP}.

In the scale-invariant case -- when $\cA_b$ and $\cT_b$ are independent of $\phi$ -- eq.~\pref{FluxQ2a} imposes a condition that relates the $\cA_b$ and $\cT_b$ ($\cT_b$ enters through $\beta$ using \pref{SUSYsoln}) in order for rugby ball solutions to exist. If $\cA_b$ were to depend on $\phi$ then scale invariance breaks and the flux quantization condition can be used to determine the value of $\phi_0$ given arbitrary choices for the $\cT_b$ or $\cA_b$. 

\subsubsection*{4D flatness and extra-dimensional relaxation}

Most remarkably, the brane action is flat ($\hat R = 0$) for {\em any} choice of brane lagrangian and in particular regardless of the value of $\cT_b$. For later purposes it is useful to see in more detail how the extra-dimensional solutions relax to achieve flat 4D geometries: $\hat R = 0$. 

The simplest way to see what happens is first to ask why the curvature is flat in the solution in the absence of branes ({\em i.e.} with $\cT = 0$ and so $\beta = 1$) \cite{SS}. The 4D scalar potential for the fields $\phi_0$ and $L$ is obtained by evaluating the action using the 2D scalar curvature $R = -2/L^2$ and the Maxwell field subject to the flux quantization condition, which implies $\int \mfg F = n$ for some integer $n$, and so $F_{mn} F^{mn} \propto n^2/L^4$. Combining the Einstein and Maxwell actions with the scalar potential\footnote{After first transforming to the 4D Einstein frame: $g_{\mu\nu} \to (1/L^2) g_{\mu\nu}$.} then gives the scalar potential for the fields $L$ and $\phi$. In the case of unit flux, $n = \pm 1$, this turns out to be a perfect square:
\bea
     V(L, \phi_0) &=&  \int \exd^2 x \sqrt{-g} \; \left( \frac{1}{2\kappa_6^2} \, R + \frac14 \, e^{-\phi_0} F_{mn} F^{mn} + \frac{2 \mfg^2}{\kappa_6^4} \, e^{\phi_0} \right) \nn\\
     &\propto& \frac{e^{\phi_0}}{L^2} \left( 1 - \frac{\kappa_6^2}{4 \mfg^2 L^2 e^{\phi_0}} \right)^2 \,,
\eea
which is minimized at a fixed value of $L^2 e^{\phi_0} = \ell^2$ -- compare this minimum with \pref{SUSYsoln} -- for which there is a flat direction along which $V = 0$ and $e^{\phi_0}/L^2 = e^{2\phi_0}/\ell^2$ is not determined.

Adding branes to this solution changes the above in two ways. First the action now includes the brane tensions coming from $S_b$. Second, the brane's gravitational field introduces a conical singularity to the 2D curvature, $\sqrt{g_2} \; R_{\rm sing} = -2 \kappa_6^2 \sum_b \cT_b \; \delta^2(x-x_b)$ localized at the brane positions, where $\cT_b$ is the brane tension. Using the curvature singularity in the Einstein action (and using the delta function to perform the extra-dimensional integral $\exd^2 x$) then gives a contribution to the action of the form $(1/2\kappa_6^2) \int \exd^2x \sqrt{g_2}\; R = -\sum_b \cT_b$, which precisely cancels the direct contribution of the brane tensions themselves. The lesson from this story is that back-reaction is crucial: this cancellation can never be seen working purely within a `probe' approximation where the brane does not perturb its environment.

\subsubsection*{More general classical solutions}

The existence of extra-dimensional solutions that allow flat 4D geometries to coexist with large 4D-lorentz-invariant energy densities does not in itself solve the cosmological constant problem. One must re-ask the cosmological constant question in the 6D context: first identify which features of the branes are required for flat brane geometries, and then ask whether {\em these} choices are stable against integrating out high-energy degrees of freedom. 

At the classical level many more general explicit solutions to these field equations are known \cite{6Dflatnophi, DiffBranes}, such as when the tensions on the two branes are not equal, and although the extra-dimensional geometry for these new solutions generically becomes warped the 4D brane geometries remain exactly flat. Nonflat solutions can also be constructed, for some of which the 4D geometry is de Sitter\footnote{de Sitter solutions to these equations are interesting in their own right as a counter-example \cite{6Dnogonot} to no-go theorems for the existence of de Sitter solutions in supergravity \cite{6Dnogo}.} rather than flat \cite{6DdS}. The nonflat solutions are obtained by allowing $\cT_b$ and/or $\cA_b$ to depend nontrivially on $\phi$. 

There is a very general argument why 4D curvature requires nontrivial dependence of $\cT_b$ and $\cA_b$ on $\phi$, since it can be proven on very general grounds that any solution to the field equations for which the near-brane limit of $r \partial_r \phi$ vanishes for all branes as $r \to 0$ (where $r$ is the proper distance from the brane) must have a flat 4D geometry.  This is proven by exploiting the scale invariance of \pref{6DScalingSym} \cite{FlatBcsScaling}. But the near-brane limit of $r \partial_r \phi$ is on very general grounds proportional to $\delta S_b/\delta \phi$, where $S_b$ is the brane action \cite{Cod2Matching}. They are related for much the same reason as the charge of a point source in electromagnetism can either be determined by differentiating the action of the point source with respect to the electrostatic potential, or by evaluating (in 2D) $r \partial_r$ of the Coulomb potential itself near the source.  

The upshot is this: for a general solution to the field equations for the action \pref{SSUSY} and \pref{SSUSYb} with maximally symmetric geometries in 4D a sufficient condition for the 4D geometry to be flat is to have none of the branes couple to the 6D dilaton. In order for quantum corrections to generate a 4D curvature they must also generate dilaton couplings to the branes.

\subsubsection*{Flux quantization vs tuning}

It is sometimes argued that in the case of scale invariant (dilaton-independent) branes the flux quantization condition \pref{FluxQ2} itself represents a fine-tuning that is ruined by quantum corrections. This is most often argued in simpler 5D models \cite{5Dversion}, where similar issues arise and for which back-reaction can also be computed explicitly. In this case closer inspection \cite{Anti5D} showed that flat solution arise due to a cancelation with branes whose presence was not explicit but required to interpret singularities that were necessary on topological grounds. 

A similar argument in 6D expresses the extra-dimensional Euler number as the sum of brane tensions plus an integral over extra dimensional curvature. For the rugby-ball geometries of interest here (with the topology of a sphere) this is equivalent to the relation between defect angle and tension given in \pref{SUSYsoln}. The situation in 5D is more similar to toroidal compactifications in 6D, for which the Euler number vanishes and so a topological condition states that the sum of brane tensions must vanish. 

Nonetheless, topological conditions are not in themselves ever an obstruction to technical naturalness (even for tori). If a tension is changed in a toroidal compactification, the extra dimensions simply curve to satisfy the topological constraint \cite{SLEDrevs}. Continuous changes cannot violate the topological condition once this is initially satisfied. For technical naturalness the real issue is to check whether the choices required for flat 4D geometries are stable against integrating out short-wavelength modes, and this is in essence a continuous procedure.
 
\subsubsection*{Robustness to quantum corrections}

Considerable effort has been invested into integrating out high-energy modes in these geometries, both from loops of high-energy bulk fields  and from high-energy brane degrees of freedom \cite{BulkLoops}. The good news is that if a brane is initially chosen to have no dilaton coupling (in the 6D Einstein frame) then no loop purely involving only brane degrees of freedom can generate a dilaton coupling. This in particular means that if Standard Model particles are localized on such a brane then the curvature of the 4D geometry is completely stable against graphs involving only Standard Model fields (which is usually the hard part in the cosmological constant problem). 

The dangerous loops are those involving the bulk fields -- the extra-dimensional graviton and its friends -- because these {\it must} both couple to the dilaton (because of supersymmetry) and couple to the brane. But for these loops supersymmetry is important in suppressing the size of quantum effects, leading to their general suppression. The interested reader is referred to the review \cite{Burgess:2013ara}. 

For these (and other) reasons SLED models provided a very promising way to shield the curvature of 4D spacetime from high-energy quantum fluctuations. But research on these models eventually tapered off, largely because although explicit calculations of loop-corrected vacuum energies did verify they were suppressed, they were never quite as suppressed as they would need to be to describe Dark Energy. The problem boiled down to the difficulty achieving sufficient precision when working with all of the complications of the full higher-dimensional theory. More progress could be made if the effective 4D EFT describing the very low energies relevant to cosmology were known.

\subsection{Below eV scales: Scaling the Supersymmetric Dark}
\label{ssec:ScalingReviz}

Motivated by the previous discussion we now ask these questions within the low-energy 4D theory directly relevant to cosmology that is valid at energies much lower than the KK scale. Rather than explicitly integrating out the higher-dimensional fields we instead work with a general 4D theory that includes the light fields and incorporates the underlying approximate symmetries.\footnote{The specific realization \cite{Burgess:2021obw} of the relaxation mechanism described in this section was initially proposed as an independent mechanism for approaching the cosmological constant problem, though it now is clear that it also captures the low-energy limit of the SLED models described above.}

One of the most important of the approximate symmetries is the scale invariance corresponding to shifts of the 6D dilaton, which descends into the effective 4D theory as an approximate scale invariance of the type outlined in \S\ref{sssec:Scaling} above. This symmetry dictates how the low-energy 4D dilaton field, $\sigma$ -- corresponding to the modulus $\phi_0$ of the 6D solution -- appears undifferentiated in the low-energy action. More precisely, given the 6D convention that weak coupling corresponds to small $e^\phi$ and the 4D convention of \S\ref{sssec:Scaling} that chooses $e^{-\sigma}$ to be small we abusively define $\sigma = - \phi_0$. With this convention the transformation law 
\be \label{sigmarule}
  e^{\sigma} \to \zeta \, e^{\sigma}
\ee
is inherited from the 6D transformation rule \pref{6DScalingSym}.

Demanding the 4D Einstein-Hilbert term $\cL_\EH = - \frac12 M_p^2 \sqrt{-g} \; g^{\mu\nu} R_{\mu\nu}$ scale as $\cL_\EH \to \zeta^2 \cL_\EH$ ({\it i.e.}~the same way as does the 6D action does under \pref{6DScalingSym}) implies the 4D Einstein-frame metric scales as\footnote{Since the focus now shifts to the 4D EFT from here on we denote the 4D Einstein-frame metric by $g_{\mu\nu}$ and relabel the 6D Einstein-frame metric -- including its 4D components -- as $\tilde g_{\ssM\ssN}$.}
\be \label{4DEFMetricScaling}
    g_{\mu\nu} \to \zeta^2 g_{\mu\nu} \,.
\ee
The kinetic energy of any other bulk fields -- such as for $\sigma$ itself or KK modes and any of their superpartners -- has the same scaling and so appears in the low-energy EFT without any prefactors of $e^{\sigma}$ in 4D Einstein frame.

These expressions show that the 4D and 6D Einstein-frame metrics are related by $g_{\mu\nu} = e^{\sigma} \, \tilde g_{\mu\nu}$, where the scaling relation $\tilde g_{\ssM\ssN} \to \zeta \, \tilde g_{\ssM\ssN}$ for the 6D metric $\tilde g_{\mu\nu}$ is as in \pref{6DScalingSym}. Comparing this to the direct dimensional reduction from 6D to 4D shows how 
\be
   e^{\sigma} \sim (M_6 L)^{2} 
\ee
is related to the extra-dimensional size, showing that large $\sigma$ incorporates the physics of large extra dimensions. If $L$ is taken as large as it can possibly be without running afoul of experiments (say $1/L \sim 0.1$ eV) and taking $M_6 \sim 100$ TeV gives $M_6 L \sim 10^{15}$ and so one can see how values as large $e^\sigma \sim (M_6 L)^2 \sim 10^{30}$ can arise.

The $\sigma$-dependence of brane-localized interactions can be obtained in a similar way. In 6D the brane tension and the localized flux terms given in \pref{SSUSYb} both scale in the same way as does the Einstein-Hilbert term, and so once written in 4D Einstein frame these must contribute to the 4D EFT in the form
\be
    \cL_{\rm pot} = - \sqrt{-g} \; e^{-2\sigma} \, U(\theta) \,,
\ee
where $\theta$ denotes any other dimensionless scalar fields. Notice the similarity with the first few terms of the lagrangian \pref{LeffScale}. 

In what follows it is useful to write $\{\theta^i\} = \{ \vartheta^u, \psi^a \}$, to keep separate track of scalars $\psi^a$ that live on the brane and those $\vartheta^u$ (including $\sigma$) that live in the bulk. This is useful because the kinetic energies for each type depend differently on $\sigma$. The kinetic energy of a bulk scalar is independent of $\sigma$ in 4D Einstein frame for the same reason no $\sigma$'s appear in the Einstein-Hilbert action in this frame. But dimensionless brane-localized scalar fields $\psi^a$ have kinetic energies of the form $\cL_{\rm bkin} \propto \sqrt{-\tilde g_4} \, \tilde g^{\mu\nu} \partial_\mu \psi\, \partial_\nu\psi$ and so $\cL_{\rm bkin} \to \zeta \, \cL_{\rm bkin}$ scales differently than other terms like the brane tension or the bulk action. In 4D Einstein frame the corresponding term in the low-energy 4D EFT must therefore have the form $\cL_{\rm bkin} \propto \sqrt{-g} \, e^{-\sigma} g^{\mu\nu} \partial_\mu \psi \partial_\nu\psi$.

These arguments lead to a lagrangian of the form \pref{LeffScale}, 
\be \label{LeffScale2}
   \cL_{\rm eff} = - \sqrt{-g} \Bigl[ M_p^4 e^{-2\sigma} U(\vartheta, \psi) + \tfrac12 M_p^2 R + \tfrac12 M_p^2 Z_{uv}(\vartheta) \, \partial \vartheta^u \, \partial \vartheta^v+ \tfrac12 M_p^2 e^{-\sigma} \, G_{ab}(\vartheta,\psi) \, \partial \psi^a \partial \psi^b + \cdots \Bigr] \,,
\ee
and its corrections. It is the Planck scale that naturally sets the dimensions in all of these terms, with other low-energy scales arising as a consequence of our currently living in the large-$\sigma$ regime. 

\subsection*{Electroweak and neutrino hierarchies}
\label{ssec:EWNuHierarchy}

It is worth asking whether a choice of $\sigma$ exists that is consistent with the hierarchies we see in nature ({\it e.g.}~explains why the electroweak scale is much smaller than $M_p$). If one of the $\psi$ fields is the Standard Model Higgs then the mass predicted for it by a $\psi^2$ term within $U(\vartheta,\psi)$ in \pref{LeffScale2} defines the electroweak scale and is of order $M_\EW \sim M_p \, e^{-\sigma/2}$. By contrast, the mass predicted for a bulk scalar by a $\vartheta^2$ term within $U(\vartheta,\psi)$ is instead of order $m_\ssB \sim M_p \, e^{-\sigma} \sim M_\EW^2/M_p$. These broadly reproduce the predictions of large extra-dimension models when $e^{\sigma/2} \sim (M_6 L) \sim 10^{14}$, with $M_\EW \sim 10$ TeV and $m_\ssB \sim 0.1$ eV being of order the KK scale.
 
Having a $\sigma$-dependence to the mass term for the Standard Model Higgs boson also implies the same $\sigma$-dependence for the Higgs expectation value, and this in turn implies the same sigma-dependence appears in the masses of all other Standard Model particles since these are all linear in the Higgs vev (so their mass {\it ratios} remain $\sigma$-independent). Whether this is a problem again requires appealing to whether a successful cosmology can be built with these choices  (though having mass ratios be field independent helps evade constraints from tests of the equivalence principle).  

Interestingly enough, the only ordinary particles not to have masses proportional to $M_\EW$ are neutrinos, which famously \cite{Weinberg:1979sa} can acquire mass through a dimension-5 effective interaction that is quadratic in the Higgs vev. If  neutrinos acquire masses in this way (or by mixing with a KK fermion in the bulk\footnote{Extra-dimensional neutrino mixing models rely on there being very light fermions in the bulk but rarely tell you why these should be present. Supersymmetric extra dimensions provide a robust answer: they are part of a generic Dark Sector. They are present and light because they are superpartners for the graviton \cite{SLEDnu}.} \cite{LEDnu}) they would be expected to have masses of size $m_\nu \sim M_p \, e^{-\sigma} \sim M_\EW^2/M_p \sim m_\ssB \sim 0.1$ eV (up to dimensionless couplings). In this framework the same value for $\sigma$ can account for both the electroweak and neutrino mass scales in a unified way. This is not entirely a surprise given that the same consistency also happens in the underlying SLED models \cite{MSLED}.

Finally, the overall size of the scalar potential is $v_{\rm eff}^4 U$ where $v_{\rm eff} \sim M_p e^{-\sigma/2} \sim M_\EW$, and so the scalar potential is technically natural but not particularly small. Although the lagrangian \pref{LeffScale2} captures the correct scaling of the brane tension it does not yet contain the extra-dimensional physics that allows the 4D solutions to remain flat in the presence of a weak-scale brane tension. What remains missing in \pref{LeffScale2} is the extra-dimensional relaxation wherein the geometry of the extra dimensions adjust and back-react to the properties of the branes.

\subsection{Natural relaxation}
\label{ssec:NaturalRelaxation}

How does extra-dimensional relaxation get communicated to the low-energy effective 4D EFT?  The idea is to `integrate in' a few of the extra-dimensional moduli whose mass (the KK scale) is high enough that they would normally not be kept within a Wilsonian action at energies low enough to be described by a 4D field theory. This is a useful artifice if one is trying to follow how quantum fluctuations involving still heavier fields (like those of SM particles) are suppressed.  

It is here that supersymmetry (of the dark sector only, as described above) might play a role. Supersymmetry is important because it provides reasons why the potential $U_{\rm tree}$ can be a perfect square (or at least approximately so), as it is for example in the lagrangian \pref{Lsusy} where $V = |\partial W/\partial \Phi|^2$. If $U_{\rm tree}$ is a perfect square then it is non-negative definite and so any places where it vanishes is necessarily a minimum. When this is true other fields in the problem will tend to seek out $U_{\rm tree} = 0$ as they minimize their energy, perhaps explaining by doing so why the Dark Energy density is so low.

This kind of approach was explored in \cite{Burgess:2021obw}, where the implications for the scaling symmetry \pref{scaling0} was explored within a framework wherein the dark sector is supersymmetric and the particle-physics sector is not. As discussed in \S\ref{sssec:SUSY}, this is a fairly generic situation in supersymmetric theories given that a gravitationally coupled sector couples to everything -- and so in particular to supersymmetry breaking -- more weakly than other sectors \cite{Burgess:2021juk}, and is realized in extra-dimensional scenarios when a supersymmetry breaking brane is localized within an otherwise supersymmetric bulk (as in SLED models \cite{SLED}). Assuming all superpartners of Standard Model particles are heavy enough to be integrated out at presently accessible energies (below 10 TeV or so) the effective theory needed to analyze this kind of situation requires coupling supergravity to non-supersymmetric matter, which is (happily) a solved problem \cite{Nilpotent}.  

In this framework $\sigma$ is in the gravity sector which is approximately supersymmetric, and so it is partnered with another scalar field that we call an axion, $\mfa$, because it has an independent shift symmetry of the form \pref{shiftabelian}. These combine into a complex scalar $T = \frac12( \tau + i \mfa)$ (called the axio-dilaton) that transforms in the standard way for a chiral multiplet under supersymmetry, where we define $\tau := e^{\sigma/2}$. 

For afficionados the model is as usual specified by a K\"ahler function $K(T,T^*,X,X^*, \Psi, \Psi^*)$ and a superpotential $W(T,X,\Psi)$ where $X$ is a nilpotent field describing the Goldstone fermion for supersymmetry breaking and $\Psi$ generically denotes all other ({\it i.e.}~Standard Model) fields. Approximate invariance under \pref{scaling0} is implemented consistent with supersymmetry by demanding
\be \label{NLRSUSY}
   e^{K/3} = T+T^* + A(X,X^*,\Psi,\Psi^*) + \frac{B(X,X^*,\Psi,\Psi^*)}{T+T^*} + \cdots \,,
\ee
where $A$ and $B$ are arbitrary functions and the ellipses denote higher powers of $1/(T+T^*)$.

The kinetic energy for this pair of scalars implied by the $T+T^*$ term in \pref{NLRSUSY} is 
\be
   \cL_{\rm kin} = - 3 M_p^2 \sqrt{-g} \,   g^{\mu\nu} \, \frac{\partial_\mu T^* \partial_\nu T}{(T+T^*)^2} = - \frac{3 M_p^2}{4\tau^2} \sqrt{-g} \,  \Bigl[ (\partial \tau)^2 + (\partial \mfa)^2 \Bigr]
\ee
up to corrections that are suppressed by additional powers of $1/\tau$. If we read off the axion decay constant as the coefficient of its kinetic term -- $\cL_{\rm kin} = - \frac12 F^2 \sqrt{-g} \,(\partial \mfa)^2$ -- (as is often done in the particle literature) it would be very small: $F \sim  M_p/\tau \sim M_p e^{-\sigma/2} \sim 0.1$ eV. The error in doing so is the assumption that $\sigma$ does not also appear in any coupling terms, like $\sqrt{-g} \, \partial_\mu \mfa J^\mu$. However both the coupling and the kinetic term share the same metric factor $\sqrt{-g} \, g^{\mu\nu}$ and so scale the same way under metric rescalings. The $\sigma$-dependence of the physical decay constant is the {\it relative} power of $e^{-\sigma}$ appearing between the kinetic and interaction terms, which ultimately depends on how $J_\mu$ scales under \pref{scaling0} and in which term in \pref{dilatonloopexpansion} this coupling appears. If, for example, $J_\mu$ scales in the same way as does $\partial_\mu \mfa$ then $F \sim M_p$ would be $\sigma$-independent (and large).\footnote{This is what actually happens in extra-dimensional UV completions in which $\mfa$ is a KK mode of the higher-dimensional metric supermultiplet and so couples with gravitational strength \cite{Brax:2022vlf}.}

The scalar potential for this class of models indeed has the form expected from \pref{dilatonloopexpansion}, which in the Einstein frame becomes 
\be \label{dilatonloopexpansion22}
   V_{\rm eff} =M_p^4 \left[ \frac{U_0}{\tau^2} + \frac{U_{ 1/2}}{\tau^3} + \frac{U_1}{\tau^4} +  \cdots \right]   \,,
\ee
with the additional information that
\be \label{U0Uhalf}
   U_0 = |w_\ssX|^2 \qquad \hbox{and} \qquad
   U_{1/2} \propto w_\ssX \,,
\ee
where $w_\ssX$ is a function of the other scalars in the problem, related to $\partial W/\partial X$. The coefficients $U_{1/2}$ and $U_1$ are calculable in terms of $W$, $A$ and $B$, but can be fairly arbitrary for the present purposes. The coefficient $U_0$ is a perfect square because it is basically an auxiliary field for supersymmetry (which is also the reason the potential in \pref{Lsusy} is a perfect square).

Now comes the main thought: because $U_0 = |w_\ssX|^2$ is a perfect square it wants to be minimized at zero, though the other terms $U_{1/2}$, $U_1$ and so on can obstruct the potential vanishing perfectly. Imagine then that the fields collectively denoted $\Psi$ above contain one non-Standard Model field, $\phi$, whose role is to seek out this minimum.\footnote{This relaxation field $\phi$ very naturally also can play another role as the inflaton in the very early universe \cite{Burgess:2022nbx, Cicoli:2024bwq}, but this is another story.} All we need assume is that $\phi$ appears in all of the $U_i$'s and all of their derivatives are order unity. In this case if only the first term of \pref{dilatonloopexpansion22} were present then minimizing $V_{\rm eff}$ would lead to $\phi = \phi_0$ where $U_0(\phi_0) = 0$. 

But the other terms {\it are} present, but subdominant in our regime of interest, where $\tau \gg 1$. This means instead we find the real minimum occurs when $\phi = \ol \phi$ where $\ol\phi - \phi_0 = \cO(1/\tau)$ and so $w_\ssX(\ol\phi) \sim \cO(1/\tau)$. Evaluated at this minimum, \pref{U0Uhalf} implies the first three terms of \pref{dilatonloopexpansion22} are all order $1/\tau^4$ and so the Dark Energy density is predicted to be order
\be \label{YogaPotential}
   V_{\rm min} = V_{\rm eff}(\ol\phi) \simeq \frac{\ol U M_p^4 }{\tau^4} = M_p^4 \, \ol U \, e^{-2\sigma} \,.
\ee
This is actually a very interesting size when written in terms of the electroweak scale, which the size of $\sigma$ was chosen to explain relative the Planck mass: $m_\EW \sim M_p \, e^{-\sigma/4}$. In terms of this the above potential minimum is 
\be \label{VminMagic}
  V_{\rm min} \sim \left( \frac{M_\EW^2}{M_p} \right)^4 \,,
\ee
which is in the right ballpark to describe Dark Energy given that $M_\EW^2/M_p \sim 0.1$ eV. 

Suggestive as this is, there are a great many things that must go right for this to be a full solution and this remains a work in progress. Here are some of the things we know so far:
\begin{itemize}
\item It is one thing to want a large value for a field like $\tau$ but if we can compute its potential we should also be able to compute its size.  If the potential for $\sigma$ is exactly as given by \pref{YogaPotential} then there is no minimum for any finite value of $\sigma$, so the present-day value of $\sigma$ is a function of the initial conditions in cosmology (whose explanation requires a theory of the earlier epochs of cosmology such as from inflation). But it is also possible that \pref{YogaPotential} is only approximate and the corrections introduce a minimum for $\sigma$. In this case its late-time value can be computed by minimizing the potential. 

A simple situation that would generate one \cite{Burgess:2021obw, Burgess:2022nbx} builds on the fact that in eq.~\pref{YogaPotential} the function $\ol U$ can acquire a weak dependence on $\sigma \sim \ln\tau$. This can happen because loop effects generically introduce logarithms of particle mass ratios everywhere and in these models particle masses in turn depend on $\tau$. So if the two particles whose masses appear in the ratio depend differently on $\sigma$ then a dependence on $\ln(m_1/m_2)$ turns into a dependence like a polynomial dependence on $\ln\tau$. 

For instance if $\ol U$ were to be a quadratic function, $a + b\ln \tau + c \ln^2\tau$, then the potential $V_{\rm min}$ can easily have a local minimum. Even better, to have this minimum give $\sigma \sim 60$ -- and so also $\tau^{1/4} \sim 10^{14}$, as required for the electroweak hierarchy -- requires only that the coefficients $a$, $b$ and $c$ in the quadratic function are themselves of order 50 or so. 
\item It is a bit of a cheat to compare the vacuum energy to the electroweak scale -- as done in \pref{VminMagic} -- since the size of $w_\ssX$ is actually dictated by the scale of supersymmetry breaking {\it in the Standard Model sector}, which cannot be smaller than $\cF \gsim (10\, \hbox{TeV})^2$. Consistency requires this to be larger than electroweak scales, since these particular superpartners were regarded as being already integrated out. Ref.~\cite{Burgess:2021obw} explores this constraint in more detail and shows that the lower bound on the size of the supersymmetry breaking scale (in the ordinary particle sector) puts a lower bound on $V_{\rm min}$ that is of order 
\be \label{VminEst}
    V_{\rm min} \gsim \frac{\epsilon^5 \cF^* \cF}{\ol\tau}  \,,
\ee
where $\epsilon \sim 1/\ln\ol\tau$ and $\ol\tau \sim 10^{28}$ is the vacuum value of $\tau$ chosen above to achieve the electroweak hierarchy. For $\ol\tau \sim 10^{28}$ (as required to reproduce the proper electroweak hierarchy) we have $(\ol\tau \ln^5\ol\tau)^{-1} \sim 10^{-37}$ and so if $\sqrt{|\cF|} \gsim 10$ TeV this gives $V_{\rm min} \gsim 10^{-93} M_p^4$. Although not as small as the value $10^{-120} M_p^4$ required for Dark Energy, this is better than any of the alternatives on the market (and is the result `out-of-the-box' inasmuch as the various inputs have not yet been seriously optimized to try to achieve the smallest possible result).
\item There is a good reason these parameters have not yet been optimized. Once the potential minimum falls below around $V_{\rm min} = v_{\rm eff}^4 \sim 10^{-80} M_p^4$ the mass of the $\sigma$ field around any minimum becomes less than of order $m_\sigma \sim v_{\rm eff}^2/M_p \sim 10^{-40} M_p \sim 10^{-13}$ eV and so the $\sigma$ Compton wavelength is longer than $m_\sigma^{-1} \sim 10^{6}$ m. In this regime $\sigma$ mediates long-range forces that can show up as deviations from GR in precision tests of gravity.  This is a generic problem for {\it any} successful proposal that gives a technically natural Dark Energy, and is a serious one. There are ways to evade such bounds, such as if macroscopic collections of atoms (like planets or stars) should couple to $\sigma$ much more weakly than would be guessed by summing the coupling strength atom-by-atom. This can happen for nonlinear couplings (and is generically called `screening' -- see {\it e.g.}~\cite{Khoury:2003aq}) but the jury is out so far on whether it can be done successfully in this case (see \cite{Brax:2022vlf, YogaScreening}). 
\item If the potential depends on other fields (such as the Higgs field, $h$, or an axion field, $\mfa$) in addition to the relaxation of the field $\phi$, then relaxation will happen locally for each value of the other fields $h$ and $\mfa$, so $\ol \phi = \ol \phi(h, \mfa)$. But this also suppresses their contribution to the energy, giving the overall potential a trough-like shape whose bottom is parameterized by $V[h,\mfa,\ol\phi(h,\mfa)]$: what minimizes a constant vacuum energy also tries to flatten the entire scalar potential for these other fields. One might (correctly) worry that this should in particular make their masses much smaller than naively expected, which at face value is a problem (at least for the higgs) whose mass is actually measured (and the predictions for this were right before relaxation).

The reason this need not actually be a problem is the relaxation is dynamical and so responds differently depending on the speed of the probe. Rapid processes like higgs particle collisions or decays occur effectively instantaneously and so $\phi$ has no time to respond. So these processes are in the `sudden' approximation and so tend just to see the curvature of the `bare' potential in the direction of the probe. This gives the mass without relaxation (as is usually assumed when computing {\it e.g.}~collider signals). But for slow processes like cosmology the evolution of $\phi$ is instead adiabatic and so has time to adapt as other fields change, leading the evolution to preferentially explore the bottom of the trough (where masses really are much smaller than their naive values).
\item Successful suppression of the vacuum energy inevitably implies the mass of the dilaton field $\sigma$ is of order the current Hubble scale, ensuring that the Dark Energy is not constant (an observation that for extra-dimensional models predates the precise formulation of SLED models \cite{Albrecht:2001xt}). Although its mass is protected by symmetries, they are not the shift symmetries usually considered \cite{naturallylight}, and its dynamics is complicated in important ways by its interactions with its axion partner.\footnote{Recall that the power-counting arguments of \S\ref{ssec:EFTs} show that it is two-derivative scalar self-interactions that like to compete at low-energies with the two-derivative interactions of GR, and that these interactions happen not to arise for single-field models. This makes exploring the low-energy implications of models with two or more fields -- such as the axio-dilaton -- interesting for tests of gravity in its own right, independent of their success or failure with the cosmological constant problem.} So far the preliminary indications continue to look good and work is underway to see whether this can persist, leading to a detailed working model. Initial indications are that cosmology can be very interesting \cite{Burgess:2021obw, Axio-Cosmo} and there is a tantalizing prospect of it pointing to a unified picture of the origins of both Dark Energy and Dark Matter \cite{Smith:2024ibv}. It is a generic feature of these models that all particle masses are field dependent and this introduces opportunities for both success and failure. In particular, the generic interactions between Dark Energy and Dark Matter to which this leads causes the Dark Matter not to evolve as simply as it does in the vanilla $\Lambda$CDM case, and this allows these cosmologies to in principle accommodate having a Dark Energy equation of state parameter with $w < -1$. It is not that the real equation of state parameter is in this range, it is just that the value inferred by observers appears to be in this range if they make the mistaken assumption that the Dark Matter evolves as it does in $\Lambda$CDM (which they inevitably do). It is the field-dependence of Standard Model particles that leads to many dangerous constraints \cite{Baryakhtar:2024rky}, though these seem survivable provided that the dilaton couplings can be made small enough also to evade solar system tests of GR.
\end{itemize}
Long-story short: this framework is promising but there are a lot of working parts that must be pinned down in order to claim real progress on the cosmological constant problem. What is interesting is that there are often superficial objections (like the ones listed above) that seem to be problems but which disappear on their own when examined more carefully. 

In my own mind the main worry is whether the phenomenology of having ordinary particle masses depend on the values of very light scalars can be ruled out based on what we know, but this is itself progress inasmuch as we trade the cosmological constant problem (which is very hard) for possibly much easier phenomenological issues to do with tests of gravity. Perhaps this is really an opportunity; if this class of models is how Nature works, it provides many observable consequences that perhaps are about to be discovered.

\section{Summary}
\label{ssec:Summary}

It is a remarkable opportunity that the long-distance physics we see in the sky seems to depend on how things work at much smaller distances; an opportunity that it behooves us to exploit. When this clue is ignored we have so many theoretical options that cosmological observations alone are unlikely to narrow our choices down sufficiently. Once this clue is included -- for the cosmological constant problem specifically -- then so far no compelling options have yet emerged at all. This shows that reconciling cosmology with high-energy physics is difficult. Should it be accomplished successfully we are likely to find an important part of how nature actually works.

These lectures have tried to make the following points.

\begin{enumerate}
\item Technical naturalness matters and is a natural consequence of the modern understanding of how classical gravitational physics fits into a broader quantum picture. Effective field theory is the key concept, designed to capture the important physics relevant at low energies when there is a large hierarchy of energy scales. 

Technical naturalness emerges as a criterion because there can be a different effective theory for every new range of scales, it should be possible to ask why a parameter is small at any scale we choose. This has two parts: why is the parameter small in the ultraviolet-complete theory at very high energies, and why does it stay small as one integrates out the lower-energy modes. Although we may not understand the answer of the first question until we get access to very high energies, the second part has implications even at low energies (and this is what makes the criterion of technical naturalness useful).  
\item Although a technically natural understanding of the small size of the Dark Energy density has proven elusive, it is argued that it is too early to despair about solving the cosmological constant problem and the rewards for doing so are very high: any such a solution is likely to have a great many low-energy tests. 
\item Personally, my own money is on low-energy approximate scale invariance being responsible for the electroweak hierarchy within which a relaxation mechanism (perhaps along the lines of \cite{Burgess:2021obw}) accounts for the small size of $\rho_{\rm vac}$. This is likely also to point to the existence of supersymmetric large extra dimensions at accessibly low energies as a UV completion. Both of these require the existence of very light dilaton and a supersymmetric dark sector, with a host of potentially observable implications for cosmology and tests of gravity. 
\end{enumerate}

These lectures argue three things, in descending order of confidence. First, EFT methods are indispensible for cosmology since they are what underpin the validity of the classical approximation -- in practice the main tool in use -- for any theory involving gravity. They are ignored at our peril.

Second, (technical) naturalness provides a useful guideline that suggests fruitful questions to ask when seeking progress in cosmology, though in the end the proof of the pudding will be in the eating. Their value is in the ideas to which they lead, and in the quality of these ideas compared with alternatives.

Finally, the cosmological constant problem is a hopeful challenge that will lead to a narrow range of viable models and not the message of despair it is usually felt to be. The Universe is a Big Place, and this fact alone may well be telling us that new physics is just around the corner, since this is required by {\em any} real solution to the cosmological constant problem. The search so far has been hard and unsuccessful, but not all avenues have been exhaustively explored and the rewards with success are very high. 

With luck the interplay between cosmology, gravity and fundamental physics will soon teach us what is really going on in the sky.

\section*{Acknowledgements}
I would like to thank the organizers of this school for their kind invitation to present these lectures to such a talented group of students in such pleasant environs. Over a lifetime I have learned much from my mentors, students and collaborators about the opportunities offered by naturalness and cosmology. I thank F. Quevedo for helpful comments on an early draft. My research has been supported in part by the Natural Sciences and Engineering Research Council of Canada. Research at the Perimeter Institute is supported in part by the Government of Canada through Industry Canada, and by the Province of Ontario through the Ministry of Research and Information.

\end{document}